%% file: ArXivVersion.tex
\documentclass[useAMS,usenatbib]{mn2e}

\usepackage{epsfig}
\usepackage{amsmath}
\usepackage{amssymb}
\usepackage{natbib}
\usepackage{threeparttable} 
\usepackage{booktabs}
\usepackage{color}
\usepackage{multirow}
\usepackage{gensymb}
\usepackage{hyperref}

\usepackage{epstopdf}
\usepackage{times}

\def \nar {NewAR}
\def \aj {AJ}
\def \mnras {MNRAS}
\def \apj {ApJ}
\def \apjs {ApJS}
\def \apjl {ApJL}
\def \aap {A\&A}
\def \nat {Nature}
\def \araa {ARAA}

\def \pasj {PASJ}
\def \aaps {AAPS}

\def \iaucirc {IAU Circ.}
\def \ssr {Space~Sci.~Rev.}
\def \apss {Astrophysics~and~Space~Science}
\def \aapr {A\&ApR}
\def \physrep {Phys.~Rep.}

\title[A new radio census of accreting NSs]{A new radio census of neutron star X-ray binaries}

\author[van den Eijnden et al.]
{J. van den Eijnden$^{1,2}$\thanks{e-mail: jakob.vandeneijnden@physics.ox.ac.uk}, 
N. Degenaar$^{2}$, 
T. D. Russell$^{3,2}$, 
R. Wijnands$^2$, 
A. Bahramian$^{4}$, 
\newauthor J. C. A. Miller-Jones$^{4}$,
J. V. Hern\'andez Santisteban$^{5}$,
E. Gallo$^{6}$,
P. Atri$^{4}$,
\newauthor R. M. Plotkin$^{5}$, 
T. J. Maccarone$^{8}$,
G. Sivakoff$^{9}$,
J. M. Miller$^{6}$,
M. Reynolds$^{6}$,
\newauthor D. M. Russell$^{10}$,
D. Maitra$^{11}$,
C. O. Heinke$^{9}$,
M. Armas Padilla$^{12,13}$,
A. W. Shaw$^{7}$
\\
$^1$ Department of Physics, Astrophysics, University of Oxford, Denys Wilkinson Building, Keble Road, Oxford OX1 3RH, UK\\
$^2$ Anton Pannekoek Institute for Astronomy, University of Amsterdam, Science Park 904, 1098 XH, Amsterdam, the Netherlands\\
$^3$ INAF/IASF Palermo, via Ugo La Malfa 153, I-90146 Palermo, Italy \\
$^4$ International Centre for Radio Astronomy Research, Curtin University, GPO Box U1987, Perth, WA 6845, Australia\\
$^5$ SUPA Physics and Astronomy, University of St Andrews, KY16 9SS, Scotland, UK\\
$^6$ Department of Astronomy, University of Michigan, 1085 S University, Ann Arbor, MI 48109, USA\\
$^7$ Department of Physics, University of Nevada, Reno, 1664 N. Virginia Street
Reno, NV 89557, USA \\
$^8$ Department of Physics \& Astronomy, Box 41051, Science Building, Texas Tech University, Lubbock, TX 79409-1051, USA\\
$^9$ Department of Physics, CCIS 4-183, University of Alberta, Edmonton, AB, T6G 2E1, Canada\\
$^10$ Center for Astro, Particle and Planetary Physics, New York University Abu Dhabi, PO Box 129188, Abu Dhabi, UAE\\
$^{11}$ Department of Physics and Astronomy, Wheaton College, Norton, MA 02766, USA\\
$^{12}$ Instituto de Astrof\'isica de Canarias, 38205 La Laguna, Tenerife, Spain\\
$^{13}$ Departamento de Astrof\'\i{}sica, Universidad de La Laguna, E-38206 La Laguna, Tenerife, Spain\\
}

\begin{document}

\date{Accepted XXX. Received YYY; in original form ZZZ}

\pagerange{\pageref{firstpage}--\pageref{lastpage}} \pubyear{2019}

\maketitle

\label{firstpage}

\begin{abstract}
We report new radio observations of a sample of thirty-six neutron star (NS) X-ray binaries, more than doubling the sample in the literature observed at current-day sensitivities. These sources include thirteen weakly-magnetised ($B<10^{10}$ G) and twenty-three strongly-magnetised ($B\geq10^{10}$ G) NSs. Sixteen of the latter category reside in high-mass X-ray binaries, of which only two systems were radio-detected previously. We detect four weakly and nine strongly-magnetised NSs; the latter are systematically radio fainter than the former and do not exceed $L_R \approx 3\times10^{28}$ erg/s. In turn, we confirm the earlier finding that the weakly-magnetized NSs are typically radio fainter than accreting stellar-mass black holes. While an unambiguous identification of the origin of radio emission in high-mass X-ray binaries is challenging, we find that in all but two detected sources (Vela X-1 and 4U 1700-37) the radio emission appears more likely attributable to a jet than the donor star wind. The strongly-magnetised NS sample does not reveal a global correlation between X-ray and radio luminosity, which may be a result of sensitivity limits. Furthermore, we discuss the effect of NS spin and magnetic field on radio luminosity and jet power in our sample. No current model can account for all observed properties, necessitating the development and refinement of NS jet models to include magnetic field strengths up to $10^{13}$ G. Finally, we discuss jet quenching in soft states of NS low-mass X-ray binaries, the radio non-detections of all observed very-faint X-ray binaries in our sample, and future radio campaigns of accreting NSs.
\end{abstract}



\section{Introduction}
\label{sec:introduction}

Accretion is a fundamental process occurring across the Universe in a plethora of objects and circumstances. These range from accreting supermassive black holes in Active Galactic Nuclei (AGN) and accreting white dwarfs, via stellar mass black holes and neutron stars in X-ray binary systems, to forming stars surrounded by proto-planetary discs. All these systems show states where the accretion of matter is observed to be accompanied by a coupled outflow of material, either in the form of wide-angled, relatively slow winds, and/or collimated and often relativistic jets. These outflows influence both the accreting systems, for instance contributing to the angular momentum loss in the accretion flow and reducing the effective accretion rate \citep[e.g.][]{tetarenko18_winds}, and the surrounding medium \citep[e.g.][]{gallo05,fabian12}. Feedback from X-ray binaries and AGN is thought to contribute to the stellar feedback regulating star formation and ionising the early Universe \citep[e.g.][]{fender05,mirabel11,justham12,fragos13,fragos12}.

The formation of jets from an accretion flow is often fundamentally attributed to either one or a combination of two mechanisms. In both those models, twisted magnetic field lines close to the compact object launch material away from the accretion flow, but an important difference lies in the origin of the twisting of the magnetic field lines. In the \citet{blandford77} mechanism, these field lines are spun up as they thread the rotating ergosphere of a black hole. A key prediction of this model, that has remained difficult to test unambiguously, is the dependence of jet power on the black hole spin \citep[e.g.][]{king13,mcclintock14,russell13}. Alternatively, for the many jet-launching systems do not contain a black hole, the \citet{blandford82} mechanism proposes that the differential rotation of the accretion flow itself tangles up the magnetic field (note that, naturally, the \citet{blandford82} mechanism can also occur in black hole systems). While this model applies to neutron stars, it predicts an upper limit on the magnetic field of neutron stars that can be spun up by an accretion flow. Therefore, it predicts that neutron stars with magnetic fields above a certain threshold should not launch jets \citep[e.g.][]{massi08,migliari11c}.

While the accretion flow in AGN and X-ray binaries typically emits strongly in the X-ray band, the jet dominates at low frequencies through the emission of synchrotron radiation. This emission results from free electrons in the jet spiralling around magnetic field lines, producing a synchroton spectrum \citep{rybicki79,longair92}. The observed spectrum of an unresolved jet depends on the jet type: discrete ejecta typically have steep spectra in the radio band, defined as radio spectral index $\alpha \approx -0.7$ (where $S_{\nu} \propto \nu^{\alpha}$) as they emit as a single, optically thin population. A compact, steadily-outflowing jet is instead observed as the superposition of synchrotron spectra from different distances downstream, resulting in a flat ($\alpha = 0$) to inverted ($\alpha > 0$) spectral shape up to the jet break frequency. The highest frequency jet emission originates from the highest energy electrons, located near the base of the jet \citep{markoff01,corbel02,markoff05,romero17,malzac13,malzac14}, while lower frequencies are emitted further down the jet \citep[e.g.][]{blandford79}. The jet synchrotron emission can extend into the sub-mm \citep{russell14,tetarenko15,diaztrigo18}, nIR, and optical \citep{russellD06,russellD07,russell13,russell13b,gandhi17,baglio2018}, and might contribute up to the X-ray band via synchrotron-self-Compton emission \citep[e.g.][]{markoff05}. With few confusing radiative processes and many sensitive observatories, the radio band is particularly suitable for jet studies.

Accreting black hole systems in their hard spectral state show a correlation between their X-ray and radio luminosity, that holds over orders of magnitudes in black hole mass from X-ray binaries to AGN, after incorporating a mass normalization term (the fundamental plane of black hole activity) and indicates a coupling between the inflow and outflow of matter
\citep{hannikainen98,corbel00,corbel03,falcke04,merloni03,gallo03,plotkin13,gallo2014}. While this sample of stellar-mass black holes is dominated by binaries with a low-mass ($\lesssim 1 M_{\odot}$) donor -- the \textit{low-mass X-ray binaries} (LMXBs) -- it includes two \textit{high mass X-ray binaries} (HMXBs) hosting black holes (Cyg X-1 and MWC 656; the candidate black hole HMXB Cyg X-3 is not included). Together, these sources follow as similar correlation between X-ray and radio luminosity, suggesting that this coupling might be independent of mass transfer / donor type for black hole systems \citep{ribo17}. Looking more closely at the X-ray -- radio correlation for stellar-mass black holes, there is evidence for both a radio-loud and radio-quiet track \citep[][]{soleri11,dincer14,meyer14,drappeau15}. Despite several possible explanations, including inclination \citep{motta18}, variable jet Lorentz factors \citep{soleri11,russellT15}, and X-ray \citep{koljonen19} and radio \citep{espinasse18} spectral shapes, these tracks remain not fully understood. Moreover, the statistical evidence of their existence remains debated \citep{gallo18}. The correlation between X-ray and radio luminosity for black hole LMXBs disappears as the system transitions via the intermediate states into the soft state: during this transition, the compact jet quenches while fast ejecta can be launched \citep{fender04,russell2020}; any remaining radio emission during the soft state is typically attributed to those ejecta, either unresolved or tracked as they move away from the LMXB \citep[e.g.][]{bright2020}.

\subsection{A brief history of neutron star jet observations}

The story is more complicated for neutron stars. Weakly-magnetized neutron stars accreting above $\sim 1$\% of the Eddington luminosity ($L_{\rm Edd}$) can be divided into two classes based on their tracks in the X-ray color-color diagram: the Z and atoll sources \citep{hasinger89}. The difference between these source classes and their various sub-classes is thought to be driven by instantaneous mass accretion rate \citep{lin09,homan10}. Z sources, accreting around the Eddington limit, are the radio brightest class of accreting neutron stars and their jets were therefore characterised first \citep{ables69,lampton71,penninx88,penninx89,hjellming1990_scox1,hjellming1990_cygx2}. These studies found different jet types along the different branches in the X-ray color-color diagram, changing in tandem with changes in accretion flow properties: from a compact jet to discrete ejecta and finally quenching at the highest mass accretion rates \citep[qualitatively similar to the black hole behaviour detailed above;][]{migliari06}. Jet studies of the X-ray and radio fainter atolls came later, with \citet{migliari03} presenting the first multi-epoch X-ray and radio campaign for such a source (4U 1728–34). 

Using the enhanced sensitivity of current day radio telescopes, neutron star LMXBs have now been studied extensively down to $L_X \sim 10^{36}$ erg/s (i.e. $\sim 1$\% of the Eddington limit for a $1.4M_{\odot}$ neutron star). Observations at lower X-ray luminosities are dominated by radio non-detections \citep{tudor17,gallo18,gusinskaia20}, although some neutron stars have been detected down in this regime as well \citep[e.g. SAX J1808.4-3658 and IGR J00291+5934, down to $\sim 2\times10^{34}$ erg/s;][]{tudor17}. The behaviour of neutron star jets at low mass acccretion rates remains poorly explored. Another open question, for jets in atolls, regards the presence and mechanism of the jet quenching seen in black hole systems \citep{fender16b}. Atolls can change (although not all do) between thermal- and Comptonisation-dominated accretion flow states: their soft and hard states, respectively. Several sources show a quenched jet in their soft spectral state \citep{migliari03,millerjones10,gusinskaia17,diaztrigo18,gusinskaia20}, as seen in  black holes \citep[e.g.][]{fender04}. Others, however, do not \citep{rutledge98,migliari04}. Finally, coordinated X-ray and radio studies have often focused on transient neutron star LMXBs, in order to probe different accretion rates -- leaving persistently accreting sources more poorly explored.

The first comprehensive investigations of the radio properties of accreting neutron stars were presented by \citet{fender00} and \citet{migliari06}. Since then, many individual accreting neutron stars have been added to the X-ray -- radio luminosity plane; see the compilations by \citet{tetarenko2016_watchdog} and \citet{gallo18}, and, for the most up to date database, Bahramian et al. (2018)\footnote{\href{https://github.com/bersavosh/XRB-LrLx_pub}{https://github.com/bersavosh/XRB-LrLx\_pub}}. All of these sources -- both atoll and Z -- contain weakly-magnetised neutron stars (e.g. $B < 10^{10}$ G), where the \citet{blandford82} mechanism can be applied. As a sample, these weakly-magnetised neutron stars are radio-fainter by a factor $\sim 20$ than stellar-mass black holes -- a difference that cannot simply be accounted for by the difference in accretor mass, bolometric X-ray corrections, or the presence of a boundary layer around the neutron star \citep{fender01,gallo18}. In the statistical analysis by \citet{gallo18}, these sources show a scatter similar to the black hole population, assuming the latter follow a single track.

The key difference between accreting black holes and neutron stars -- with respect to jet formation -- is the presence of a solid stellar surface in the latter. This comes with additional differences, such as anchored magnetic fields and a compact object spin that can be measured via pulsations for neutron stars. As shown by X-ray pulsations detected in a subset of accreting neutron stars, their magnetic fields can dynamically alter the geometry of the inner accretion flow, where jets are launched from. The neutron star magnetic fields can be measured directly through the detection of the cyclotron resonance scattering feature \citep[hereafter cyclotron lines; see][for a recent review]{staubert19}. Alternatively, and more indirectly, one can measure the inner disc radius and use that to constrain the magnetic field \citep{cackett2008_iron,degenaar17,ludlam2019}, or constrain the magnetic field strength from the relation between the spin evolution and mass accretion rate \citep[e.g.][]{ghosh1978,campana02,strohmayer2018}. Neutron star spins are measured directly through X-ray pulsations or nearly-coherent oscillations during thermonuclear bursts on their surface \citep{patruno12,staubert19}. 

Despite decades of radio studies, jets were until recently not detected in strongly-magnetised neutron stars. A number of sample studies between the 1970s and 2000s reported only non-detections \citep{duldig79,nelson88,fender00,migliari06,migliari11}, while the radio detection of the HMXB X-ray pulsar GX 301-2 by \citet{pestalozzi09} could be attributed to the radio emission from the stellar wind. The young neutron star X-ray binary, Cir X-1 \citep{heinz13}, is known to launch strong jets \citep{stewart93,fender98,tudose06,heinz07,soleri09,coriat2019}. However, despite claims of a strong magnetic field \citep[see, e.g.][]{schulz20}, its field strength has not been measured directly. The series of radio non-detections for strongly-magnetised neutron stars also formed the basis for the theoretical reasoning from \citet{massi08}, explaining why the \citet{blandford82} mechanism should not operate in this magnetic field regime. 

Recently, the first jet from a strongly-magnetised neutron star was observed, contrary to this theoretical expectation \citep{vandeneijnden2018_swj0243}. The slow (spin period exceeding $1$ second) X-ray pulsar Swift J0243.6+6124, that accretes from a Be star \citep{kouroubatzakis17}, launched a jet both during its so-called giant outburst in 2017/2018, and during X-ray re-brightenings in the outburst decay \citep{vandeneijnden2019_reb}. While this jet is not expected to be launched via the \citet{blandford82} mechanism, it remains unknown what alternative process could be responsible. Additionally, the strongly-magnetised neutron stars Her X-1 and GX 1+4 were detected in radio \citep{vandeneijnden2018_gx,vandeneijnden2018_her}, although the origin of this emission was not conclusively attributed to a jet. All three sources were detected at faint radio flux densities below $\sim 100$ $\mu$Jy, which explains the radio non-detections of this source class in earlier decades.

\subsection{An extended parameter space for neutron star jets}
\label{sec:introduction2}

The inclusion of strongly-magnetised neutron stars into the class of jet-launching sources, greatly expands the parameter space to study (neutron star) jets. Firstly, in addition to the larger range in neutron star magnetic field, a much greater spin range can now be accessed: while weakly-magnetised neutron stars show spins in the millisecond range \citep{patruno17}, their strongly-magnetised counterparts reach spins up to thousands of seconds \citep{staubert19}. Secondly, all confirmed HMXB neutron stars are, when measured, strongly magnetised\footnote{Note that the opposite is not true: not all strongly-magnetised neutron stars have high-mass companions. Instead, a handful of them either reside in LMXBs, or accrete from the stellar wind of wide-orbit, evolved low-mass stars in Symbiotic X-ray binaries (SyXRBs), or have an intermediate mass donor.}, with typical magnetic fields of the order $>10^{12}$ G. Therefore, these systems probe a much wider range of donor types, binary periods, eccentricities, and mass transfer mechanisms, than is accessible through only weakly-magnetised neutron stars. As the vast majority of HMXBs contains a neutron star, probing the effect of these binary and donor properties on jet formation was also barely possible with black hole systems.

At the same time, the origins of radio emission in HMXBs can be more complicated to untangle. Unresolved radio emission from a stellar-mass black hole or weakly-magnetised neutron star with a low mass donor is often automatically assumed to originate from a jet. In systems with a high-mass donor, the donor's stellar wind can also contribute to the radio emission. Studies of isolated O/B giants and Be-stars have shown them to be radio emitters \citep{lamers98,lamers98b,gudel02}, implying that the companion's wind cannot be ignored when interpreting the radio emission from HMXBs: depending on properties such as mass-loss rate, clumpiness, and density, it can show either thermal (Bremsstrahlung) or non-thermal (shock) radio emission \citep{wright75,blomme97,dougherty00}. These processes can lead to radio luminosities up to several times $10^{27}$ erg/s in C-band, where jets are often studied, depending on the exact wind properties (see Section \ref{sec:disc_wind_1} for more details and discussion). Estimating the required wind properties to explain a HMXB radio detection, and having (multiple) coordinated radio and X-ray observations, preferably with spectral or polarization constraints, are therefore important tools to distinguish the stellar winds from the jet. 

While a larger parameter space for jet studies is now accessible in terms of both neutron star magnetic field strength and spin frequency, it remains unclear what model can explain the existence of jets such as observed in Swift J0243.6+6124. Possible models, including jets powered by opened neutron star field lines \citep{parfrey16} or magnetic propellers \citep{romanova09}, typically predict a dependence of jet power on magnetic field and spin frequency. Searches for such dependencies have been performed earlier for samples of weakly-magnetised neutron stars \citep{migliari11}, or samples including both neutron stars and black holes \citep{king13}. However, these studies were typically inconclusive or unable to find strong evidence for these relations, partially due to the limited spin range covered by weakly-magnetised neutron stars \citep{patruno17}. With a sample of neutron stars including a sufficient number of strongly-magnetised, slowly spinning sources, such studies can be revisited and extended -- assuming, fundamentally, a common jet launching mechanism across the entire sample.

In this paper, we present a detailed study of a large set of new radio and X-ray observations of accreting neutron stars. Our sample includes 13 weakly-magnetised and 23 strongly-magnetised targets, more than doubling the total number of neutron stars in the radio/X-ray--luminosity plane observed at current radio sensitivities. Preliminary results from the weakly-magnetised sample were reported in \citet[][which included, in total, 41 neutron stars]{gallo18}, while here we provide the full data analysis and the most up-to-date results. With this large set of new observations, we present the first systematic study of differences between the jets of weakly- and strongly-magnetised neutron stars where sources from both categories are detected, and aim to observationally constrain the neutron star jet formation mechanism. 

\section{Observations and data analysis}
\label{sec:observations}

In this section, we will first introduce the sample of observed neutron star X-ray binaries and the observing campaigns in radio and X-rays. In Sections 2.2 and 2.3, we will give a general introduction into the reduction and analysis of the radio and X-ray observations, respectively. Given the large number of analysed sources and the wide variety in setup of both the radio and X-ray campaigns, we discuss all details per source in the Online Supplementary Materials.

\subsection{Targets and observing campaigns}
\label{sec:targets}

The sample of targets presented in this paper consists of 13 weak-magnetic field neutron stars and 23 strong-magnetic field neutron stars, where we define weak and strong magnetic fields as $B \leq 10^{10}$ G and $B > 10^{10}$ G, respectively. For most strong-magnetic field sources, the detection of a cyclotron line provides direct, robust magnetic field measurements \citep[e.g.][]{staubert19}. For the remaining targets in this category, from a combination of the neutron star spin (evolution) and comparisons with slow pulsars with measured field strengths, their strong magnetic fields have been inferred (although therefore no actual measurement is performed). For weakly-magnetised accreting neutron stars, on the other hand, direct magnetic field measurements are not available. Instead, indirect measurements based on reflection spectroscopy \citep{cackett2008_iron,degenaar17,vandeneijnden2018_igr,ludlam2019}, modeling of the evolution of the spin frequency \citep{patruno12c}, and magnetic propeller states \citep[e.g.][]{mukherjee15} imply typical magnetic fields between $\sim 10^7$ and $\sim 10^9$ G. In Table 2 of the Online Supplementary Materials, we list magnetic field measurements and estimates for all sources, alongside their spin and orbital period, when known. We want to stress again that these $B$-field measurements can be rather uncertain and affected by systematic effects -- hence our broad classification in weak and strong magnetic fields.

The observations of all but one of the weak-magnetic field sources – IGR J17379-3747 – were already included in the statistical analysis of neutron star LMXBs by \citet{gallo18}. However, that work only focused on the overall sample properties without focusing on individual systems. Moreover, it did not include details on the radio and X-ray data analysis, which are presented in the current paper. To the comparison literature sample, we have also added the recently published observations of two weakly-magnetised neutron stars: IGR J16597-3704 \citep{tetarenko18_igrj16597} and IGR J17591-2342 \citep{russell18,gusinskaia20_igrj17591}, for the latter assuming the $7.6$ kpc distance reported by \citet{kuiper20}. Given the novelty of radio detections of the strongly-magnetised neutron stars, in the studied sample we include the few recently published observations of such sources: the radio detections of GX 1+4 \citep{vandeneijnden2018_gx} and Her X-1 \citep{vandeneijnden2018_her}, and the multi-epoch monitoring of Swift J0243.6+6124 \citep{vandeneijnden2019_reb}.

An overview of all sources, divided based on neutron star magnetic field, is shown in Tables \ref{tab:sample} and \ref{tab:sample_strong}. These sources cover different, and overlapping, source classes: the thirteen weakly-magnetised neutron stars include (i) atolls; (ii) accreting millisecond X-ray pulsars (AMXPs), hosting a neutron star whose X-ray pulsations reveal a spin of several hundreds of Hz \citep{patruno17}; (iii) ultra-compact X-ray binaries (UCXBs), systems with an orbital period less than one hour; and (iv) very-faint X-ray binaries (VFXBs), where the neutron star persistently emits below $\sim 10^{36}$ erg/s \citep{wijnands16}. In the VFXB category, we include the quasi-persistent source XMMU J174716.1-281048, that was likely continuously active between 2003 and 2011 \citep[see e.g.][]{degenaar11} but was not detected in X-rays during the 2014 radio observations reported in this work. These four source classes combined tackle different, relatively unexplored science cases: the jet properties of neutron star LMXBs that are persistently accreting  (i and iii), are in the soft state (i), or are X-ray faint (ii, iii, and iv). We note, finally, that our source sample does not include any Z-sources: given their radio brightness, these sources have been studied in detail previously, and this work's approach of single or a small numbers of observations per source, would not significantly contribute to their understanding. Hence, we decided not to perform observations of such targets and, lacking new observational results, will not further discuss this source class in much detail in this work.

The strongly-magnetised neutron stars (Table \ref{tab:sample_strong}) include one candidate and two confirmed symbiotic X-ray binaries, where the neutron star accretes from the stellar wind of an evolved low-mass donor in a wide orbit; three LMXBs (of which one is a UCXB); one intermediate-mass X-ray binary (IMXB); and sixteen HMXBs. Based on the donor type, the latter are categorised either as Be/X-ray binaries (BeXRBs), or as Super-giant X-ray binaries (SgXBs). For an overview of the differences between these source classes, see \citet{reig11}. Finally, 3A 1239-599 is simply denoted as HMXB as it is unknown in what subcategory it falls. The sources in this class were mainly targeted to study the poorly-understood jet properties of strongly-magnetised neutron stars and explore the effect of other radio emission mechanisms, such as their stellar winds. The four new BeXRBs (i.e. all but Swift J0243.6+6124) were targeted specifically to probe their radio properties at very low accretion rates, close to or in their propeller regimes. 

The latter class also includes GRO J1744-28, known as the Bursting Pulsar: a LMXB where the neutron star has an intermediate spin frequency of 2.1 Hz \citep{cui97}, causing it to fall somewhat between the two source categories. Its magnetic field is likely lower than typically seen in strongly-magnetised systems (e.g. $B\geq 10^{12}$ G): using different methods, it has been claimed to lie between $2\times10^{10}$ and $7\times10^{11}$ G \citep{cui97,rappaport97,bildsten97,degenaar14,younes15}. For this paper, we include it in the strong-magnetic field class, although ultimately, this classification does not affect our conclusions significantly: as shown by the preliminary results published by \citet{russell17}, the Bursting Pulsar is not detected at radio frequencies, with a relatively unconstraining upper limit due its proximity to the Galactic centre. 

Where possible, we used parallax measurements from \textit{Gaia} Early Data Release 3 \citep{gaia2020} to constraint the distance. This was done for sources where (i) the detected \textit{Gaia} counterpart has (ii) a positive parallax measurement with (iii) a signal-to-noise-ratio larger than three. To assess the affect of choice of prior in converting the parallax into distance, we then compared inferred distances from the priors in \citet{atri2019} (for Galactic LMXBs), \citet{bailerjones18}, and \citet{bailerjones2020}. We found these to be consistent within errors and use the \citet{atri2019} prior in the analysis. For sources where \textit{Gaia} was not used, we searched the literature for distance measurements instead. Finally, we note that using literature (non-\textit{Gaia}) distances for all sources does not alter the main conclusions of this work.

\subsection{Overview of radio data analysis}
\label{sec:radioanalysis}

The radio observations of our neutron star sample were performed with the Karl G. Jansky Very Large Array (hereafter VLA) for most sources with declinations above $\sim -40\degree$, and the Australia Telescope Compact Array (ATCA) for the remaining Southern targets (negative declinations). ATCA was in the most extended 6-km configurations during all campaigns, while the VLA changed configuration between observations. All raw data sets are publicly available under VLA programme codes 13A-352, 14A-163, 17B-136, 17B-406, 17B-420, SD0134, 18A-456, and 18B-104, and ATCA programmes C3108, C3184, C3243, and CX379 (see table 1 in the Online Supplementary Materials).  

To flag, calibrate, and image the observations, we used the \textsc{Common Astronomy Software Application} \citep[\textsc{casa};][]{mcmullin07} package v4.7.2. We removed RFI using a combination of automatic flagging routines and visual inspection. We performed imaging using the multi-scale multi-frequency \textsc{clean} task, with a Briggs robust parameter adjusted to the target field in order to balance sensitivity and confusion. We then fitted an elliptical Gaussian with Full-Width Half Maxima equal to the synthesized beam size of the observation using the \textsc{casa}-task \textsc{imfit}. We measured the root-mean-square variability over a nearby region devoid of sources in case the target was detected, or over the target region for a non-detection. In the latter case, we set the $3\sigma$ upper limit to three times this RMS measurement. Target-specific details, such as VLA configuration, primary and secondary calibrator, and beamsizes, are listed per source in Table 1 of the Online Supplementary Materials.    

For a subset ofd neutron stars, data was recorded at two frequencies. In those cases, we calculated the radio spectral index $\alpha$, where $S_\nu \propto \nu^\alpha$, between the two bands. The error on $\alpha$ is estimated through a propagation of the uncertainties on the radio flux densities measured at each frequency, using Monte-Carlo simulations. In case radio emission is detected in only the lower frequency band (2 sources), we follow the Monte-Carlo approach of \citet{vandeneijnden2019_reb} to estimate an upper limit on the spectral index. For those sources, we show the diagnostic figures of this method in Section 4 of the Online Supplementary Materials. In this work, we will refer to negative and positive spectral indices as steep and inverted spectra, respectively. 

\subsection{Overview of X-ray data analysis}
\label{sec:xrayanalysis}

We measured unabsorbed X-ray fluxes of our targets using pointed observations of the \textit{Neil Gehrels Swift Observatory} \citep[\textit{Swift}][]{gehrels04} or monitoring observations with the \textit{Monitor of All-Sky X-ray Image/Gas Slit Camera} \citep[\textit{MAXI/GSC}][]{matsuoka09}. Details about the X-ray analysis per source, such as observations used and spectral fit parameters, can be found in the second section of the Online Supplementary Materials. We aimed to use observations taken on the same day as the radio observation; in the cases where such observations did not exist, we used the closest X-ray observation in time. In those cases, we used longer-term X-ray monitoring, combined with typical time-scales of state changes, to ensure that the source did not change its state or X-ray flux significantly around the radio observation. We preferentially used pointed \textit{Swift} X-ray Telescope observations, either measuring the flux directly from the spectrum, or for very faint sources converting the count rate or count rate upper limit into the flux. These \textit{Swift} analyses where permormed in the $0.5$--$10$ keV range. When no pointed observations were available, we measured the flux from the (multi-day) \textit{MAXI} spectrum (fitted between 2 and 10 keV) or converted the detected \textit{MAXI} 2-10 keV count rate into a flux.For all sources that are systematically undetected in \textit{MAXI}, we ensured that \textit{Swift} observations were performed. 

In order to convert either \textit{Swift} or \textit{MAXI} count rates into unabsorbed fluxes, we used the \textsc{webpimms} tool\footnote{\href{https://heasarc.gsfc.nasa.gov/cgi-bin/Tools/w3pimms/w3pimms.pl}{https://heasarc.gsfc.nasa.gov/cgi-bin/Tools/w3pimms/w3pimms.pl}}. When the source was in a state with a typical and known X-ray spectrum, we used the literature to model the spectrum used in the count rate conversion. Otherwise, we used the Crab conversion following the \textit{NuSTAR} measurement of the Crab spectrum by \citet{madsen17}. Note that, whether a spectral model was assumed/fitted (see below) or the Crab spectrum was used, we use the full flux and do not attempt to distinguish between different spectral components; see \citep{miller12} for a study where such effects are taking into account.

All measured fluxes were calculated in the $0.5$--$10$ keV range; hence, we note that we extrapolated the model fitted to the \textit{MAXI} spectra down to lower energies, thereby possibly introducing extra uncertainy in the flux measurement due to interstellar absorption. While differences exist in the shape of the X-ray spectrum between LMXBs and HMXBs, we used the same energy band to enable direct comparison between the source classes and with the literature. We fitted X-ray spectra using \textsc{xspec} v.12.10.1 \citep{arnaud96}, setting the ISM abundances and cross-sections to \citet{wilms00} and \citet{verner96}, respectively. As some spectra contain few counts, we used C-statistics to find the best fit \citep{cash1979}. All spectra were modelled with three models, combining interstellar absorption (\textsc{tbabs}) with power law or/and blackbody models: \textsc{tbabs*powerlaw}, \textsc{tbabs*bbodyrad}, and \textsc{tbabs*(powerlaw + bbodyrad)}. We picked the best-fitting model of the former two, based on the lowest test statistic, and then compared this with the latter, combined model using a f-test. We selected the combined model if the f-test preferred it at a $>5\sigma$ probability over the best-fitting single-component model. Finally, we measured the flux and its uncertainty by convoluting the selected model with \textsc{cflux}. While this approach is phenomenological, it suffices to measure the X-ray flux even for observations with low numbers of total counts.

For two radio observations, no X-ray information from either monitoring or pointed observations was available sufficiently close in time, compared to the source's typical time scale of variability and state transitions. These observations were the second radio observation of GX 1+4 and the second radio observation of 4U 1954+31. The former was detected in radio during this epoch, while the latter was not. Given the lack of X-ray data, we do not include these observations in the X-ray -- radio luminosity diagrams later in this paper.  

\section{Results}
\label{sec:results}

We list all sources, radio flux densities and X-ray fluxes in Tables \ref{tab:sample} (the 13 weakly-magnetised sources) and \ref{tab:sample_strong} (the 23 strongly-magnetised sources). Out of the weakly-magnetised category, a radio counterpart is detected for four targets for the first time: the persistent atolls GX 3+1, GS 1826-24, and 4U 1702-429, and the AMXP IGR J17379-3747. From the 23 strongly-magnetised neutron stars, nine sources are detected: the symbiotic X-ray binaries GX 1+4 and 4U 1954+31, the intermediate-mass X-ray binary Her X-1, the Supergiant X-ray binaries 1E 1145.1-6141, 4U 1700-37, Vela X-1, IGR J16318-4848 and IGR J16320-4751, and the Be/X-ray binary Swift J0243.6+6124. The radio detections of the latter and Her X-1 were already reported \citep{vandeneijnden2018_her,vandeneijnden2018_swj0243} but we include these as they were not compared with a larger sample of strongly-magnetised neutron stars yet. GX 1+4 was also presented before \citep{vandeneijnden2018_gx}, but here we add three more detections in new observations at different observing frequencies. 

\begin{figure*}
	\includegraphics[width=\textwidth]{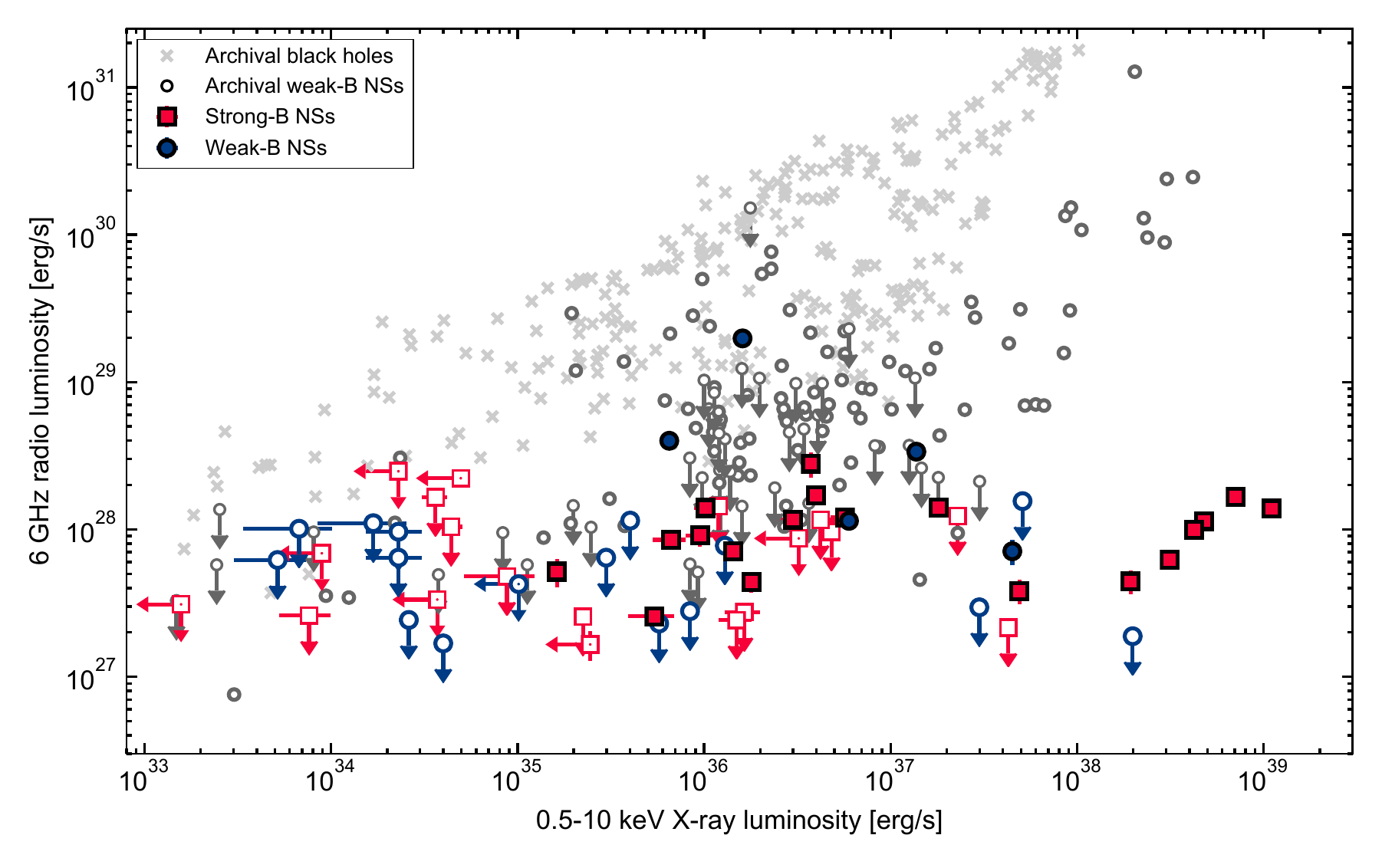}
    \caption{The X-ray -- radio luminosity plane for X-ray binaries. Archival observations of black holes are shown as the light grey crosses, while those of weakly-magnetised neutron stars are shown as dark grey circles. The observations presented in this work are shown in blue (weakly-magnetised) and red (strongly-magnetised). Upper limits are shown in open markers, while detections are plotted as filled markers. Apart from GX 1+4 and Her X-1, we do not show previous observations of strongly-magnetised neutron stars, as these are all upper limits at typically unconstraining levels (although we discuss several exceptions in Section \ref{sec:disc_deep}). Radio luminosities have been converted to $6$ GHz, using a measured spectral index if available, or assuming a flat spectrum otherwise.}
    \label{fig:permagfieldclass}
\end{figure*}

In Figure \ref{fig:permagfieldclass}, we show the X-ray -- radio luminosity plane for black hole and neutron star X-ray binaries, in order to search for coupling between the X-ray emitting accretion flow and radio-emitting jets. For now, we do not yet attempt to distinguish between sources where the radio emission is clearly attributable to a jet, or other processes that might contribute; see Section \ref{sec:discuss} for an extensive discussion on this topic. The newly added sources from our sample are shown per magnetic-field class as the blue (weak magnetic field) and red (strong magnetic field) data points. We originally compiled the comparison sample, shown with grey crosses for black holes and black circles for neutron stars, for the statistical study in \citet{gallo18}, and complemented it by two neutron stars discovered since (see Section \ref{sec:targets}). We stress that this comparison sample does not include soft state atolls, while our weakly-magnetised sample does. We plot the $6$-GHz radio luminosity, which we calculated by first estimating the $6$-GHz flux density $S_\nu$ using the spectral shape where known or otherwise assuming a flat radio spectrum. Then we calculated $L_R = 4\pi\nu S_\nu D^2$, where $\nu = 6$ GHz and D is the distance to the source. All X-ray luminosities $L_X$ are calculated in a similar fashion in the 0.5-10 keV range using $L_X = 4\pi F_X D^2$, where $F_X$ is the measured, unabsorbed X-ray flux. 

\begin{figure*}
	\includegraphics[width=\textwidth]{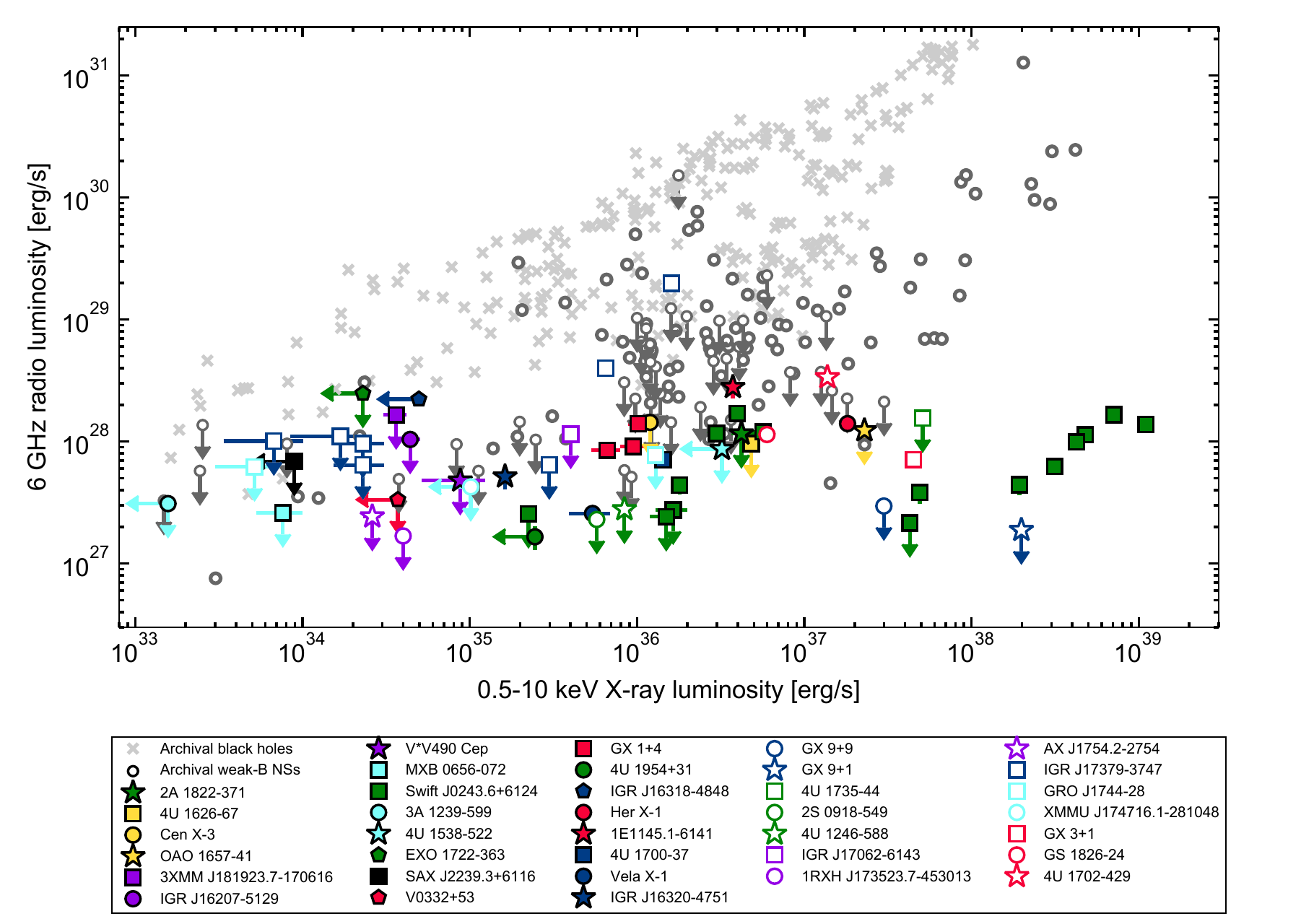}
    \caption{The same as Figure \ref{fig:permagfieldclass}, now labelling individual sources in our sample. Filled markers indicate strongly-magnetised neutron star, while the open ones show weakly-magnetised sources.}
    \label{fig:persource}
\end{figure*}

We briefly note that in the X-ray -- radio luminosity plane, we only plot statistical errors on both luminosities. Underlying assumptions, for instance a flat spectrum for single-frequency observations, and issues such as non-simultaneity of observations, uncertainties on distances, and errors in absolute flux calibration, cause systematic uncertainties in these comparisons. We do not include those in Figures \ref{fig:permagfieldclass} and \ref{fig:persource} as the literature sample similarly uses statistical errors only and these systematics are challenging to constrain accurately. However, one should keep their existence in mind when interpreting X-ray -- radio luminosity diagrams.

From Figure \ref{fig:permagfieldclass}, it is immediately apparent that only few neutron stars are detected in the radio band below an X-ray luminosity of $L_X \approx 5\times10^{34}$ erg/s. Conversely, above $L_X \approx 10^{36}$ erg/s, most of the radio observations of accreting neutron stars yield a detection at current sensitivities. As discussed in detail in Section \ref{sec:disc_part1}, none of the strongly-magnetised neutron stars reach above $L_R \approx 3\times10^{28}$ erg/s, independent of the X-ray luminosity. This faint apparent maximum radio luminosity -- corresponding to $\sim 170$ $\mu$Jy for a typical distance of $5$ kpc -- explains why these sources remained undetected in previous observing campaigns at lower radio sensitivity \citep{duldig79,nelson88,fender00}.

In Figure \ref{fig:persource}, we again show the X-ray -- radio luminosity plane, however now we individually label each source in our sample. Filled markers correspond to strongly-magnetised sources, while open markers show weakly-magnetised neutron stars. Showing individual sources reveals how the strongly-magnetised sample is dominated, especially above $L_X \approx 2\times10^{37}$ erg/s ($\geq 10\%$ $L_{\rm Edd}$ for a neutron star), by Swift J0243.6+6124: the only active transient HMXB in our sample and therefore the only HMXB with radio coverage across multiple orders of magnitude in X-ray luminosity   \citep{vandeneijnden2019_reb}. At the faint X-ray luminosity end of the diagram ($L_X < 10^{35}$ erg/s), it is clear that none of the four VFXBs (1RXH J173523.7-453013, AX J1754.2-2754, XMMU J174716.1-281048, and IGR J17062-6143) and the four faint BeXRBs (V*V490 Cep, MXB 0656-072, SAX J2239.3+6116, V0332+53) were detected in the radio band, despite sensitive VLA observations. Similarly, none of the UCXBs in our sample (2S 0918-549, 4U 1246-588, 4U 1626-67, and, again, IGR J17062-6143) were detected at radio frequencies, independent of their magnetic fields. In the past, sources in the VFXB and UCXB classes have been detected at similar X-ray luminosities \citep{millerjones11b,bogdanov18,bahramian18,li20}, while such X-ray faint BeXRBs have not. 

In the sample of sources in Figure \ref{fig:persource}, two subtle points might be easily missed. Therefore, we point them out explicitly here: firstly, two strongly-magnetised neutron stars, the SyXRBs 4U 1954+31 and the HMXB IGR J16318-4848, were only detected in radio and not in X-rays. Secondly, we plot the detected X-ray luminosity of 2A 1822-371; however, this source is likely viewed at high inclination, causing the inner flow to be obscured; the intrinsic X-ray luminosity might exceed the Eddington limit \citep{burderi10,baknielsen17}, which would move it to $L_X \geq 2\times10^{38}$ erg/s.

\begin{figure*}
	\includegraphics[width=\textwidth]{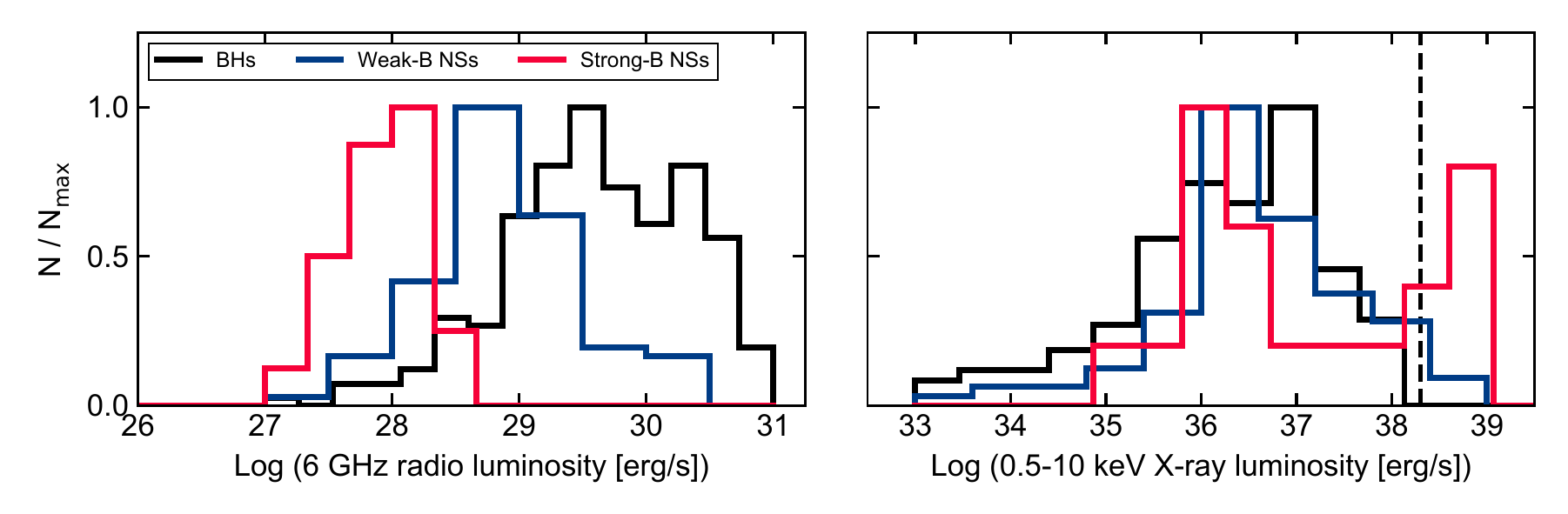}
    \caption{Histograms of the radio (left) and X-ray (right) luminosity for the full (archival plus this work) samples of black hole, weakly-magnetised, and strongly-magnetised X-ray binaries. Note that we plot all observations, including when multiple observations of the same source are available. Coloring is black, blue, and red, respectively, following Figure \ref{fig:permagfieldclass}. We only show detected sources and do not take upper limits into account. The histograms are normalised by their peak values to ease comparison of their shapes. The dashed line in the right panel indicates the Eddington limit for a $1.4$ $M_{\odot}$ neutron star.}
    \label{fig:hists}
\end{figure*}

As noted before, the strongly-magnetised neutron stars were not detected at radio luminosities above $\sim 3\times10^{28}$ erg/s. To compare this limiting radio luminosity to the other neutron stars and black holes, we show a normalised histogram of the radio luminosities of detected X-ray binaries in Figure \ref{fig:hists} (left panel; note that we plot all detection, including multiples from the same source). We do not make any selections in X-ray luminosity. From the radio luminosity histogram it is evident that, as has been noticed several times in the literature, neutron stars are in general radio fainter than accreting black holes \citep{fender01,migliari06,gallo18}. However, it also appears that strongly-magnetised neutron stars are systematically radio fainter than weakly-magnetised neutron stars. Indeed, a Kolmogorov-Smirnov test comparing the strongly- and weakly-magnetized neutron star samples returns a p-value of $p = 7^{-11}$ for the hypothesis that both are drawn from the same underlying distribution. Alternatively, the Anderson-Darling test consistently finds $p < 10^{-3}$ (we note that this limit is due to the implementation of this test in \textsc{python/scipy}: the measured test statistic greatly exceeds the critical value corresponding to $p=10^{-3}$). 

We note that a bias can be introduced by differences in the X-ray luminosity distribution, as the three different source classes might be dominated by different X-ray luminosities, translating in different dominant radio luminosities if a coupling exists between both \citep[see e.g.][for a detailed discussion]{gallo18}. To assess this effect, we also show the X-ray luminosity histograms of each source class in the right panel. While there are differences in the distributions, these are minor compared to the differences in radio luminosity. In fact, the only major difference is the peak at super-Eddington X-ray luminosities for strongly-magnetised sources, attributable completely to Swift J0243.6+6124. However, this difference only emphasises the striking radio faintness of strongly-magnetised neutron stars: the super-Eddington peak in X-rays should shift the radio luminosity distribution of strongly-magnetised sources to higher values, if these sources would show a black hole-like X-ray--radio coupling (\textit{without} an apparent maximum radio luminosity). In other words, the strongly-magnetic neutron stars are radio faint, despite the dominance of the X-ray bright Swift J0243.6+6124 in the HMXB sample. 

\begin{table*}
	\centering
	\caption{Radio and X-ray observations of 13 weakly-magnetised accreting neutron stars.  For source type acronyms, see Section \ref{sec:targets} -- hard and soft for atolls refer to their state during the radio observation. *IGR J17062-6143 is also an AMXP and VFXB. Upper limits in radio are listed at the $3\sigma$ level, while X-ray upper limits are shown at the $90\%$ level. All other uncertainties are $1\sigma$. The X-ray fluxes are unabsorbed fluxes in the $0.5$--$10$ keV band. Distance references: $^a$poorly constrained, see \citet{savolainen09}, $^b$poorly constrained, see \citet{vandenberg17}, $^c$\citet{vandenberg14}, $^d$\citet{thompson05}, $^e$\citet{iaria16}, $^f$\citet{galloway2008}, $^g$\citet{atri2019}, $^h$\citet{intzand08}, $^i$\citet{keek17}, $^j$\citet{degenaar11}, $^k$\citet{degenaar10}, $^l$\citet{chelovekov07}, $^m$\citet{chelovekov06}.} 
	\label{tab:sample}
	\begin{tabular}{llccccccc} %
\multicolumn{9}{c}{Weak magnetic field neutron stars} \\ \hline
\multirow{2}{*}{Source name} & \multirow{2}{*}{Source type} & \multirow{2}{*}{Epoch} & Radio & Radio flux & Spectral & X-ray flux & X-ray & distance \\
& & & freq. [GHz] & density [$\mu$Jy] & index $\alpha$ & [erg/s/cm$^2$] & obs. & [kpc] \\ \hline
GX 9+9 & Atoll (soft) & & 9.0 & $< 16.5$ & & $(1.0 \pm 0.1)\times10^{-8}$ & \textit{MAXI} & 5.0$^a$ \\
GX 9+1 & Atoll (soft) & & 9.0 & $< 10.5$ & & $(6.6 \pm 0.7)\times10^{-8}$ & \textit{MAXI} & 5.0$^b$ \\
GX 3+1 & Atoll (soft) & & 9.0 & $26.6 \pm 5.1$ & & $(1.0 \pm 0.1)\times10^{-8}$ & \textit{MAXI} & 6.1$^c$ \\
GS 1826-24 & Atoll (hard) & & 9.0 & $63.7 \pm 3.7$ & & $(2.0 \pm 0.2)\times10^{-9}$ & \textit{MAXI} & 5.0$^d$ \\ \cline{2-9}
\multirow{2}{*}{4U 1702-429 } & \multirow{2}{*}{Atoll (soft)} & & 5.5 & $161 \pm 9.5$ & \multirow{2}{*}{$-0.8\pm0.3$} & \multirow{2}{*}{$(3.94 \pm 0.02)\times10^{-9}$} & \multirow{2}{*}{\textit{Swift}} & \multirow{2}{*}{5.4$^e$}\\
& & & 9.0 & $110 \pm 9.5$ & & & & \\ \cline{2-9}
\multirow{2}{*}{4U 1735-44} & \multirow{2}{*}{Atoll (hard)} & & 5.5 & $< 30$ & & \multirow{2}{*}{$(5.9 \pm 0.2)\times10^{-9}$} & \multirow{2}{*}{\textit{MAXI}} & \multirow{2}{*}{8.5$^f$} \\
 & & & 9.0 & $< 28.5$ & & & & \\ \cline{2-9}
\multirow{2}{*}{2S 0918-549} & \multirow{2}{*}{UCXB} & & 5.5 & $< 20$ & & \multirow{2}{*}{$(3.0 \pm 0.1)\times10^{-10}$} & \multirow{2}{*}{\textit{Swift}} & \multirow{2}{*}{4.0$^g$} \\
& & & 9.0 & $< 20$ & & & & \\ \cline{2-9} 
\multirow{2}{*}{4U 1246-588} & \multirow{2}{*}{UCXB} & & 5.5 & $< 21$ & & \multirow{2}{*}{$(3.8 \pm 0.1)\times10^{-10}$} & \multirow{2}{*}{\textit{Swift}} & \multirow{2}{*}{4.3$^h$} \\
& & & 9.0 & $< 24$ & & & & \\ \cline{2-9}
IGR J17062-6143 & \multicolumn{2}{l}{UCXB / AMXP / VFXB} & 5.5+9 & $< 30$ & & $(6.3 \pm 0.5)\times10^{-11}$ & \textit{Swift} & 7.3$^i$ \\
XMMU J174716.1-281048 & VFXB & & 10 & $< 8.4$ & & $< 1.2\times10^{-11}$ & \textit{Swift} & 8.4$^j$ \\
1RXH J173523.7-354013 & VFXB & & 10 & $< 2.6$ & & $(3.7 \pm 0.4)\times10^{-12}$ & \textit{Swift} & $<$9.5$^k$ \\
AX J1754.2-2754 & VFXB & & 10 & $< 4.0$ & & $(2.6 \pm 0.3)\times10^{-12}$ & \textit{Swift} & 9.2$^l$ \\ \cline{2-9}
\multirow{9}{*}{IGR J17379-3747} & \multirow{9}{*}{AMXP} & \multirow{2}{*}{1} & 4.5 & $431\pm7.0$ & \multirow{2}{*}{$-0.04\pm0.05$} & \multirow{2}{*}{$(2.1\pm0.1)\times10^{-10}$} & \multirow{2}{*}{\textit{Swift}} & \multirow{9}{*}{8.0$^m$}\\
& & & 7.5 & $422\pm7.0$ & & & & \\
& & \multirow{2}{*}{2} & 4.5 & $87\pm11$ & \multirow{2}{*}{$0.1\pm0.4$} & \multirow{2}{*}{$(8.5\pm0.7)\times10^{-11}$} & \multirow{2}{*}{\textit{Swift}} & \\
& & & 7.5 & $92\pm9$ & & & & \\
& & 3 & 4.5+7.5 & $< 14$ & & $(3\pm1)\times10^{-12}$ & \textit{Swift} & \\
& & 4 & 4.5+7.5 & $< 14$ & & $(3.9\pm0.3)\times10^{-11}$ & \textit{Swift} & \\
& & 5 & 4.5+7.5 & $< 21$ & & $(3\pm1)\times10^{-12}$ & \textit{Swift} & \\
& & 6 & 4.5+7.5 & $< 22$ & & $(8.8\pm4.4)\times10^{-13}$ & \textit{Swift} & \\
& & 7 & 4.5+7.5 & $< 24$ & & $(2.2\pm1.1)\times10^{-12}$ & \textit{Swift} & \\ 
\hline 	\end{tabular}
\end{table*}

\begin{table*}
	\centering
	\caption{Radio and X-ray observations of 23 strongly-magnetised accreting neutron stars. For column definitions, see Table \ref{tab:sample}. 3A 1239-599, with source type denoted with *, has not been identified as either SgXB or BeXRB. The X-ray flux of Her X-1 is not measured directly, see Van den Eijnden et al. (2018). The X-ray fluxes are unabsorbed fluxes in the $0.5$--$10$ keV band. Distance references: $^a$\citet{hinkle06}, $^b$\citet{atri2019}, $^c$\citet{qiu17}, $^d$\citet{chakrabarty98}, $^e$\citet{fender00}; note that this distance is poorly constrained and might be up to 20 kpc, $^f$\citet{chakrabarty02}, $^g$\citet{gimenez15}, $^h$\citet{court18}.}
	\label{tab:sample_strong}
	\begin{tabular}{lcccccccc}
\multicolumn{9}{c}{Strong magnetic field neutron stars} \\ \hline
\multirow{2}{*}{Source name} & \multirow{2}{*}{Source type} & \multirow{2}{*}{Epoch} & Radio & Radio flux & Spectral & X-ray flux & X-ray & distance \\ 
& & & freq. [GHz] & density [$\mu$Jy] & index $\alpha$ & [erg/s/cm$^2$] & obs. & [kpc]\\ \hline
\multirow{7}{*}{GX 1+4} & \multirow{7}{*}{SyXRB} & 1 & 10 & $105.3 \pm 7.3$ & $-0.7 \pm 3.3$ & $(4.6 \pm 0.6)\times10^{-10}$ & \multirow{2}{*}{\textit{MAXI}} & \multirow{7}{*}{4.3$^a$} \\
& & \multirow{2}{*}{2} & 4.5 & $37 \pm 15$ & \multirow{2}{*}{$0.9 \pm 0.8$} & \multirow{2}{*}{--} & \multirow{2}{*}{--} & \\
& & & 7.5 & $59 \pm 9$ & & & & \\
& & \multirow{2}{*}{3} & 4.5 & $69 \pm 11$ & \multirow{2}{*}{$-0.6 \pm 0.6$} & \multirow{2}{*}{$(4.3 \pm 0.7)\times10^{-10}$} & \multirow{2}{*}{\textit{MAXI}} & \\
& & & 7.5 & $50 \pm 8 $ & & & & \\
& & \multirow{2}{*}{4} & 4.5 & $64 \pm 8$ & \multirow{2}{*}{$0.2 \pm 0.4$} & \multirow{2}{*}{$(3.0 \pm 0.6)\times10^{-10}$} & \multirow{2}{*}{\textit{MAXI}} & \\ 
& & & 7.5 & $70 \pm 8$ & & & & \\ \cline{2-9}
\multirow{2}{*}{4U 1954+31} & \multirow{2}{*}{SyXRB} & 1 & 4.5+7.5 & $21.2 \pm 4.8$ & -- & $(1.9 \pm 0.1)\times10^{-10}$ & \textit{Swift} & \multirow{2}{*}{3.3$^b$} \\
& & 2 & 4.5+7.5 & $< 19.0$ & -- & -- & -- & \\
\cline{2-9}
\multirow{2}{*}{3XMM J181923.7-170616} & \multirow{2}{*}{SyXRB candidate} & & 5.5 & $<36$ & \multirow{2}{*}{--} & \multirow{2}{*}{$(4.7 \pm 0.8)\times10^{-12}$} & \multirow{2}{*}{\textit{Swift}} & \multirow{2}{*}{8$^c$} \\
& & & 9 & $<180$ & & & & \\ 
\multirow{2}{*}{2A 1822-371} & \multirow{2}{*}{LMXB} & & 5.5 & $<33$ & \multirow{2}{*}{--} & \multirow{2}{*}{$(7 \pm 1)\times10^{-10}$} & \multirow{2}{*}{\textit{MAXI}} & \multirow{2}{*}{7.0$^b$} \\
& & & 9 & $<30$ & & & & \\
\multirow{2}{*}{4U 1626-67} & \multirow{2}{*}{LMXB / UCXB} & & 5.5 & $<21$ & \multirow{2}{*}{--} & \multirow{2}{*}{$(6.3 \pm 0.5)\times10^{-10}$} & \multirow{2}{*}{\textit{MAXI}} & \multirow{2}{*}{8$^d$} \\
& & & 9 & $<22$ & & & & \\
Her X-1 & IMXB & & 9 & $38.7 \pm 4.8$ & $-0.7 \pm 5.3$ & $\sim 3\times10^{-9}$ & \textit{MAXI} & 7.1$^b$ \\
\multirow{2}{*}{1E1145.1-6141} & \multirow{2}{*}{SgXB} & & 5.5 & $56.6\pm11$ & \multirow{2}{*}{$<0.38$} & \multirow{2}{*}{$(4.5 \pm 0.5)\times10^{-10}$} & \multirow{2}{*}{\textit{MAXI}} & \multirow{2}{*}{8.3$^b$} \\
& & & 9 & $<31.5$ & & & & \\
\multirow{2}{*}{Cen X-3} & \multirow{2}{*}{SgXB} & & 5.5 & $< 42$ & \multirow{2}{*}{--} & \multirow{2}{*}{$(2.1 \pm 0.3)\times10^{-10}$} &  \multirow{2}{*}{\textit{MAXI}} & \multirow{2}{*}{6.9$^b$} \\ 
& & & 9 & $< 48$ & & & & \\ 
\multirow{2}{*}{3A 1239-599} & \multirow{2}{*}{HMXB*} & & 5.5 & $< 27$ & \multirow{2}{*}{--} & \multirow{2}{*}{$< 8.2\times10^{-13}$} & \multirow{2}{*}{\textit{Swift}} & \multirow{2}{*}{4$^e$} \\
& & & 9 & $< 27$ & & & & \\
\multirow{2}{*}{OAO 1657-41} & \multirow{2}{*}{SgXB} & & 5.5 & $< 42$ & \multirow{2}{*}{--} & \multirow{2}{*}{$(4.7 \pm 0.2)\times10^{-9}$} & \multirow{2}{*}{\textit{MAXI}} & \multirow{2}{*}{6.4$^f$} \\
& & & 9 & $< 30$ & & & & \\
\multirow{2}{*}{4U 1538-522} & \multirow{2}{*}{SgXB} & & 5.5 & $< 36$ & \multirow{2}{*}{--} & \multirow{2}{*}{$< 8\times10^{-10}$} & \multirow{2}{*}{\textit{MAXI}} & \multirow{2}{*}{5.8$^b$} \\
& & & 9 & $< 30$ & & & & \\
\multirow{2}{*}{4U 1700-37} & \multirow{2}{*}{SgXB} & & 5.5 & $484 \pm 13$ & \multirow{2}{*}{$0.46\pm0.16$} & \multirow{2}{*}{$(5.3 \pm 0.1)\times10^{-9}$} & \multirow{2}{*}{\textit{Swift}} & \multirow{2}{*}{1.5$^b$} \\
& & & 9 & $551 \pm 9$ & & & & \\
\multirow{2}{*}{EXO 1722-363} & \multirow{2}{*}{SgXB} & & 5.5 & $< 54$& \multirow{2}{*}{--} & \multirow{2}{*}{$< 3\times10^{-12}$} & \multirow{2}{*}{\textit{Swift}} & \multirow{2}{*}{8$^g$} \\
& & & 9 & $< 45$ & & & & \\
\multirow{2}{*}{Vela X-1} & \multirow{2}{*}{SgXB} & & 5.5 & $92.4 \pm 10.5$ & \multirow{2}{*}{$0.56\pm0.36$} & \multirow{2}{*}{$(1.2 \pm 0.3)\times10^{-9}$} & \multirow{2}{*}{\textit{MAXI}} & \multirow{2}{*}{1.97$^b$} \\
& & & 9 & $121.7 \pm 10.0$ & & & & \\
\multirow{2}{*}{IGR J16207-5129} & \multirow{2}{*}{SgXB} & & 5.5 & $< 39$ & \multirow{2}{*}{--} & \multirow{2}{*}{$(9.9 \pm 1.5)\times10^{-12}$} & \multirow{2}{*}{\textit{Swift}} & \multirow{2}{*}{6.1$^g$} \\
& & & 9 & $< 33$ & & & & \\
\multirow{2}{*}{IGR J16318-4848} & \multirow{2}{*}{SgXB} & & 5.5 & $239 \pm 13$ & \multirow{2}{*}{$1.06 \pm 0.35$} & \multirow{2}{*}{$< 3.2\times10^{-11}$} & \multirow{2}{*}{\textit{Swift}} & \multirow{2}{*}{3.6$^g$} \\
& & & 9 & $404 \pm 13$ & & & & \\
\multirow{2}{*}{IGR J16320-4751} & \multirow{2}{*}{SgXB} & & 5.5 & $59 \pm 13$ & \multirow{2}{*}{$< 1.0$} & \multirow{2}{*}{$(1.1 \pm 0.1)\times10^{-10}$} & \multirow{2}{*}{\textit{Swift}} & \multirow{2}{*}{3.5$^g$} \\
& & & 9 & $< 37.5$ & & & & \\ \cline{2-9}
V*V490 Cep & qBeXRB & & 10 & $< 12$ & -- & $(1.3 \pm 0.5)\times10^{-11}$ & \textit{Swift} & 7.5$^b$ \\
MXB 0656-072 & qBeXRB & & 10 & $< 11$ & -- & $(1.9 \pm 0.6)\times10^{-12}$ & \textit{Swift} & 5.7$^b$ \\
SAX J2239.3+6116 & qBeXRB & & 10 & $< 18$ & -- & $< 1.4\times10^{-12}$ & \textit{Swift} & 7.3$^b$ \\
V 0332+53 & qBeXRB & & 10 & $< 15$ & -- & $< 1\times10^{-11}$ & \textit{Swift} & 5.57$^b$ \\ \cline{2-9}
\multirow{2}{*}{GRO J1744-28} & \multirow{2}{*}{LMXB} & 1 & 5.5+9 & $< 30$ & & $(3.0\pm0.1)\times10^{-10}$ & \textit{Swift} & \multirow{2}{*}{4--8$^h$} \\ 
& & 2 & 5.5+9 & $< 24$ & & $(1.2\pm0.5)\times10^{-12}$ & \textit{Swift} & \\ \cline{2-9}
Swift J0243.6+6124 & BeXRB & \multicolumn{7}{l}{Data taken from \citet{vandeneijnden2018_swj0243} and \citet{vandeneijnden2019_reb}} \\
\hline
	\end{tabular}
\end{table*}

    

\section{The origin of radio emission from accreting neutron stars}
\label{sec:discuss}

\subsection{Comparing low-mass and high-mass X-ray binaries}
\label{sec:disc_part0}

Radio emission observed in Roche-lobe overflowing LMXBs -- whether the primary is a black hole or a weakly-magnetised neutron star -- is typically attributed to synchrotron processes in a relativistic jet \citep[e.g.][]{corbel00,dhawan2000,stirling2001,fender04,migliari06,gallo18}. Our samples of weakly and strongly-magnetised neutron stars contain 16 LMXBs: all weakly-magnetised neutron stars, plus the slow pulsars 2A 1822-371, 4U 1626-67, and GRO J1744-28 (here, we ignore the SyXRBs, which have a low-mass donor but accrete from the stellar wind; see Section \ref{sec:syxrbs}). Out of these sixteen, four sources are detected (e.g. Table \ref{tab:sample}): the Atolls GX 3+1, GS 1826-24, and 4U 1702-429, and the AMXP IGR J17379-3747. Their radio spectral shapes and positions on the X-ray -- radio luminosity diagram fit with a jet identification for accreting neutron stars, being consistent with the larger sample of sources in the literature \citep{russell13,gallo18}. In the undetected sources, the non-detections can typically be attributed to either their spectral state or faint X-ray luminosity. For a detailed discussion, we refer the reader to Section \ref{sec:disc_part3}, where we will comment on individual (detected and undetected) sources.

The identification of the radio emission origin is less straightforward for high-mass and symbiotic X-ray binaries. As alluded to in the introduction, these sources are more complicated for two reasons: firstly, their donors launch stellar winds that could contribute to the detected radio flux via various mechanisms. Secondly, the jet properties of strongly-magnetised neutron stars are more poorly explored and therefore existing literature offers few comparison studies. In the remainder of this section, we will review six options for the origin of the radio emission: emission from the donor star itself, emission from stellar winds and their interaction with other components of the binary system, coherent processes, emission from a slow, wide-open outflow from a propeller-type mechanism, and relativistic jets. 

Three options can be excluded directly. While stars do emit in the radio band, such emission is typically seen in coronally-active low-mass stars (type F or later). In addition, these stars reach maximum X-ray luminosities of $\sim 10^{32}$ erg/s, orders of magnitude below the X-ray luminosities of our radio-detected targets \citep{gudel93,gudel02,kurapati17}. Secondly, a wide-open, slow gas outflow driven by a magnetic propeller-type mechanism, as for instance seen in simulations of accretion onto magnetised (neutron) stars by \citet{romanova09} and \citet{parfrey17}, is similarly unlikely: as discussed also in Section \ref{sec:disc_part2}, none of our radio-detected targets resided in the propeller regime during the observations.

Thirdly, coherent processes have been inferred the radio emission of types of accretion magnetic white dwarfs \citep[e.g.][see also Section \ref{sec:jets}]{barrett17}. These processes, namely an electron-cyclotron maser or gyrosynchrotron emission, are unlikely to operate in the systems considered here: firstly, electron-cyclotron maser emission is associated with high ($\sim 100$\%) levels of circular polarization. While the ATCA observations studied here were not set up to measure circular polarization, the earlier VLA studies of Her X-1, GX 1+4 and Sw J0243, did not show any circular polarization \citep[e.g.][]{vandeneijnden2018_gx}. Secondly, gyrosynchrotron emission is expected to show lower levels of polarization. Such emission would originate from the magnetosphere of the neutron star. However, in actively accreting HMXBs, the magnetospheric radius does not, for reasonable neutron star parameters and accretion rate, exceed $\sim 10^4$ $R_g$ \citep{tsygankov17}. In comparison, the typical minimum emission size of the detected radio emission is roughly two orders of magnitude higher (see Section \ref{sec:jets} and equation \ref{eq:size}).

\subsection{Winds from massive donor stars and their interactions?}
\label{sec:disc_wind}

\subsubsection{Stellar winds}
\label{sec:disc_wind_1}
Stellar winds and their emission properties have been studied extensively for both single and binary stars in the past decades \citep[see e.g.][and references therein]{lamers98,lamers98b,gudel02,vanloo2007}. Through radio and IR observations, fundamental wind properties such as mass loss rate and terminal velocity can be inferred, probing stellar feedback and the effect on stellar evolution. Wind radio emission is typically attributed to one of two emission processes: either thermal Bremsstrahlung emission from the ionised gas in the wind, or non-thermal emission from shocks in the outer wind. As shown by \citet{wright75}, the thermal emission is expected to have a positive ($\alpha \approx 0.6$) radio spectral index, which is indeed often observed in isolated O stars \citep[e.g.][]{vanloo2007}. Non-thermal shocks locally create steep spectra, i.e. with negative spectral indices $\alpha$. But, taking into account that the shocks both weaken and peak at lower frequencies as they move away from the star, their cumulative spectrum for single stars also has a positive spectral index \citep{dougherty00,vanloo2007}.

Negative spectral indices have been measured in several stellar winds from high-mass stars, however. In those cases, it is thought that these systems are (massive) binary systems, and the shock between the two stellar winds (that does not move outwards as it would for a single star) is observed \citep{moran89,churchwell92,dougherty96,dougherty00,williams97,chapman99,contreras97,ortiz2011,blomme2014,blomme2017}. At the most extreme end, where both stars are early-type / Wolf-Rayet stars with strong winds, these sources are observed as colliding wind binaries, which are known sources of X-ray and non-thermal radio emission \citep[e.g.][]{dubus13}. 

For thermal wind emission, the expected radio flux densities can be estimated using the formalism derived by \citet[][note how similar results were derived around the same time by \citealt{olnon1975} and \citealt{panagia1975}]{wright75}. Based on the observing frequency $\nu$, electron temperature $T_e$, wind mass loss rate, the mean atomic weight per electron $\mu_e$, the terminal wind velocity $v_{\infty}$, and the distance $d$, we can estimate the flux density as:

\begin{equation}
\begin{split}
S_{\nu} = 200 \left(\frac{\nu}{5.5 \rm GHz}\right)^{0.6} &\left(\frac{T_e}{10^4 \rm K}\right)^{0.1}\left(\frac{\dot{M}}{10^{-6} M_{\odot}/\rm yr}\right)^{4/3}\\
&\left(\frac{\mu_e v_{\infty}}{100 \rm km~s^{-1}}\right)^{-4/3}\left(\frac{d}{\rm 5 kpc}\right)^{-2} \mu\rm Jy
\end{split}
\label{eq:wind}
\end{equation}

\noindent Winds from massive stars are known to be clumped, which affects their observational appearance. However, the clumpiness decreases with distance from the star, and hence, the radio emission is the least affected by such effects. As a result, we can apply the above formalism, which was formulated for smooth plasmas \citep{puls2006}. The brightness temperature of these thermal winds is typically $10^4$-$10^5$ K \citep{longair92}; however, even with VLBI-like resolution, none of the targeted HMXBs are bright enough to reject the thermal wind hypothesis based on a minimum brightness temperature argument (see also Section \ref{sec:disc_part4}). For the non-thermal, shocked emission, the brightness is more difficult to predict, but observational constraints exist. 

Such observational constraints on the radio emission of stellar winds (in single and binary stars) have been obtained through many studies. For our comparison, the most relevant are the studies of OB supergiants. The most constraining and sensitive study of this kind was recently performed with ATCA and ALMA, observing the stellar cluster Westerlund 1 \citep{fenech18,andrews19}. Located at approximately $5$ kpc \citep{clark19}, Westerlund 1 hosts at least 100 OB supergiants \citep{negueruela10}. However, at current sensitivities, only seven of those stars (i.e. at most a few percent) are detected at GHz frequencies. Combining the ATCA and ALMA observations reveals that all radio detected OB supergiants have steep spectra, indicating that these are likely binary star systems \citep{andrews19}. Therefore, \citet{andrews19} state that at current sensitivities, they `would not expect to detect any emission from purely thermal stellar wind emitters in radio'. Importantly, these results imply that the radio emission from single OB supergiants is difficult to detect and might be overpredicted by, for instance, the \citet{wright75} formalism. 

\begin{figure}
	\includegraphics[width=\columnwidth]{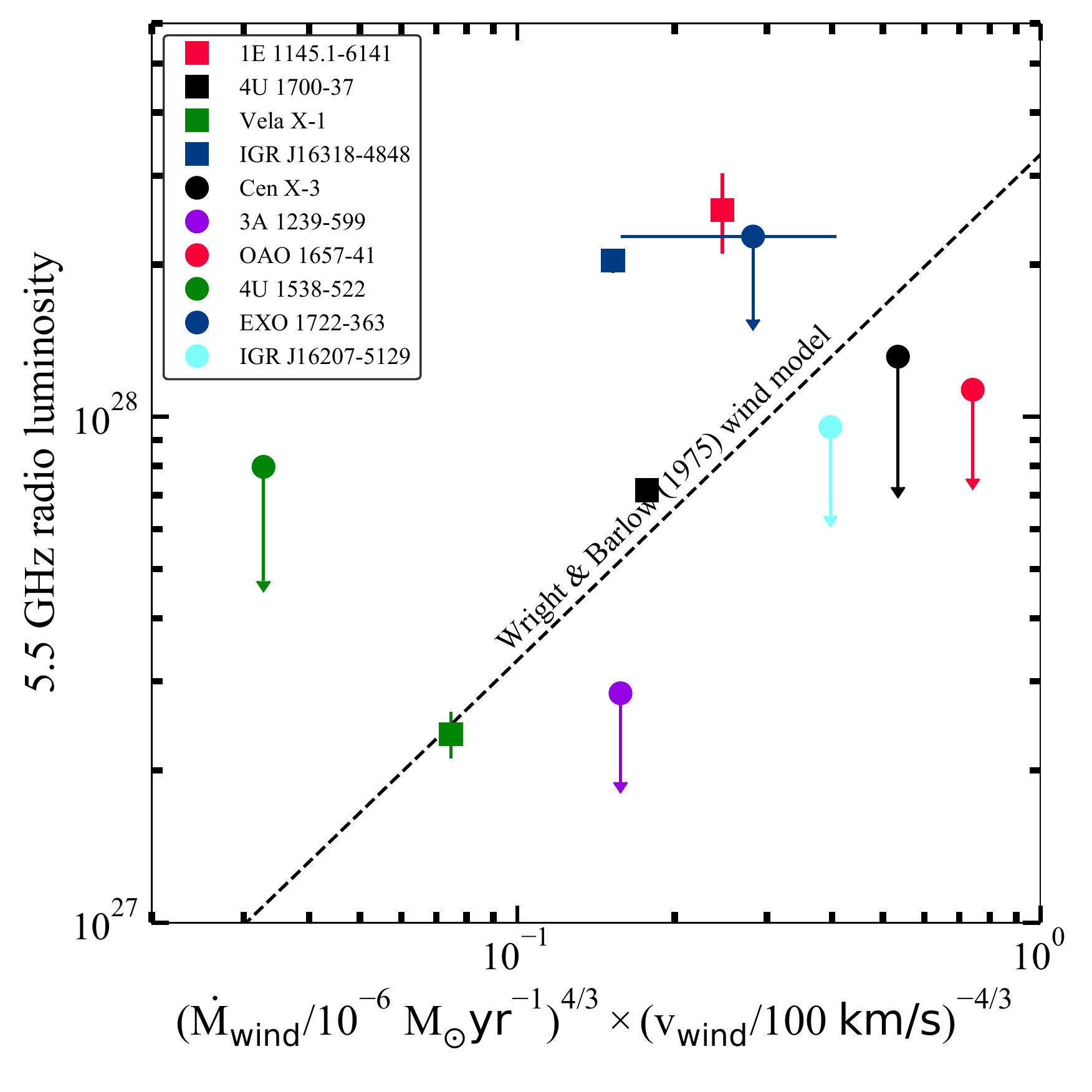}
    \caption{A visual comparison of the measured radio luminosities of the SgXBs in our sample and their predicted \citet{wright75} stellar wind radio luminosity. The dashed line indicates the theoretical wind prediction, as a function of the parameterization of wind velocity and mass loss rate in Equation \ref{eq:wind}. Most undetected target lie below the predicted relation, while three our of four detected sources are too radio-bright, suggesting a second/other dominant emission mechanism.}
    \label{fig:windflux}
\end{figure}

The studies of Westerlund 1, and in fact all radio studies on massive stars in the literature, focus on single and binary massive, nondegenerate stars. Similarly, the \citet{wright75} model was not developed for massive stars in X-ray binaries, but for single massive stars. Therefore, this model does not include the effect of X-rays emitted by the accretion flow on the stellar wind properties, and its resulting radio luminosity. However, the picture painted by the comparison between Westerlund 1 and the model predictions, fits with our results on OB supergiants with a neutron star companion. We show this graphically in Figure \ref{fig:windflux}: we plot the measured radio luminosity (upper limits) as a function of the wind mass loss rate and velocity as parameterized in Equation \ref{eq:wind}, alongside their relation according to that equation. In our calculations, we ignore the clumpy nature of winds in HMXBs \citep{martineznunez17,grinberg_cygx1,grinberg17}, for the reasons discussed above.

As is apparent from Figure \ref{fig:windflux}, four sources out of the non-detected HMXBs (Cen X-3, 3A 1239-599, OAO 1657-41, IGR J16207-5129) should have been detected given the ATCA sensitivity, stellar wind properties, distance, and equations for thermal wind radio emission. A fifth source, EXO 1722-363, would be above the detection limit, while the remaining non-detected SgXBs are too distant for the wind to be detected. The non-detection of the five named sources above supports the notion that OB supergiants are radio fainter than in the simple Wright \& Barlow estimate and challenging to detect in radio, unless they are in stellar (not X-ray) binary systems.

One can consider a scenario wherein the apparent theoretical over-prediction of the radio flux density for the single stars in Westerlund 1 would result from incomplete ionisation. However, given the temperatures of O/B stars, such an explanation appears unlikely. Moreover, in the X-ray binary estimates above, we assumed that the literature mass loss rate corresponds fully to ionised material. Comparing the radio non-detected with detected HMXBs in our sample, we do not find a systematic difference in X-ray luminosity or orbital period; therefore, we do not necessarily expect a systematic difference in the degree of X-ray ionisation of the wind. Hence, we expect that even if a partial ionisation of the wind could explain their radio non-detections, this possibility should also exist for the radio-detected systems discussed below. 

In our sample, five Supergiant X-ray binaries are detected with ATCA. For two of those sources (1E 1145.1-6141, IGR J16318-4848), their donor (wind) properties predict a radio flux density substantially below the detected levels, assuming the wind is fully ionised (e.g Figure \ref{fig:windflux}). Of those, only IGR J16318-4848 has a spectral shape that may be consistent with such stellar wind emission ($\alpha = 1.06 \pm 0.35$). 4U 1700-37 is detected at slightly higher radio luminosity than predicted, although systematic uncertainties in this comparison might account for that difference. For the fourth source, IGR J16320-4751, no wind characteristics are known, preventing a flux density estimate. For Vela X-1, both the flux density estimates and spectral index fit with the \citet{wright75} description. Therefore, we conclude that for Vela X-1 and 4U 1700-37, the stellar wind might have a substantial contribution to the observed emission (again with the caveat that thermal winds appear difficult to detect for OB supergiants), while it likely contributes less in the other detected sources. All details on these estimates can be found in the third section of the Online Supplementary Materials.

\subsubsection{Intrabinary wind shocks}

Could the stellar wind interact with the pulsar wind, causing the observed radio emission? It is commonly assumed that, in accreting systems, the radio pulsar mechanism is suppressed as the magnetosphere is filled with ionised material, implying that no pulsar wind is launched. However, with that in mind, we can briefly compare the radio properties of the strongly-magnetised sources in our sample with the known types of systems where the pulsar and stellar wind interact. For instance, shocks between the stellar and pulsar winds, similar to those in stellar binaries resulting in negative spectral indices in the wind radio emission, can occur in binary systems where no accretion takes place. In such systems, this shock creates non-thermal emission that dominates the spectrum from radio to gamma-rays \citep{dubus13}. As the spectral energy distribution of these systems peaks above 1 MeV, they are referred to as $\gamma$-ray binaries. Only a small number of $\gamma$-ray binaries are known to date \citep{paredes19,corbet19}. None of our detected sources are $\gamma$-ray binary candidates; indeed, the typical radio luminosities of $\gamma$-ray binaries are $\sim$two orders of magnitude larger \citep[see e.g.][]{dubus13} than the apparent maximum radio luminosity of neutron star HMXB of $\sim 3\times10^{28}$ erg/s we find in our results.

$\gamma$-ray binaries are one example of shock interaction between the pulsar wind and surrounding material; similar shocks can be seen in pulsar wind nebulae (PWNe), systems where the pulsar wind of an isolated neutron star interacts with surrounding supernova matter or the interstellar medium. The magnetic wind of relativistic electrons and positrons can carry away the majority of energy lost in the spin down of the pulsar \citep{rees74,michel82,kennel84}. The shock can be detected through radio synchrotron emission as the electrons gyrate around the magnetic field lines. If we again ignore, for the sake of the comparison, the common assumption that accreting neutron stars do not launch a pulsar wind, we can ask the following question: could a similar interaction be at play in HMXBs -- creating a dialed-up version of a PWN, or a dialed-down version of the extreme $\gamma$-ray binaries -- to explain our radio detections? The most quantitative constraints follow from the radio spectral shape. PWNe typically show power-law radio spectra with an index $\alpha \approx -0.3$ \citep{weiler78,gaensler00}. The best spectral constraints in our sample of HMXBs (i.e. Vela X-1, 4U 1700-37, IGR J16318-4848) show $\alpha > 0$ instead. 

The energetics of the shocks can also provide constraints. The radio luminosity of an individual system's shock is challenging to predict, as the spin down energy of the isolated pulsar population spans a wide range \citep[e.g. $10^{28}$--$10^{39}$ erg/s;][]{gaensler00}, and the fraction transferred into radio emission can vary between PWNe. Furthermore, in accreting strongly-magnetised neutron stars, the pulse frequency evolution is regulated by the interactions with the accretion flow. Therefore, a measurement of the spin period and its derivative do not translate into a spin-down energy estimate, as it does for isolated pulsars. However, using magnetic field and spin measurements, one can estimate what the corresponding spin-down energy would be in the absence of accretion (ignoring the effects of accretion on the strength of the magnetic field trough, for instance, burial). Assuming a typical NS mass of $1.4$ $M_{\odot}$ and radius of $10$ km, one can combine the spin down energy $\dot{E}_{\rm spin} = 4\pi^2 I \dot{P}/P^3$ and the estimate of the magnetic field $(B/10^{12} {\rm G}) \geq 3.2\times10^7 (P\dot{P} / 1{\rm s})^{1/2}$ to derive:

\begin{equation}
    \dot{E}_{\rm spin} \leq 4\times10^{31} \left( \frac{B}{10^{12} {\rm G}} \right)^2  \left( \frac{P}{1 {\rm s}} \right)^{-4} {\rm erg/s}
\end{equation}

For the sources in our sample, we find a wide range of values between $\sim 3\times10^{14}$ erg/s to $10^{32}$ erg/s, with just six sources overlapping with the low end of the distribution of the PWNe population. More importantly, there are no systematic differences between radio-detected and non-detected targets, and for five out of the seven detected sources where both $B$ and $P$ are known, $\dot{E}_{\rm spin}$ is significantly lower than the observed radio luminosity. Therefore, the energetics argue against contributions from intrabinary shocks. Also, we again stress that this comparison only holds if accreting neutron stars do launch pulsar winds, which is not usually assumed. 

We are left with four detected, strongly-magnetised sources that we have not yet discussed in this context; for Swift J0243.6+6124, a wind was excluded by \citet{vandeneijnden2018_swj0243,vandeneijnden2019_reb}, based on its luminosity, spectral change, and variations in those two properties throughout the source's giant outburst in 2017/2018. Secondly, Her X-1 has an intermediate-mass, Roche-lobe overflowing donor, which does not launch strong stellar winds. Therefore, it is unlikely that the radio emission in this system is attributable to such a mechanism. Finally, we observed and detected two SyXRBs, where the neutron star accretes from the stellar wind of an evolved low-mass donor. Given the different wind properties of these sources compared to HMXBs, we separately review them in Section \ref{sec:syxrbs}.

\subsection{Relativistic jets from strongly-magnetized neutron stars?}
\label{sec:jets}
Relativistic jets, as observed in weakly-magnetised neutron star and black hole X-ray binaries, emit radio emission through synchrotron processes. Given the broad range of observed spectral and brightness properties of jets in X-ray binaries, those observations pose few stringent constraints on what the emission can look like. However, we can make the comparison with these properties. Depending on the type of jet, this emission is either optically thick ($\alpha \geq 0$) or thin \citep[$\alpha < 0$, typically $\alpha \approx -0.7$;][]{fender04,russell13} for a compact, steady jet or discrete ejecta, respectively. In terms of the observed spectrum, the detected strongly-magnetised neutron stars fit within the expectation for a compact jet. However, with the wide range of indices possibly generated by jet synchrotron processes, this hardly implies the emission necessarily originates from a jet.

Secondly, we can consider the observed radio luminosity, given the X-ray luminosity. No unique prediction, from either theory or observations, exists for the radio luminosity based on source state, mass accretion rate, or X-ray luminosity. However, it is known observationally that weakly-magnetised neutron star X-ray binaries are typically substantially radio fainter than accreting black holes \citep{fender01,migliari06,gallo18}. All radio-detected, strongly-magnetised neutron stars are radio-fainter than the black hole tracks in the X-ray - radio luminosity plane, fitting with our expectations for relativistic jets.

Finally, we can consider the constraints from the Compton limit on the brightness temperature for synchrotron radiation of $T=10^{12}$ K. The angular size of the emitting region $\theta$ can be calculated as \citep[e.g.][]{longair92}:
\begin{equation}
    \left(\frac{\theta}{1{\rm "}}\right) = \sqrt{1.36 \left(\frac{\lambda}{1{\rm cm}}\right)^2  \left(\frac{T}{1{\rm K}}\right)^{-1} \left(\frac{S_{\nu}}{1{\rm mJy}}\right)}
    \label{eq:size}
\end{equation}
For a typical flux density of $100$ $\mu$Jy at $5$ GHz (i.e. $\sim 6$ cm), setting $T<10^{12}$ K yields $\theta > 2$ $\mu$as. At a typical distance of $5$ kpc, this angular size sets a minimal physical size of the emitting region of $>10^{11}$ cm, or $>7\times10^5$ $R_g$ for a neutron star. Therefore, the observed flux densities are consistent with jet synchrotron emission, as the radio-emitting regions of the jet typically lie further out (i.e. $10^7$--$10^9$ $R_g$). 

In this comparison with other source classes, we can also briefly compare the strongly-magnetised neutron stars with jet-launching white dwarfs. On the one hand, these systems are far apart in their fundamental physical properties such as magnetic field, accretor size. However, they share an important property: a $10^{12}$ G, $1.4$ $M_{\odot}$ neutron star, accreting from a disc at $L_X \approx 10^{36}$ erg/s, has a magnetospheric radius of $R_m \sim 6600$ km ($3.2\times10^{3}$ gravitational radii). This scale is comparable to the typical size of a white dwarf. Therefore, jet formation mechanisms at play in the accretion discs of accreting white dwarfs, might play a role beyond $R_m$ in the disc of strongly-magnetised neutron stars as well. 

What types of accreting white dwarfs launch jets? Similar to strongly-magnetised neutron stars, accreting white dwarfs were long thought not to launch jets \citep{livio97,livio99}. However, radio observations of SS Cyg in the past two decades reveal a jet launched by this famous accreting white dwarf, whose magnetic field is weak enough that the accretion flow extends to its surface \citep{kording08,russell16,fender19}. SS Cyg is an example of a non-magnetic Cataclysmic Variable (CV), a white dwarf accreting from a low-mass donor via Roche-lobe overflow\footnote{Other non-magnetic CVs had been detected at radio frequencies before, but those observations were not interpreted in a jet framework at that time \citep{benz83,benz89}.}. Subsequent observations of persistent (i.e. nova-likes) and transient CVs in outburst (i.e. dwarf novae) mostly resulted in radio detections as well, consistent with jet emission \citep{coppejans15,coppejans16}. However, in those systems, alternative mechanisms could not be ruled out as confidently as in SS Cyg. Therefore, while it appears that non-magnetic CVs are capable of launching jets, the increase in detected sources has introduced many remaining questions \citep{coppejans20}, a development that repeats in this study of strongly-magnetised neutron stars. 

More strongly magnetised accreting white dwarfs, where the accretion flow is magnetically truncated (i.e. intermediate polars and polars), have been detected in the radio band \citep{chanmugam82,wright88,abada-simon93,pavelin94}. However, their radio properties, such as circular polarization \citep{barrett17} and flaring \citep{dulk83,chanmugam87}, suggest a gyrosynchrotron or cyclotron maser, instead of a jet, origin \citep{mason07}. Finally, highly-accreting white dwarfs in super-soft sources have had jet detections \citep{crampton96,cowley98,motch98,becker98}.

Comparing non-magnetic CVs with our sample more quantitatively, one finds that the radio luminosities of non-magnetic CVs lie significantly below those reported here for strongly-magnetised neutron stars: $\leq 4\times10^{26}$ erg/s \citep{russell16,coppejans15,coppejans16,coppejans20} versus $\sim 2\times10^{27}$--$3\times10^{28}$ erg/s, respectively. However, the former are detected at lower X-ray luminosities as well ($L_X < 10^{33}$ erg/s), which are observable due to the smaller distances to the observed targets: the accretion flux is not dominated by the X-ray band to the extent that it is in X-ray binaries. To finish, we note that this comparison assumes the strongly-magnetised neutron star accretes from a disc -- a separate comparison with the wind-accreting white dwarfs in symbiotic stars is made below (Section \ref{sec:syxrbs}). 

The recent radio campaigns on the 2017/2018 giant outburst of Swift J0243.6+6124 have demonstrated that strongly-magnetised neutron stars can launch jets \citep{vandeneijnden2018_swj0243,vandeneijnden2019_reb}. Combined with the above issues with explaining the observed radio properties purely through stellar winds, we therefore conclude that it is possible that the radio emission observed from the neutron star HMXBs in our sample is dominated by synchrotron emission from a relativistic jet, and will assume so in the following section of the discussion. The exception to this interpretation are Vela X-1 and 4U 1700-37, which are the sources where the theoretical stellar wind predictions are similar to the observed radio emission. We will also assume such a jet origin for the emission in Her X-1, although we stress that the origin of its radio emission could not be unambiguously identified \citep{vandeneijnden2018_her}.

\subsection{The case of symbiotic X-ray binaries}
\label{sec:syxrbs}

Finally, we turn to the SyXRBs. All but one \citep{shaw20} of known SyXRBs \citep[nine confirmed and three candidate systems; e.g.][]{bahramian14,bahramian17,qiu17,kennea17,bozzo18} host a strongly-magnetised neutron star. Therefore, they are interesting analogues to (wide) wind-fed HMXBs. Within the wind capture radius, their accretion flows and possible jets could be similar, while the donor wind itself can be quite different. Particularly, the wind velocities of late-type giants are significantly lower, at typically $\sim 100$ km/s, while their mass loss rates tend to be lower and span a large possible range of $\sim 10^{-10}$--$10^{-5}$ $M_{\odot}$/yr \citep{espey08,enoto14}. This combination of lower velocity, boosting possible thermal wind radio emission, and wide range in mass loss rate, complicates the comparison of the jet and wind scenario for the three SyXRBs in this study: GX 1+4, 4U 1954+31, and the candidate 3XMM J181923.7-170616. 

We detect two out of the three SyXRBs systems in the radio, namely GX 1+4 and 4U 1954+31. The former is detected during four epochs at levels between $\sim 40$--$100$ $\mu$Jy, while the latter is detected in one of two observations at $21 \pm 5$ $\mu$Jy. During its second observation, no X-ray information was available, although it is likely that 4U 1954+31 remained in the same faint X-ray state as during the first observation. In that scenario, the radio non-detection is likely due to a poorer sensitivity during the second observation (see also Table 1 in the Online Supplementary Materials). Finally, 3XMM J181923.7-170616 was not detected, with a $3$--$\sigma$ upper limit of $36$ $\mu$Jy at $5.5$ GHz. 

The wind velocity in GX 1+4 is inferred to be $\sim 100$ km/s \citep{chakrabarty97,hinkle06}, while no direct measurements are available for the other two sources. Assuming all three have similar wind velocities, we can invert Equation \ref{eq:wind} to estimate the required mass loss rates to explain their observed radio properties. This yields $\dot{M}_{\rm wind} \approx (1.4$--$2.7)\times10^{-7}$ $M_{\odot}$/yr for GX 1+4, $\dot{M}_{\rm wind} \approx 1.1\times10^{-8}$ $M_{\odot}$/yr for 4U 1954+31, and $\dot{M}_{\rm wind} \leq 4.5\times10^{-7}$ $M_{\odot}$/yr for 3XMM J181923.7-170616. All these estimates are consistent with the large range observed for late-type giants. However, via independent methods based on the donor star properties, \citet{vandeneijnden2018_gx} estimate an upper limit of $\dot{M}_{\rm wind} \leq 7\times10^{-8}$ $M_{\odot}$/yr on the mass loss rate in GX 1+4. While estimating wind mass loss rates is challenging, if true, that estimate would rule out a wind origin of the radio emission of GX 1+4. 

During the first (and likely during the second) radio observation, 4U 1954+31 was observed to be in a faint X-ray state by \textit{Swift}/BAT and \textit{MAXI} monitoring. This source shows X-ray variability on a wide range of time scales, from flares lasting hundreds of seconds pointing to a very clumpy, inhomogeneous wind in density and ionisation, to slower variations on 200-400 day time scales \citep{masetti07a,enoto14}. The origin of the latter, slow evolution is unknown, but its long time scale makes it unlikely to be related to local inhomogeneities in the wind. Instead, it might relate to more large-scale changes in wind velocity, mass loss rate, or ionisation, due to the orbital phase or changes in the donor. In such a scenario, the low-flux state where we caught 4U 1954+31 might result from lower ionisation or mass loss rate, or high wind velocity. All three of those factors decrease the expected radio luminosity of the wind. 

No radio spectral information was available for 4U 1954+31 and 3XMM J181923.7-170616, while it was for the final three observations of GX 1+4. Of those, only the second observation shows a spectral shape ($\alpha = -0.6 \pm 0.6$) inconsistent with the expected thermal wind spectrum at $1\sigma$, although the uncertainty is large and $\alpha$ varies by more than $1\sigma$ between the observations. 

We can also briefly compare the observed radio properties of the three SyXRBs with a jet scenario. The measured spectral indices for GX 1+4 are consistent with a synchrotron-emitting compact radio jet, while the radio luminosities of the two detected SyXRBs fit with the distribution seen from other strongly-magnetised neutron stars. Therefore, a jet could also plausibly explain the radio emission. Without a better theoretical understanding of jets from strongly-magnetised neutron stars (see Section \ref{sec:disc_part1}), there are no further tests of this scenario that we can perform with the current data. However, once the wind has been captured by the neutron star, we do not expect significant differences with jet-launching HMXBs.

The non-detection of 3XMM J181923.7-170616 is not surprising in either the wind or jet interpretations. Firstly, its $8$-kpc distance yields an upper limit of $L_{\rm R} \leq 2\times10^{28}$ erg/s, while we find that no strongly-magnetised neutron stars appear above a similar maximum luminosity. In addition to its large distance, we do not find a strong constraint on the wind mass loss rate corresponding to the upper limit: it is realistic that the stellar wind indeed follows $\dot{M}_{\rm wind} \leq 4.5\times10^{-7}$ $M_{\odot}$/yr \citep{espey08}. Finally, a large orbital size might imply a low wind capture rate and ionisation, possibly decreasing the wind and jet luminosities. However, GX 1+4 likely has a 1161 day period \citep{hinkle06}, and for this explanation, 3XMM J181923.7-170616 would require an even larger orbit. 

The white dwarf analogues of SyXRBs, symbiotic stars, are much more numerous and have been characterised in detail in radio over the past decades, allowing for an interesting comparison. Seminal early studies by, e.g., \citet{seaquist84}, \citet{seaquist90}, and \citet{seaquist93} detected unresolved radio emission from a large fraction of symbiotic stars at radio luminosities reaching up to $10^{30}$ erg/s (converted to $10$ GHz). This emission is attributed typically to thermal wind radiation, requiring mass loss rates of the order of $10^{-7}$--$10^{-10}$ $M_{\odot}$/yr. More recent studies have confirmed that those symbiotic stars that show hydrogren shell burning on their surface show unresolved thermal wind emission at luminosities between $10^{28}$ -- $10^{30}$ erg/s, while the non-shell-burning systems tend to be a factor $10$-$100$ fainter in radio \citep{weston16a,weston16b}. Possibly, this difference might arise from a difference in mass accretion rate, which could correspond to lower mass loss rate and therefore wind radio luminosity. Alternatively, the X-ray photons from shell burning might enhance the ionisation of the stellar wind \citep[see][for a recent review]{sokoloski17}. 

Symbiotic stars also launch jets: MWC 560 launches an unresolved radio jet \citep{lucy19}, while in several close-by symbiotic stars, resolved jets have also been detected \citep{padin85,dougherty95,ogley02,brocksopp03,brocksopp04,karovska10}. The radio luminosities of these jets lie in the range of $10^{27}$ - $10^{30}$ erg/s, overlapping with, but up to higher luminosities than, the range for strongly-magnetised neutron stars. A comparison of stellar wind properties between symbiotic stars and SyXRB might be obvious due to the shared donor star types, but the inner accretion flow has similarities too: as argued before, these non-magnetic white dwarfs have sizes similar to the magnetospheric radius of a $10^{12}$ G, $1.4$ $M_{\odot}$ neutron star accreting at $L_X \approx 10^{36}$ erg s$^{-1}$. Therefore, a jet comparison between SyXRBs and symbiotic stars can be valid as well. 

What does a radio comparison with symbiotic stars reveal about SyXRBs? Firstly, the wind loss rates required to explain the SyXRB radio emission are consistent with those seen in symbiotic stars. More interestingly, despite similar donor wind properties, symbiotic stars reach up to two orders of magnitude higher radio luminosities. What could explain such a difference? Those high radio luminosities are seen in shell-burning systems, which have a continuous, additional source of ionising photons. This can maintain a higher degree of ionisation in the wind and thereby increase its radio emission. Similar thermonuclear burning is possible on neutron stars, but is not sustained for similar lengths of time and does not occur on strongly-magnetised neutron stars \citep[e.g.][]{galloway2008}. In addition, the accretion flux is significantly softer and therefore more ionising in symbiotic stars comared to SyXRBs. As an alternative to ionisation, we might simply observe an effect of small number statistics: the number of known SyXRB is of the order of ten, while the number of known and candidate symbiotic stars is of the order of thousands \citep{belczynski00,corradi08,corradi10}. Therefore, we might simply not know of any SyXRBs with high enough mass loss rates to reach higher radio luminosity. Finally, the difference in maximum radio luminosity might arise from a maximum jet luminosity for strongly-magnetised neutron stars (see Section \ref{sec:disc_part1}), if a jet dominates the radio emission from SyXRBs. 

Concluding our discussion of SyXRBs, we find that their radio properties fit either a jet or stellar wind interpretation. Symbiotic stars offer an interesting comparison, but do not make allow us to disentangle these scenarios for SyXRBs. Following our approach for Vela X-1 and 4U 1700-37, we will in the remainder of this discussion assume that a jet contributes to the observed radio emission, in order to derive as complete as set of constraints on jet physics from strongly-magnetised neutron stars as possible.

\section{Implications for neutron star jet physics}
\label{sec:disc_II}

After discussing the possible physical origins of the detected radio emission in the previous section, we will now turn to the implications for jet physics. We stress that for this section, we assume that all detected radio emission originates in a jet in order to use the most information possible. However, as detailed in the previous section, it is not certain that all detected sources indeed launch a jet.

\subsection{The radio brightness of neutron stars' jets}    
\label{sec:disc_part1}

As shown in Figure \ref{fig:hists}, despite some overlap, the typical radio luminosities of accreting BHs, weakly, and strongly magnetized NSs differ substantially: black holes are, as a sample, radio brighter than weakly-magnetised neutron stars, which are in turn radio brighter than strongly-magnetised neutron stars. This difference was known already between the first two source classes \citep[e.g.][]{fender01,migliari06}, and is further confirmed by our new observations \citep[e.g.][which already include some of our preliminary results]{gallo18}. We have extended this comparison to stronger magnetic field sources, which also possess systematically slower spin periods (e.g. $\nu \lesssim 1 $ Hz) than their weakly-magnetised counterparts (e.g. $\nu > 100$ Hz for AMXPs). We will investigate the role of magnetic field and spin in the next section, but here we will first discuss the radio luminosity trends in broader terms. 

Comparing the black hole and neutron star X-ray binaries, the former appear to be radio brighter over the entire explored range in X-ray luminosity (e.g. $\sim 10^{33}$--$10^{39}$ erg/s). However, comparing the two classes of accreting neutron stars in more detail, the story is more complicated. While generally, strongly-magnetised neutron stars are fainter, this effect establishes itself most clearly above X-ray luminosities of $\sim 10^{37}$ erg/s (see Figure \ref{fig:permagfieldclass}). Between $\sim 10^{35}$ and $10^{37}$ erg/s (i.e. $\sim 0.1$ -- $10$\% of the Eddington luminosity for neutron stars), the strongly-magnetised neutron stars overlap with the fainter detections of their weakly-magnetised counterparts. In the highest X-ray luminosity ranges, the comparison is dominated by Swift J0243.6+6124, as that parameter space is poorly explored for HMXBs. However, we do not expect that these sources -- or, at least, the persistent sources -- can be much radio brighter than Swift J0243.6+6124, as otherwise they would have been detected more easily in earlier decades. Therefore, it appears as if the strongly-magnetised neutron stars reach a ceiling radio luminosity of $\sim 3\times10^{28}$ erg/s, while the weakly-magnetised neutron stars continue a scattered but broadly-coupled increase between X-ray and radio luminosity. This difference becomes apparent above $10^{37}$ erg/s.

One can then ask whether the overlap between the two neutron star samples between $L_X \sim 10^{35}$ and $10^{37}$ erg/s is merely a coincidence or hints towards similarities in the launching process. For instance, at lower X-ray luminosity, the characteristic radii for the jet launching process might be located further away from the neutron star -- whether these radii are set by the size of the Comptonising medium, the inner radius of the accretion disc, or the location and height of the jet base. At such launching radii, less gravitational potential energy is available to tap for the launch of an outflow. As the accretion rate and X-ray luminosity increase, and the characteristic radii move inwards, more energy becomes available. However, in the strongly-magnetised neutron star case, the magnetospheric radius, where accretion flow and magnetosphere pressures are equal, will be significantly larger and could, in this scenario, create a maximum power available to launch a jet (where we assume that radio luminosity and power are directly related). This description is purposefully vague, as we discuss more detailed models in the next section. However, while we cannot exclude a coincidental overlap in radio luminosity and this scenario ignores the significant differences in accretion flow structure between Roche-lobe and wind-accreting systems, this scenario might qualitatively account for the observed divergence at higher accretion rates.

\subsection{The role of neutron star magnetic field and spin}
\label{sec:disc_part2}

In this section, we will discuss in more detail whether the neutron star magnetic field and/or spin can affect the jet power; or, phrased more accurately, whether our observations contain evidence for such an effect of the neutron star's properties on the radio jet. The brief answer is that we find no such evidence, beyond the maximum radio luminosity of strongly-magnetised neutron stars that might be attributable to their strong magnetic fields, as hypothesised in the previous section. 

\subsubsection{Fast-rotating, weakly-magnetized versus slowly-rotating, strongly-magnetized neutron stars}

The weakly- and strongly-magnetised neutron star samples overlap in radio luminosity in the X-ray luminosity range between $\sim 10^{35}$--$10^{37}$ erg/s. We cannot exclude that this similarity in spectral shape and radio brightness is completely due to coincidence, while the jets of strongly and weakly magnetised neutron stars would be launched by completely independent mechanisms. For instance, such a scenario might be at work if the radio observing band probes regions down the jet where the emission properties are uncoupled from the exact launching mechanism. However, assuming a relation between launch mechanism, jet power, and jet luminosity (but see below), this scenario still requires a similar jet power by coincidence. 

Alternatively, the overlap in radio luminosity could result from a similar jet launching mechanism for both classes of accreting neutron stars. In that case, the enormous differences in magnetic field of three to four orders of magnitude, in neutron star spin of up to six orders of magnitude, and in accretion flow geometry between Roche-lobe-overflowing and wind-accreting systems, do not have a large effect on the radio luminosity of the jet (beyond the apparent maximum for strongly-magnetised systems). This lack of effect might imply that the jet launching mechanism does not depend, in any significant and detectable manner, on the neutron star properties or mass transfer type. On the other hand, these properties might have an effect, but degeneracies between their influence could mean that our sample remains too small to disentangle these. Finally, there is the recurring challenge in radio observations of jets that these might not directly probe the jet power. What we measure in this work is the radio flux at maximally two frequencies, while the conversion to jet luminosity depends for instance on the complete, unmeasured, broadband jet spectrum. Our limited information about the actual jet power therefore further washes out any effects that magnetic field and spin might have -- not only in our sample, but also in the archival comparison data.

\subsubsection{Among slowly-rotating, strongly-magnetized neutron stars}

Prior searches of spin and magnetic field effects, such as by \citet{migliari11b}, have always focused on the class of weakly-magnetised neutron stars. With the first sample of radio detections from strongly-magnetised neutron stars, we can search for such effects within this source class as well. Focusing on this subgroup circumvents the question whether different jet launch mechanisms exist for the weakly and strongly-magnetised types, as we will assume that within this subgroup, only a single mechanism is responsible for jet launching. 

While neutron stars in HMXBs tend to have similar magnetic fields -- or at least within an order of magnitude, between $10^{12}$--$10^{13}$ G -- they show a spread in their radio luminosities at similar $L_X$ (e.g. Figure \ref{fig:hists}). Most notably, there is the difference between detected and non-detected targets that is not merely due to differences in distance, despite their similar neutron star properties and X-ray luminosity. What could explain these variations between sources? Firstly, it does not clearly correlate with subclass, as a BeXRB, SgXBs, SyXRBs, and an IMXB are all detected. Secondly, it could simply be another example of the spread in radio luminosity that is also observed in LMXB jets, regardless of the accretor nature \citep{gallo18}: both black hole and neutron star X-ray binaries show such spread at a similar level. While this spread remains poorly understood \citep{tudor17,espinasse18,motta19} both for individual sources and the sample as a whole, it shows that jet launching mechanisms do not create a one-to-one relation between X-ray and radio luminosity. Given the small difference between maximum and sensitivity-limited radio luminosity for strongly-magnetised neutron stars (i.e. $3\times10^{27}$ to $3\times10^{28}$ erg/s), a measurement of the scatter in the radio luminosities would be heavily biased by the sensitivity limit, artificially posing an upper limit on the amount of observable scatter in the detected sources.  

\begin{figure*}
	\includegraphics[width=\textwidth]{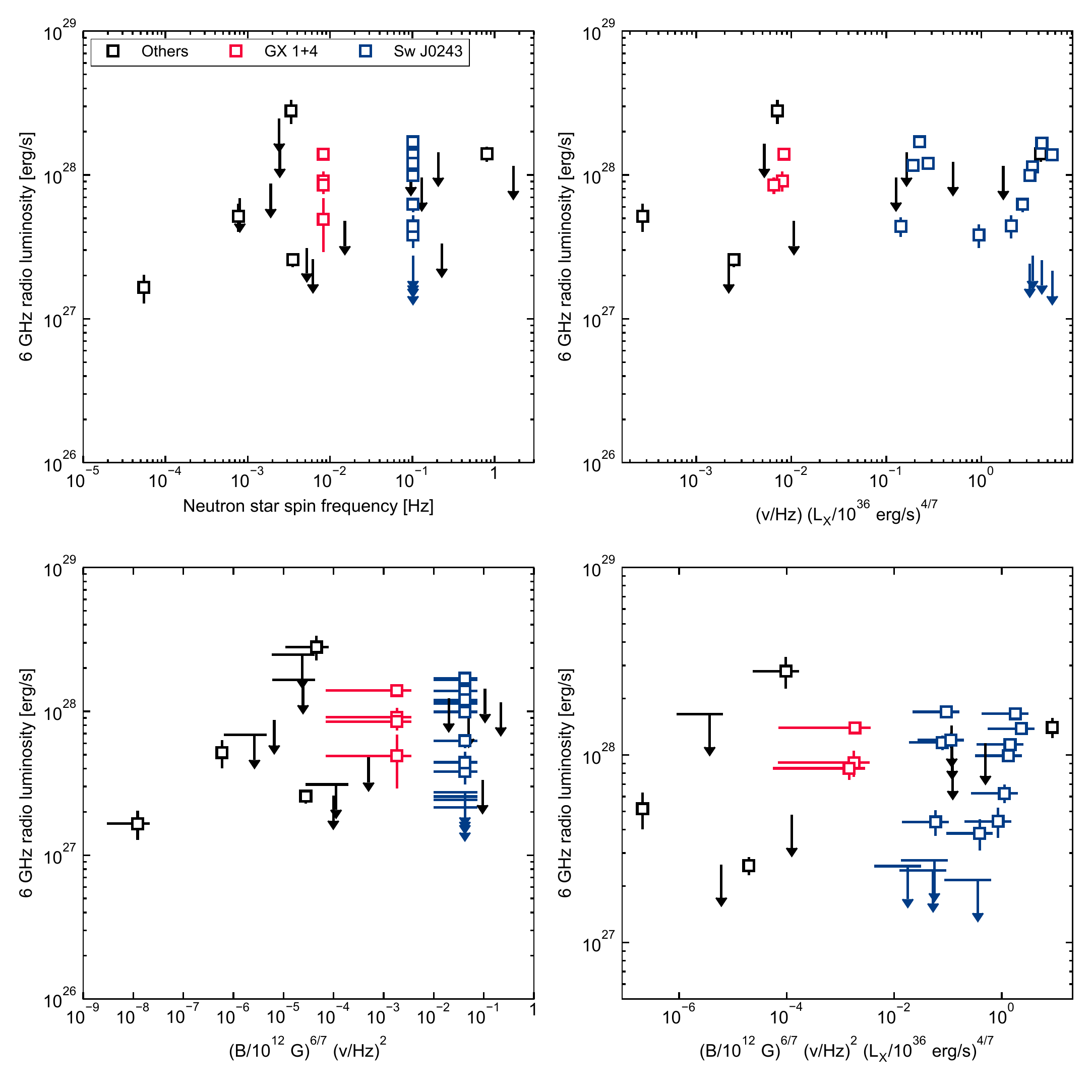}
    \caption{The relation between radio luminosity and neutron star parameters for strongly-magnetised neutron stars. In all panels, Swift J0243.6+6124 (referred to as Sw J0243) and GX 1+4 are highlighted separately, as these are the only two sources with multiple data points. \textit{Top left}: radio luminosity plotted versus neutron star spin $\nu$. \textit{Top right}: radio luminosity plotted versus neutron star spin, corrected for mass accretion rate, parameterised by X-ray luminosity $L_X$ following the prescription in \citet{parfrey16}. \textit{Bottom left:} radio luminosity versus the jet power scaling with magnetic field $B$ and spin of the model by \citet{parfrey16}. \textit{Bottom right:} same as bottom left, again correcting for mass accretion rate following \citet{parfrey16}.}
    \label{fig:all_four}
\end{figure*}

Could the neutron star magnetic field or spin contribute to or explain the spread in observed radio luminosity \citep[e.g.][]{migliari11b}? For this, we consider several scenarios (where we, again, simplistically use radio luminosity as a direct proxy for jet power). Firstly, the radio luminosity does not scale with the magnetic field in our sample: all strongly-magnetised accreting neutron stars have similar magnetic fields (when measured), which cannot account for the variation between sources that we observe. The targets in our sample span a much larger range in spin frequencies, between $<10^{-3}$ and $>1$ Hz. However, plotting the radio luminosity at $6$ GHz versus neutron star spin in Figure \ref{fig:all_four} (top left panel) does not reveal a clear relation -- in fact, the radio brightest object, 1E 1145.1-6141, is amongst the most slowly rotating neutron stars in our sample ($3.367\times10^{-3}$ Hz), while the radio detected but faint Vela X-1 has a similar spin of $3.5\times10^{-3}$ Hz. Of course, searching for a scaling with only spin and not taking the effect of mass accretion rate differences into account, is a naive approach. However, no scaling between X-ray and radio luminosity is apparent from the data, nor have we yet considered any model of these effects. 

Such models, that take into account magnetic field, spin, and mass accretion rate, do exist. For instance, we can consider whether the launching mechanism might be related to a magnetic propeller. In the magnetic propeller regime, the radius where the accretion flow's and magnetosphere's pressures are equal (\textit{the magnetospheric radius $R_{\rm m}$}), lies far outside the radius where the neutron star's and accretion flow's rotational velocities are equal \citep[\textit{the co-rotation radius $R_{\rm co}$;}][]{illarionov75,dangelo12}. As the Keplerian velocity of the disc decreases with distance to the neutron star, this implies that the magnetosphere rotates faster than the flow and expels material in an outflow, instead of channeling it to the magnetic poles. Whether a source sits in this regime depends on a careful balance of magnetic field strength and mass accretion rate, which set $R_{\rm m}$, with the neutron star spin, which sets $R_{\rm co}$. 

During our radio observations, none of the detected strongly-magnetised neutron stars were at luminosities which would have been likely to place them in the propeller regime, regardless of the calculation assumptions about the accretion flow structure (i.e. disc or spherical accretion). This supports the hypothesis put forward in \citet{vandeneijnden2019_reb} to explain the sudden turn on of the jet of Swift J0243.6+6124 during an X-ray re-brightening: depending on the exact magnetic field strength, this turn on could be matched to the transition out of the propeller regime, allowing material to reach closer to the neutron star into a possible jet launching region. Therefore, in such a scenario, no jet is expected for sources residing in their propeller regimes. 

A more extensive model for the launch of neutron star jets, regardless of magnetic field strength, is the model proposed by \citet{parfrey16}. Both analytically and through simulations, it investigates the possibility that the accretion flow opens up magnetic field lines, releasing energy to launch outflows. The strength of the outflow in this scenario scales with magnetic moment to the power $6/7$ and with the spin frequency squared. Therefore, we plot the radio luminosity as a function of this scaling in Figure \ref{fig:all_four} (bottom left panel) in arbitrary units: $(B/10^{12} G)^{6/7} \times (\nu/1\rm{ Hz})^2$. For sources where the magnetic field is not measured from the cyclotron line, we plot a range corresponding to $B = 10^{12}$ -- $10^{13}$ G, as is typical for these slow pulsars. As in the top left panel in the same figure, we again highlight Swift J0243.6+6124 and GX 1+4, which have multiple observations. However, this exercise again reveals no obvious scaling or trends. 

So far, we have neglected the effects of variations in mass accretion rate. While the global X-ray -- radio luminosity diagram does not clearly reveal such a correlation in the strongly-magnetised sample, we can consider it for the \citet{parfrey16} model: this model also includes a scaling with X-ray luminosity, as a probe of the mass accretion rate, with a power of $4/7$. Therefore, we add this scaling in the two right panels of Figure \ref{fig:all_four}. We have removed the two sources that were only detected in radio and therefore do not have a measured X-ray luminosity (i.e. 4U 1954+31 and IGR J16318-4848). However, including this correction, we again find no clear trend or correlation. 

So, to conclude, our radio sample study of strongly-magnetised accreting neutron stars does not show any obvious evidence for a coupling between radio luminosity and neutron star spin, magnetic field, or mass accreting rate. The only possible effect of magnetic field strength, revealed through the comparison with weakly-magnetised neutron stars, is the apparent presence of a ceiling in jet luminosity for strong-magnetic field neutron stars. Beyond that, the scatter in radio luminosities is similar in the strongly and weakly-magnetised neutron stars, as is it in the black hole LMXBs.   

\subsection{Individual sources and source classes}
\label{sec:disc_part3}

\subsubsection{Radio non-detected source classes: UCXBs, VFXBs, and qBeXRBs}
\label{sec:nondets}
For three source classes, we do not obtain any new radio detections: UCXBs, VFXBs, and BeXRBs. The former two are both classes of LMXBs, that overlap: UCXBs are binary systems with orbital periods shorter than $1$ hour, while the VFXBs are persistent or transient X-ray binaries with maximum X-ray luminosities of $\sim 10^{36}$ erg/s \citep{wijnands16}. A possible explanation for their faintness is that the VFXBs are UCXBs with small accretion disk. Alternatively, in neutron star systems, a dynamically active magnetic field could also inhibit efficient accretion and high X-ray luminosities \citep[e.g.][]{wijnands08,heinke15,degenaar17}. We observed four confirmed UCXBs (4U 1626-67 with a strong magnetic field, and 2S 0918-549, 4U 1246-588, and IGR J17062-6143 with weak magnetic fields) and four VFXBs (XMMU J174716.1-281048, 1RXH J173523.7-354013, AX J1754.2-2754, and, again, IGR J17062-6143), not detecting any of them in radio.

Starting with the UCXBs, the deepest constraints on radio emission are found for 2S 0918-549 and 4U 1246-588: both are detected below $10^{36}$ erg/s in X-rays while their radio luminosity does not exceed $\sim 3\times 10^{27}$ erg/s -- an order of magnitude lower than typical radio-detected weakly-magnetised neutron stars at similar X-ray luminosities. From the perspective of jet formation, UCXBs might offer an interesting view given their non-standard disc composition: while ordinary LMXBs have a hydrogen-dominated disc, the discs of UCXBs are typically hydrogen-deficient \citep{nelemans10,hernandezsantisteban19}. In the case of a white dwarf donor, the disc might instead be formed predominantly of helium, with high abundances of carbon, oxygen, neon, and/or magnesium \citep{juett01}. Such a different composition changes the ionisation properties of the disc \citep{ludlam2019} as well as the charge-to-mass ratio, possibly affecting the interaction between the disc and magnetic field and the formation of outflows. 

Reported radio observations of UCXBs typically detect a jet, assuming that the source's X-ray luminosity exceeds $L_X \sim 10^{36}$ erg/s. Comparing the donor types of these reported UCXBs \citep[e.g.][]{rappaport87,nelemans04,madej13,homer96,dieball05,galloway02,sanna17,sanna2018} and the four in our sample \citep{hemphill19,intzand05,vandeneijnden2018_igr,hernandezsantisteban19} reveals no systematic difference between the two classes: both groups are dominated by white dwarf donors, of various types when identified. Therefore, differences in accreted material or donor type do not appear to explain the difference in radio brightness. 

The radio behaviour of VFXBs is interesting for a different reason: several authors have previously suggested that these systems might be transitional millisecond pulsars, or tMSPs \citep{heinke15,degenaar17}. These systems are neutron star binaries switching between an X-ray binary and radio pulsar state \citep{archibald09,papitto13}, of which only three are known and which are thought to form an evolutionary link between X-ray binaries and radio millisecond pulsars. At higher X-ray luminosities than we targeted for the VFXBs XMMU J174716.1-281048, 1RXH J173523.7-354013, and AX J1754.2-2754, the tMSP M28i appears relatively radio bright within the neutron star sample \citep[e.g][]{deller2015,jaodand18}. Our radio upper limits on the three VFXBs lie firmly below the radio luminosities of the tMSPs at low X-ray luminosities. That argues that either these three VFXBs are not tMSPs\footnote{Indeed, \citet{shaw20} suggest that a fraction of the VFXB population could instead be SyXRB systems, i.e. with an evolved donor. Alternative models for the VFXB population include ultracompact orbits \citep{heinke15} or magnetic inhibition of the accretion flow \citep{wijnands08,degenaar17}.}, or that no single tMSP X-ray -- radio correlation exists.

The final undetected class of sources in our new radio observations are the BeXRBs. The specific aim of observing these targets was to search for radio signatures of the propeller regime: with their low accretion rates and strong magnetic fields, BeXRBs would be prime targets to observe this state. Out of the four sources, V 0332+53 clearly resided in the propeller state during the radio observations: we measure an X-ray luminosity upper limit of $L_X < 3\times10^{34}$ erg/s, while the maximum X-ray luminosity of the propeller state is estimated to be $L_{\rm lim} = 3\times10^{35}$ erg/s \citep[following][and assuming disk accretion, i.e. $k\equiv0.5$. See Table 2 in the Online Supplementary Materials for the used source parameters]{tsygankov18}. Therefore, we can place an upper limit of $L_R < 3\times10^{27}$ erg/s on the radio emission of any type of propeller outflow that might be launched in this system. 

The two X-ray detected BeXRBs, V*V490 Cep and MXB 0656-072, did not reside in their propeller states during out observations: we detect these sources at respectively $L_X = 8.9\times10^{34}$ erg/s and $7.6\times10^{33}$ erg/s, above their respective propeller limits of $L_{\rm lim} = 4.6\times10^{33}$ erg/s and $5.0\times10^{32}$ erg/s. Finally, SAX J2239.3+6116 was not detected at $L_X < 8\times10^{33}$ erg/s, while its propeller regime is expected below $L_{\rm lim} = 6\times10^{29}$ erg/s -- significantly below the observed upper limit. However, at the order of magnitude of the propeller regime limit, the source is likely not actively accreting. In addition, both MXB 0656-072 and SAX J2239.3+6116 spin at frequencies below $0.01$ Hz. In the cold disk model by \citet{tsygankov17}, only neutron stars spinning faster than $0.01$ Hz are capable of entering the propeller state; otherwise, the disk recombines into its un-ionised state before decreasing in X-ray luminosity below the propeller limit. Therefore, we conclude that these three BeXRBs were not in their propeller states. 

\subsubsection{Soft state neutron stars: jet quenching and spectral shapes}
\label{sec:softstates}
The previous subsection discussed sources that were radio non-detected at low X-ray luminosity. However, radio non-detections can also occur at higher X-ray luminosities, in particular in the soft state, where the inner disc is expected to be geometrically thin. This phenomenon, where the radio emission from the compact jet quenches as the source transitions between hard to soft state, is observed in all black hole LMXBs with good monitoring across the transition or in both states \citep{tananbaum72b,harmon95,fender99,gallo03,fender09,millerjones12}. This state transition is also often accompanied with optically thin radio flaring, likely associated with the launch of discrete ejecta \citep{fender04}. Once the black hole has fully entered the soft state, no compact core radio emission is detected anymore \citep{coriat11}; most extremely, upper limits on the compact core emission have been measured more than 3.5 order of magnitude below the hard state flux in the black hole LMXB MAXI J1535-571 \citep{russell2019_1535}. Once black hole LMXBs later return to the hard state, at lower X-ray luminosity, the compact jet re-establishes \citep{fender04}.

For neutron star X-ray binaries, this picture is more complex (note: we only focus on LMXBs here, where hard and soft states equivalent to black hole systems can be identified). In three neutron star X-ray binaries, the radio emission has been studied in both the hard and the soft state. Comparing these two states, \citet{migliari03} find only marginal evidence of radio quenching in the persistent X-ray binary 4U 1728-34. For the transient Aql X-1, \citet{millerjones10} report quenching by at least an order of magnitude in the soft state \citep[although a more recent campaign also caught a much more radio bright soft state of Aql X-1; see][and below]{diaztrigo18}. Finally, \citet{gusinskaia17} observed how the transient 1RXS J180408.9-342058 quenched as well during its transition to the soft state. Considering these three sources, one notices that the difference between hard state detections and soft state limits is not as large as in black hole LMXBs, as a result of the faintness of the neutron star systems. Four more atoll sources have been observed in radio only in their soft state: three of those (4U 1820-30 and Ser X-1 in \citealt{migliari04}, and MXB 1730-355 in \citealt{rutledge98}) were detected while GX 9+9 was not \citep{migliari11c}. Whithout a hard state observation, however, it is unclear whether the three detected sources significantly differ in radio luminosity between states. All combined, the picture emerges that some neutron stars quench while others do not, and the origin of this  difference remains debated \citep[e.g.][]{fender16,gusinskaia17}. 

\begin{figure}
	\includegraphics[width=\columnwidth]{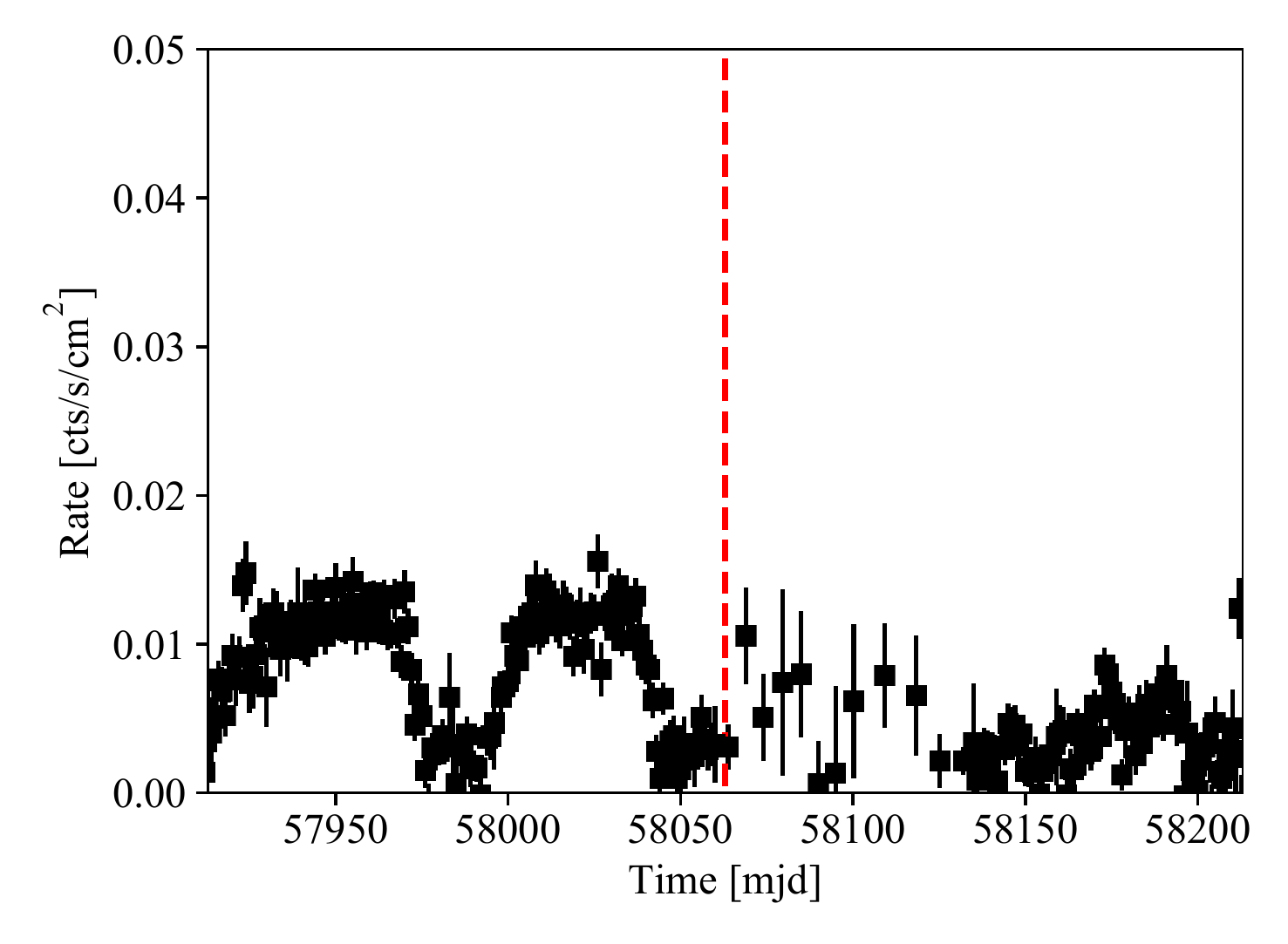}
    \caption{\textit{Swift}/BAT 15--50 keV monitoring light curve of 4U 1702-429. The red dashed line show the time of the radio observation. Two distinct states, with count rates around 0.015 and below 0.01 ct/s/cm$^2$ can be distinguished: the hard and soft state, respectively. During the radio observation, 4U 1702-429 was in the soft state. Data between MJD 58061 and 58130 have been re-binned by a factor 5 to increase signal-to-noise.}
    \label{fig:4u1702}
\end{figure}

The sample of $13$ neutron star LMXBs presented in this paper contains several sources in their soft state: GX 9+1, GX 9+9, and GX 3+1 all persistently reside in the soft state. The persistent atoll 4U 1702-429 switches between hard and soft states. While no simultaneous, high-quality X-ray data was taken during its radio observation, we think it is likely that 4U 1702-429 also resided in a soft state: during a \textit{Swift} observation taken 8 days before the radio epoch, the source showed an X-ray luminosity of $\sim 1.4\times10^{37}$ erg/s, or close to $10\%$ of the Eddington limit for neutron stars. For comparison, when the source was observed to be in its hard state with \textit{NuSTAR} by \citet{ludlam2019}, its X-ray luminosity was a factor 10 lower. The XRT spectrum also requires the addition of a soft thermal component, not seen in this source's hard state, at a significance of $3.7\sigma$ ($p = 0.0001$). This interpretation is supported by \textit{Swift}/BAT monitoring, shown in Figure \ref{fig:4u1702}, wherein 4U 1702-429 switches between brighter ($\sim 0.015$ ct/s/cm$^2$) and fainter states ($< 0.01$ ct/s/cm$^2$). As \textit{Swift}/BAT is sensitive at hard X-ray energies, i.e. $15$--$50$ keV, the brighter state corresponds to the source's hard state. At the time of the radio observation, 4U 1702-429 was in the faint \textit{Swift}/BAT state, that had already started during the \textit{Swift}/XRT observation. Finally, as discussed more extensively later on in this section, it shows a steep radio spectral index, which is not typically seen in neutron star hard states (compare with, for instance, IGR J17379-3747 in our sample). 

What do these four soft state atolls tell us? Firstly, we do not detect radio emission from GX 9+9, confirming the earlier results from \citet{migliari11c}, and GX 9+1, adding an extra example of a radio-undetected soft-state-only neutron star. However, to complicate matters, GX 3+1, which is very similar to the other two sources in terms of its X-ray properties, \textit{is} detected in radio. While black holes appear to always quench in the soft state, neutron stars show two types of behaviour, as highlighted by these three sources. Finally, 4U 1702-429 is detected as well, with a steep radio spectrum between the two ATCA bands at $5.5$ and $9$ GHz. 

As none of the four sources above has been observed in radio during a hard state, we stress that a radio detection does not automatically exclude quenching: if GX 3+1 were to be a factor $>10$ times radio brighter in the hard state (i.e. $> 7\times10^{28}$ erg/s), this would be consistent with other hard state atolls (see Figure \ref{fig:persource}). 4U 1702-429, on the other hand, would be among the radio brightest neutron stars in its hard state (i.e. $L_R > 4\times10^{29}$ erg/s) in the case for a similar soft state radio quenching factor of 10.

Neutron stars, both in the literature and our sample, thus show a range in soft state radio behaviour. To better understand this variance, we can consider the nature of the radio emission during the soft state, if present. Does it originate from the compact jet, or do we observe discrete ejecta? When the spectral index has changed, do we observe a change in jet break frequency? To answer such questions, we need to consider sources with multi-band observations, such as 4U 1702-429: how does its negative spectral index compare to other (soft state) neutron star X-ray binaries?

The best jet spectral shape constraints for neutron stars are the four cases where the jet spectral break is measured: 4U 0614+091, a persistent atoll and UCXB observed in its hard state with spectral break frequency $\nu_b \approx (1-4)\times10^{13}$ Hz \citep{migliari10}; 1RXS J180408.9-342058, the aforementioned transient with a hard state $\nu_b \sim 5\times10^{14}$ Hz \citep[][see also \citealt{diaztrigo17}]{baglio16}; the persistent source 4U 1728-34, where a $\nu_b$ of $(1.3-11)\times10^{13}$ Hz was measured during the transition from its hard to soft state \citep{diaztrigo17}; and the transient Aql X-1 \citep{diaztrigo18}. In the latter, the break frequency was constrained in four epochs, with a break frequency in the range of $30$--$100$ GHz during the hard state but a break frequency below the ATCA band ($\nu_b < 5.5$ GHz) during the soft state. Finally, deep ATCA+ALMA observations of the persistent UCXB 4U 1820-30 in its soft state did not detect a spectral break, implying a inverted radio-to-sub-mm spectrum.

In our sample, we cannot infer a spectral shape for the two non-detected soft-state sources, and GX 3+1 is too faint for a meaningful spectral shape measurement ($26.6\pm 5.1$ $\mu$Jy). 4U 1702-429, we measure a spectral index of $\alpha = -0.8 \pm 0.3$ between $5.5$ and $9$ GHz. The source likely transitioned into its soft state $15$ days prior (e.g. Figure \ref{fig:4u1702}), suggesting relatively steady soft state emission: while it is possible we caught an optically thin radio flare associated with the launch of individual ejecta, that would be highly coincidental. Instead, we deem it more likely that we are observing a more steady radio source with a spectral break frequency below the ATCA bands.

Prior to this work, \citet{diaztrigo18} suggested that the soft state spectral shapes (up to the ALMA band) of neutron stars might depend on their transient or persistent nature: the transient Aql X-1 shows a steep soft-state radio spectrum, while the persistent 4U 1820-30 shows an inverted spectrum in that state. This also fits with the results for 4U 1728-34, which showed a spectral break in the nIR, although it was only transitioning into the soft state \citep{diaztrigo17}. However, the spectral index measurement in 4U 1702-429 counters this suggestion, showing once more that the differences and similarities in soft state radio behaviour of neutron stars remain challenging to summarise. If the negative spectral index has a similar origin as that in Aql X-1, it suggests that the spectral break might be located at higher frequencies during its hard state. More observations of 4U 1702-429, particularly in its hard state, would be valuable to see how the spectrum and flux evolves between states. Finally, we remind the reader of an important caveat: during one of its hard state epochs with an inverted broad-band spectrum between the ATCA and ALMA bands, Aql X-1 also showed a lower radio flux at $9$ than at $5.5$ GHz \citep{diaztrigo18}. The spectral index measurement for 4U 1702-429 using only ATCA might suffer the same effect, although our data does not allow us to test this option.

\subsubsection{The nature of the compact object in 4U 1700-37}
\label{sec:4u1700}

4U 1700-37 is a well-known HMXB system that was discovered by the \textit{Uhuru} mission \citep{jones73}. It is located relatively nearby at $\sim 1.9$ kpc \citep{ankay01} and its companion OB supergiant HD 153919 is the hottest and most luminous HMXB donor star known \citep{vandermeer04}. The nature of the compact object -- neutron star or black hole -- remains unconfirmed. While its X-ray spectrum resembles that of accreting X-ray pulsars, no pulsations or a cyclotron line have been unambiguously detected \citep[although see][for a possible cyclotron line detection at $\sim 37$ keV]{reynolds97}. The mass of the compact object is possibly larger than $2M_{\odot}$, consistent with a heavy neutron star, similar to Vela X-1, or a low-mass black hole \citep{rubin96,clark02,falanga15}. 

In terms of radio flux density, 4U 1700-37 is the brightest HMXB in our sample. However, as it is also the most proximate, its position in the radio -- X-ray luminosity plane is consistent with the neutron star sources (see Figure \ref{fig:persource}). This position is completely inconsistent with the radio-bright track for black hole X-ray binaries, and roughly an order of magnitude below the radio-faint track of these sources. While not many black hole HMXBs are known, all radio observations of these systems currently place them at the radio-brightest part of the correlation seen in LMXBs \citep{ribo17}. Therefore, while we stress that it does not amount to direct proof, the radio luminosity of 4U 1700-37 is strongly suggestive of a neutron star primary. 

\subsubsection{Deep radio constraints in the literature: 4U 2206+54, A 0535+262, and X Per}
\label{sec:disc_deep}

Finally, we briefly turn to the deepest radio non-detections of strongly-magnetised neutron stars in the literature. With our survey, we have probed down radio luminosities of approximately $10^{27}$ erg/s. Combined with the apparent maximum radio luminosity of these sources of $L_R \approx 3\times10^{28}$ erg/s, this limited range implies that we merely scratched the surface in terms of the full population. What can be expected in future studies probing deeper in radio luminosity?

Before the VLA and ATCA sensitivity upgrades in the past decade, after which all our observations were taken, deep radio observations were made of the neutron star HMXB 4U 2206+54 \citep{blay05}. A non-detection of $F_R < 39$ $\mu$Jy at $8.4$ GHz, given a source distance of $\sim 3$ kpc, translates into an upper limit of $L_R < 3.5\times10^{27}$ erg/s -- still comparable with our sensitivity. While no simultaneous X-ray detection was obtained, two \textit{INTEGRAL} observations within a few weeks places the source at a $4$--$12$ keV luminosity of $4.5$--$8.5\times10^{35}$ erg/s. Westerbork Synthesis Radio Telescope observations of A 0535+262, at a distance of only $2$ kpc, similarly resulted in a non-detection of $L_R < 3.8\times10^{27}$ erg/s \citep{migliari11}. While no X-ray luminosity is reported, the \textit{MAXI} monitoring count rate suggests a $2$--$20$ keV luminosity around $6\times10^{36}$ erg/s, assuming a Crab spectrum for simplicity. 

The deepest existing radio observation of a HMXB neutron star is, to our knowledge, that of the persistent source X Per, with a spin period of $~835$ seconds \citep{white76}. The Westerbork Synthesis Radio Telescope did not detecting the source down to $L_R \lesssim 8\times10^{25}$ erg/s \citep{nelson88}. This exceptionally deep upper limit, due to the small distance ($\sim 800$ pc) to X Per, shows that our radio non-detections are not necessarily due to limited sensitivity alone. Unfortunately, no X-ray information is reported for the time of the radio non-detection. Therefore, it remains unclear what accretion state and rate X Per showed around the radio observation. However, assuming X Per was active, its radio upper limit shows that other accreting, strongly-magnetised neutron stars might also not be detected with future instruments, such as the \textit{next-generation Very Large Array} \citep[see Section \ref{sec:disc_part4}][]{selina18,coppejans18}, because their jets are either incredibly weak or simply do not exist.

\section{Conclusions and future outlook}

\label{sec:disc_part4}

We have presented radio observations of a sample of 36 accreting neutron stars. Based on these observations, we draw the following eight conclusions:
\begin{enumerate}
    \item Strongly-magnetised accreting neutron stars can be detected at radio frequencies; the previously detected source Swift J0243.6+6124 is no outlier (Section \ref{sec:results}).
    \item Strongly-magnetised accreting neutron stars are, as a sample, radio fainter than their weakly-magnetised counterparts, which are in turn fainter than black holes. This difference is established especially at high X-ray luminosities, e.g. $L_X \gtrsim 10^{37}$ erg/s, as the radio luminosity of strongly-magnetised systems appears to reach a ceiling value around $3\times10^{28}$ erg/s (Sections \ref{sec:results} and \ref{sec:disc_part1}).
    \item No new very-faint X-ray binaries, ultracompact X-ray binaries, or faint Be/X-ray binaries were detected in radio (Sections \ref{sec:results} and \ref{sec:nondets}).
    \item Strongly-magnetised accreting neutron stars, as a sample, do not show a clear correlation between X-ray and radio luminosity, although this might be caused by sensitivity limits (Section \ref{sec:results}).
    \item Accretion-powered jets explain the radio detections of weakly-magnetised neutron stars, and are the likely explanation for the radio emission from strongly-magnetised neutron stars. The exceptions to the latter statement are Vela X-1, 4U 1700-37, and the symbiotic X-ray binaries, where the stellar wind possibly contributes (Section \ref{sec:disc_part0}). 
    \item Strongly-magnetised accreting neutron stars do not show any obvious scaling between their radio luminosity and their spin, magnetic field strength, and accretion rate, or combination thereof, as predicted by existing models (Section \ref{sec:disc_part2}).
    \item The radio properties of 4U 1700-37 suggest a neutron star primary (Section \ref{sec:4u1700}).
    \item Like their transient counterparts, persistent neutron star LMXBs can show steep soft-state radio spectra (Section \ref{sec:softstates}).
\end{enumerate}

Finally, we will end this work by briefly turning to the future. As this work is an observational study, we will focus mainly on the observational developments. Firstly, since the first systematic comparison of accreting black holes and neutron stars in the X-ray -- radio luminosity plane \citep{migliari06}, tens of neutron stars have been observed or monitored in radio, amounting to a sample of over sixty X-ray binaries \citep[e.g.][and this work]{gallo18}. It has however proven difficult to obtain monitoring over a large range in X-ray luminosity, due to the inherent faintness of neutron star jets and difficulty in coordinating X-ray and radio observations. Such monitoring would be especially valuable for strongly-magnetised neutron stars, where the only monitored outburst (of Swift J0243.6+6124) has revealed surprising behaviour that will be interesting to compare to other binaries. 

The 23 strongly-magnetised neutron stars with observations at current-day sensitivities have merely scratched the surface; these constitute only a small fraction of all such sources. Observing a larger sample, especially down to lower radio luminosities, will be essential to probe whether or not a scattered, but global correlation between X-ray and radio luminosity exists, that is currently cut off by the detection limit. Future arrays, such as the next-generation VLA (ngVLA) with its planned sensitivity of $\sim 0.23$ $\mu$Jy at $8$ GHz in one hour of observing time \citep{selina18}, will probe 50 times deeper than the current VLA, bringing X-ray binaries in M31 above sensitivity limits. In addition, monitoring of transient sources will complement the persistent sources that currently dominate the sample, while probing the jet's turn-on and fade. 

What are the prospects for high spatial resolution observations of neutron star jets? Among the newly-added neutron star X-ray binaries -- and the strongly-magnetised neutron stars in general -- many persistent sources might have launched jets long enough to create feedback structures, despite their low radio luminosity. At even higher resolution, very-long baseline observations mightresolve the compact jets of neutron stars depending on the size of the radio emitting region: at the high end of the typical range, $\sim 10^7$--$10^9$ $R_g$, this size corresponds to a resolvable $\sim 7$ mas at 2 kpc. However, if the radio emission regions are located towards $10^7$ $R_g$, these will not be resolvable. Discrete ejecta in transient sources, however, might be resolved using the newly-employed MeerKAT radio telescope, following such detections in black hole systems \citep[e.g.][]{russell2019_1535,bright2020}.

For a sufficiently radio bright neutron star, VLBI observations could help distinguish between radio emission from a stellar wind or jet through brightness temperature estimates. Stellar winds typically have brightness temperatures of $T_b \approx 10^4$--$10^5$ K \citep{rybicki79,longair92,russell16}. At $6$ cm, the \textit{Very Long Baseline Array} can reach a spatial resolution of $1.4$ mas. If an accreting neutron star, with a flux density $\geq 0.5$ mJy at $5$ GHz, is unresolved at that resolution, it implies a minimum brightness temperature of $T_b \gtrsim 10^7$ K, arguing against a stellar wind origin. Beyond the flux density required, this constraint does not depend on distance. Assuming the apparent maximum radio luminosity of $\sim 3\times10^{28}$ erg/s holds for all neutron star HMXBs, such a measurement is therefore feasible for sources within $\sim 3$ kpc. 

Two promising avenues that remain poorly explored for neutron star jets are eclipse mapping and polarimetry. Recently, \citet{maccarone20} presented model calculations for the eclipse of the jet by the donor star, showcasing how such measurements can constrain the geomtrical and radiative properties of the jet. Due to the large extend and density profile of the stellar wind in high-mass systems, this technique is most suitable for neutron star LMXBs. Polarimetry could help constrain the origin of the radio emission in HMXBs, as radio jets can be linearly polarized while stellar winds are not expected to be. However, the level of polarization, especially in compact jets, can be lower (i.e. few per cent) than current sensitivities allow to detect \citep{vandeneijnden2018_swj0243}. 

From a multi-wavelength perspective, the recent detection of a jet spectral break in an accreting neutron star with ALMA \citep{diaztrigo18} has shown such studies are possible for the brightest neutron star X-ray binaries. Strongly-magnetised neutron stars would be an interesting target for ALMA or ngVLA, going up to $\sim 100$ GHz, as their jets are likely launched further out, possibly resulting in lower jet break frequencies. The low luminosity of these sources would pose a challenge for ALMA, but are realistically detectable with the \textit{ngVLA}. At higher frequencies, in for instance the infrared band, studies of strongly-magnetised neutron stars will often be limited by the bright emission of their high-mass companion stars. However, in sources with low or intermediate-mass donors, such as Her X-1, nIR observations could also probe the jet launch regions. 

As a final remark, all above suggestions for future studies are observational. These studies will add to the broad set of observational constraints on jet formation already existing. From the side of new jet launch models, testable predictions, such as scaling relations between radio luminosity and magnetic field, spin, and mass accretion rate, are essential to match to the observational constraints and guide the design of new observing campaigns. Similarly, predictions from either analytical theory or GRMHD simulations for the effect of binary orbit properties, mass transfer type, and donor star type on the jets might in the future be testable given the large variety in types of HMXBs. 


\section*{Acknowledgements}
We thank the anonymous referee for their report, which improved the quality of this work. JvdE is supported by a Lee Hysan Junior Research Fellowship awarded by St. Hilda's College, and, together with ND, by an NWO Vidi grant awarded to ND. JvdE is grateful for the hospitality of the International Centre for Radio Astronomy Research at Curtin University and the University of Michigan, where part of this research was performed. JvdE thanks K. Parfrey, R. Fender, K. Pottschmidt, and N. Rea for interesting discussions, and V. Tudor for performing ATCA observations. COH is supported by NSERC Discovery Grant RGPIN-2016-04602. The authors thank the directors and schedulers of both ATCA and \textit{Swift} for accepting and rapidly performing several Director's Discretionary Time observations reported in this work. The authors acknowledge the use of public data from the Swift data archive. This research has made use of MAXI data provided by RIKEN, JAXA and the MAXI team. The National Radio Astronomy Observatory is a facility of the National Science Foundation operated under cooperative agreement by Associated Universities, Inc. The Australia Telescope Compact Array is part of the Australia Telescope National Facility which is funded by the Australian Government for operation as a National Facility managed by CSIRO. We acknowledge the Gomeroi people as the traditional owners of the ATCA observatory site. This research has made use of data and software provided by the High Energy Astrophysics Science Archive Research Center (HEASARC) and NASA's Astrophysics Data System Bibliographic Services.

\section*{Data availability}

The data and data products presented in this paper are available at the following DOI: 10.5281/zenodo.4624807.

\input{output.bbl}

\appendix

\section{Radio analysis details}
\label{sec:appendixA}

In this appendix, we show additional information and analysis details for each source. In Table \ref{tab:app_radio}, we list the technical radio observation details, such as observatory, date, configuration, calibrators, \textsc{clean} robust parameters, and synthesised beam. In Table \ref{tab:app_sourcespars}, we list the magnetic field (estimates), spin frequencies, and binary orbital period of all sources, including references. In Table \ref{app:positions}, we list the fitted radio positions of all detected targets or point to the reference where these positions can be found. 

\begin{table*}
	\centering
	\renewcommand\thetable{A1} 
	\caption{Background information about the radio observations presented in this paper. An * in the ATCA primary calibrator indicates that 0823-500 was also observed but not used the calibration. The observing frequencies are listed in Tables 1 and 2 in the main paper. When two robust parameters are listed, these correspond to the two (increasing) observing frequencies. The synthesized beam parameters correspond to the lowest observing frequency when observations were performed at two frequencies simultaneously. Shortened source names: XMMU J174716.1 $\equiv$ XMMU J174716.1-281048, 1RXH J173523.7 $\equiv$ 1RXH J173523.7-354013, and 3XMM J181923.7 $\equiv$ 3XMM J181923.7-170616. When multiple dates are listed for one epoch, the data were combined in the analysis. **Not included due to the presence of bright, proximate background source, yielding any flux density upper limit physically unconstraining.} 
	\label{tab:app_radio}
	\begin{tabular}{lcccccccccc}
Source & Observatory & Date & Config- & \multicolumn{2}{c}{Calibrators} & Robust & \multicolumn{3}{c}{Synthesized beam} & Program\\
& & & uration & Primary & Secondary & & Major & Minor & Angle &  \\ \hline
\multicolumn{11}{c}{Weakly-magnetized: $B<10^{10}$ G}  \\ \hline
GX 9+9 & VLA & 2013-06-14 & C & 3C 286 & J1733-1304 & \multicolumn{4}{c}{-- -- --  not detected at $3\sigma$  -- -- --} & 13A-352  \\
GX 9+1 & VLA & 2013-06-09 & C & 3C 286 & J1820-2528 & \multicolumn{4}{c}{-- -- --  not detected at $3\sigma$  -- -- --} & 13A-352 \\
GX 3+1 & VLA & 2013-06-10 & C & 3C 286 & J1820-2528 & 0 & 4.32" & 1.78" & -171.82$\degree$ & 13A-352 \\
GS 1826-24 & VLA & 2013-06-10 & C & 3C 286 & J1820-2528 & 0 & 3.93" & 1.91" & -13.54$\degree$ & 13A-352 \\
4U 1702-429 & ATCA & 2017-11-07 & 6A & 1934-638* & J1714-397 & 1 & 5.32" & 1.67" & -28.85$\degree$ & C3184 \\
4U 1735-44 & ATCA &  2017-11-07/8 & 6A & 1934-638* & 1714-397 & \multicolumn{4}{c}{-- -- --  not detected at $3\sigma$  -- -- --} & C3184 \\
2S 0918-549 & ATCA & 2017-12-11/12 & 6C & 1934-638* & J0852-5755 & \multicolumn{4}{c}{-- -- --  not detected at $3\sigma$  -- -- --} & C3184 \\
4U 1246-588 & ATCA & 2017-12-04/5 & 6C & 1934-638* & J1217-55 & \multicolumn{4}{c}{-- -- --  not detected at $3\sigma$  -- -- --} & C3184 \\
IGR J17062-6143 & ATCA & 2016-08-24 & 6C & 1934-638* & J1721-6154 & \multicolumn{4}{c}{-- -- --  not detected at $3\sigma$  -- -- --} &  C3108 \\
SAX J1712.6-3739 & ATCA & 2017-12-13/14/29/30 & 6C & 1934-638* & J1714-397 & \multicolumn{4}{c}{-- -- --    Not included**     -- -- --} & C3184\\
XMMU J174716.1 & VLA & 2014-02-26/03-05 & A & 3C 286 & J1744-3116 & \multicolumn{4}{c}{-- -- --  not detected at $3\sigma$  -- -- --} & 14A-163  \\
& & 03-27/04-20 & & & & & & & & \\
1RXH J173523.7 & VLA & 2014-02-28/03-17 & A & 3C 286 & J1744-3116 & \multicolumn{4}{c}{-- -- --  not detected at $3\sigma$  -- -- --} & 14A-163  \\
& & 04-11/04-19 & & & & & & & & \\
AX J1754.2-2754 & VLA & 2014-02-20/03-18 & A & 3C 286 & J1744-3116 & \multicolumn{4}{c}{-- -- --  not detected at $3\sigma$  -- -- --} & 14A-163 \\
& & 04-01/04-13 & & & & & & & & \\
IGR J17379-3747 & VLA & 2018-03-22 & C & 3C 286 & J1733-3722 & 0 & 1.10" & 0.30" & -3$\degree$ & 18A-194 \\
& VLA & 2018-03-30 & C & 3C 286 & J1733-3722 & 0 & 1.61" & 0.30" & -22$\degree$ & 18A-194 \\
& VLA & 2018-04-04 & C & 3C 286 & J1733-3722 & \multicolumn{4}{c}{-- -- --  not detected at $3\sigma$  -- -- --} & 18A-194 \\
& VLA & 2018-04-07 & C & 3C 286 & J1733-3722 & \multicolumn{4}{c}{-- -- --  not detected at $3\sigma$  -- -- --} & 18A-194 \\
& VLA & 2018-04-19 & C & 3C 286 & J1733-3722 & \multicolumn{4}{c}{-- -- --  not detected at $3\sigma$  -- -- --} & 18A-194 \\
& VLA & 2018-04-23 & C & 3C 286 & J1733-3722 & \multicolumn{4}{c}{-- -- --  not detected at $3\sigma$  -- -- --} & 18A-194 \\
& VLA & 2018-04-27 & C & 3C 286 & J1733-3722 & \multicolumn{4}{c}{-- -- --  not detected at $3\sigma$  -- -- --} & 18A-194 \\ 
\hline
\multicolumn{11}{c}{Strongly-magnetized: $B>10^{10}$ G} \\ \hline
GRO J1744-28 & ATCA & 2017-02-17 & 6D & 1934-638 & J1741-312 & \multicolumn{4}{c}{-- -- --  not detected at $3\sigma$  -- -- --} & CX379 \\  & ATCA & 2017-03-06 & 6D & 1934-638 & J1741-312 & \multicolumn{4}{c}{-- -- --  not detected at $3\sigma$  -- -- --} & CX379 \\ 
4U 1954+31 & VLA & 2019-01-01 & C & 3C 286 & J1953+3537 & 0 & 3.66" & 2.57" & $-79.45\degree$ & 18B-104 \\
 & VLA & 2019-01-11 & C & 3C 286 & J1953+3537 & \multicolumn{4}{c}{-- -- --  not detected at $3\sigma$  -- -- --} & 18B-104 \\
GX 1+4 & VLA & 2013-06-16 & C & 3C 286 & J1751-2524 & 0 & 4.05" & 1.80" & 1.59$\degree$ & 13A-352  \\
& VLA & 2018-12-14 & C & 3C 286 & J1751-2524 & 0.5 & 5.09" & 2.17" & 19.96$\degree$ & 18B-104  \\
& VLA & 2019-01-06 & C & 3C 286 & J1751-2524 & 0.5 & 8.24" & 2.61" & -33.66$\degree$ & 18B-104 \\
& VLA & 2019-01-08 & C & 3C 286 & J1751-2524 & 0.5 & 6.81" & 2.88" & 26.01$\degree$ & 18B-104 \\
3XMM J181923.7 & ATCA & 2018-05-15 & 6D & 1934-638* & J1600-48 & \multicolumn{4}{c}{-- -- --  not detected at $3\sigma$  -- -- --} & C3234  \\
2A 1822-371 & ATCA & 2017-12-17 & 6C & 1934-638 & J1759-39 & \multicolumn{4}{c}{-- -- --  not detected at $3\sigma$  -- -- --} & C3184 \\
4U 1626-67 & ATCA & 2017-12-13 & 6C & 1934-638* & J1619-680 & \multicolumn{4}{c}{-- -- --  not detected at $3\sigma$  -- -- --} & C3184 \\
Her X-1 & VLA & 2013-06-06 & C & 3C 286 & J1635+3808 & 0 & 3.24" & 1.8" & 8.57$\degree$ & 13A-352 \\
1E 1145.1-6141 & ATCA & 2018-05-13 & 6D & 1934-638* & J1214-60 & 0 & 4.53" & 1.05" & -10.06$\degree$ & C3234 \\
Cen X-3 & ATCA & 2018-05-12 & 6D & 1934-638* & J1129-58 & \multicolumn{4}{c}{-- -- --  not detected at $3\sigma$  -- -- --} & C3234 \\
3A 1239-599 & ATCA & 2018-05-12 & 6D & 1934-638* & J1214-60 & \multicolumn{4}{c}{-- -- --  not detected at $3\sigma$  -- -- --} & C3234 \\
OAO 1657-41 & ATCA & 2018-05-11 & 6D & 1934-638 & J1714-397 & \multicolumn{4}{c}{-- -- --  not detected at $3\sigma$  -- -- --} & C3234\\
4U 1538-522 & ATCA & 2018-05-16 & 6D & 1934-638* & J1600-48 & \multicolumn{4}{c}{-- -- --  not detected at $3\sigma$  -- -- --} & C3234\\
4U 1700-37 & ATCA & 2018-05-12 & 6D & 1934-638* & J1714-397 & 0.5 & 6.23" & 1.20" & -20.95$\degree$ & C3234\\
EXO 1722-363 & ATCA & 2018-05-13 & 6D & 1934-638* & & \multicolumn{4}{c}{-- -- --  not detected at $3\sigma$  -- -- --} & C3234\\
Vela X-1 & ATCA & 2018-05-15/16 & 6D & 1934-638* & J0906-47 & 0.5/0 & 8.63" & 1.20" & -14.75$\degree$ & C3234\\
IGR J16207-5129 & ATCA & 2018-05-15 & 6D & 1934-638* & J1600-48 & \multicolumn{4}{c}{-- -- --  not detected at $3\sigma$  -- -- --} & C3234\\
IGR J16318-4848 & ATCA & 2018-05-17 & 6D & 1934-638 & J1646-50 & 0.5/0 & 6.64" & 1.27" & -19.04$\degree$ & C3234\\
IGR J16320-4751 & ATCA & 2018-05-16 & 6D & 1934-638* & J1646-50 & 0.5/0 & 5.92" & 1.22" & -20.73$\degree$ & C3234\\
V*V490 Cep & VLA & 2017-09-10 & C & 3C 286 & J2123+5500 & \multicolumn{4}{c}{-- -- --  not detected at $3\sigma$  -- -- --} & 17B-136 \\
MXB 0656-072 & VLA & 2017-10-07 & C & 3C 286 & J0653-0625 & \multicolumn{4}{c}{-- -- --  not detected at $3\sigma$  -- -- --} & 17B-136 \\
SAX J2239.3+6116 & VLA & 2017-09-16 & C & 3C 286 & J2148+6107 & \multicolumn{4}{c}{-- -- --  not detected at $3\sigma$  -- -- --} & 17B-136 \\
V 0332+53 & VLA & 2017-09-17 & C & 3C 286 & J0346+5400 & \multicolumn{4}{c}{-- -- --  not detected at $3\sigma$  -- -- --} & 17B-136 \\
Swift J0243.6+6124 & VLA & \multicolumn{6}{c}{Data taken from \citet{vandeneijnden2018_swj0243} and \citet{vandeneijnden2019_reb}} & \multicolumn{3}{r}{17B-406, 17B-420, 18A-456} \\
\hline 

	\end{tabular}
\end{table*}

\begin{table*}
	\centering
	\renewcommand\thetable{A2}
	\caption{Magnetic field strengths, spin frequencies, and orbital periods of the X-ray binaries in this work. Sources where the magnetic field has not been measured/inferred, are classified as weakly or strongly magnetized based on their (combination of) bursting properties, pulse evolution, X-ray behavior, and/or evolutionary state. We note that all parameters, especially spin frequencies, evolve over time, and therefore refer to the listed references for more details. Question marks indicate that a measurement might be debated. Again, more details are found in the referenced work(s).}
	\label{tab:app_sourcespars}
	\begin{tabular}{lllll}
Source & Magnetic field & Spin  & Orbital & References \\
& strength [G] & freq. [Hz] & period [h] & \\ \hline
\multicolumn{5}{c}{Weakly-magnetized: $B<10^{10}$ G}  \\ \hline
GX 9+9 & unknown & unknown & 4.19 & \citet{kong06} \\
GX 9+1 & unknown & unknown & unknown & N/A \\
GX 3+1 & $<6.7\times10^8$, $<6\times10^8$ & unknown & unknown & \citet{ludlam2019}, \citet{mondal19} \\
GS 1826-24 & unknown & unknown & 2.25? & \citet{meshcheryakov10} \\
4U 1702-429 & $3.3\times10^8$ & 330 & unknown & \citet{ludlam2019}, \citet{markwardt99} \\
4U 1735-44 & unknown & unknown & unknown & N/A \\
2S 0918-549 & unknown & unknown & 0.29 & \citet{zhong11} \\
4U 1246-588 & unknown & unknown & Poss. $<1$ & \citet{bassa06}, \citet{intzand08} \\
IGR J17062-6143 & $>4\times10^8$, & 163.165 & 0.633 & \citet{degenaar17}, \citet{vandeneijnden2018_igr}, \\
& $(2.5 \pm 1.1)\times10^8$ & & & \citet{strohmayer2018} \\
XMMU J174716.1-281048 & unknown & unknown & unknown & N/A \\
1RXH J173523.7-354013 & unknown & unknown & unknown & N/A \\
AX J1754.2-2754 & unknown & unknown & unknown & N/A \\
IGR J17379-3747 & $(0.4-23)\times10^8$ & 468 & 1.88 & \citet{sanna2018}, \citet{strohmayer2018_1737} \\
\hline
\multicolumn{5}{c}{Strongly-magnetized: $B>10^{10}$ G} \\ \hline
GRO J1744-28 & $(2-70)\times10^{11}$ & 2.1 & 11.8 & \citet{cui97}, \citet{rappaport97}, \citet{bildsten97},\\
& & & & \citet{degenaar14}, \citet{younes15}, \citet{finger96} \\
4U 1954+31 & unknown & $5.46\times10^{-5}$ & unknown & \citet{kuranov15} \\
GX 1+4 & $10^{12}$--$10^{14}$ & $\sim 0.0083$ & 27864 & \citet{rea05}, \citet{ferrigno07}, \citet{cui04} \\
& & & & \citet{lewin71}, \citet{hinkle06}, \citet{cutler86} \\
3XMM J181923.7-170616 & unknown & $2.5\times10^{-3}$ & unknown & \citet{qiu17} \\
2A 1822-371 & $5\times10^{10}$ & 1.69 & 5.568 & \citet{staubert19} \\
4U 1626-67 & $3.8\times10^ {12}$ & 0.13 & 0.7 & \citet{staubert19}, \citet{middleditch81} \\
Her X-1 & $3\times10^{12}$ & 0.806 & 40.8 & \citet{staubert19}, \citet{leahy14} \\
1E 1145.1-6141 & unknown & $3.367\times10^{-3}$ & 134.4 & \citet{white80}, \citet{ilovaisky82} \\
Cen X-3 & $2.9\times10^{12}$ & 0.207 & 50.16 & \citet{staubert19}, \citet{clark88} \\
3A 1239-599 & unknown & $5.2\times10^{-3}$ & unknown & \citet{blair85} \\
OAO 1657-41 & $3\times10^{12}$? & $9.6\times10^{-2}$ & 912 & \citet{staubert19} \\
4U 1538-522 & $2\times10^{12}$ & $1.9\times10^{-3}$ & 89.52 & \citet{staubert19} \\
4U 1700-37 & unknown & unknown & unknown & N/A \\
EXO 1722-363 & unknown & $2.4\times10^{-3}$ & 233.76 & \citet{thompson07} \\
Vela X-1 & $2.6\times10^{12}$ & $3.5\times10^{-3}$ & 215.04 & \citet{staubert19} \\
IGR J16207-5129 & unknown & unknown & unknown & N/A \\
IGR J16318-4848 & unknown & unknown & unknown & N/A \\
IGR J16320-4751 & unknown & $7.7\times10^{-4}$ & 215 & \citet{corbet05}, \citet{rodriguez06} \\
V*V490 Cep & $2.5\times10^{12}$ & $1.5\times10^{-2}$ & 500 & \citet{staubert19} \\
MXB 0656-072 & $3\times10^{12}$ & $6.2\times10^{-3}$ & 2424 & \citet{staubert19}  \\
SAX J2239.3+6116 & unknown & $8\times10^{-4}$ & 6288? & \citet{intzand00}, \citet{intzand01}, \citet{reig17} \\
V 0332+53 & $2\times10^{12}$ & 0.228 & 816 & \citet{staubert19} \\
Swift J0243.6+6124 & $10^{12}-10^{13}$ & 0.102 & 662.1 & \citet{tsygankov18}, \citet{wilson18}, \\ 
& & & & \citet{jenke17} \\ \hline

	\end{tabular}
\end{table*}

\begin{table*}
	\centering
	\renewcommand\thetable{A3}
	\caption{Radio positions as measured with the \textsc{casa}-task \textsc{imfit}. The uncertainty is calculated as the beamsize divided by the S/N ratio, limited by $10\%$ of the beam size for the maximum accuracy as suggested by the VLA. The position with the best accuracy, out of all frequency bands and observations of the source, is shown.}
	\label{app:positions}
	\begin{tabular}{lll}
Source & RA & Declination  \\ \hline
GX 3+1 & $17$h$47$m$56.11$s$\pm0.03$s & $-26\degree33$'$49.23$"$\pm0.87$" \\
GS 1826-24 & $18$h$29$m$28.11$s$\pm0.01$s & $-23\degree47$'$48.92$"$\pm0.34$" \\
4U 1702-429 & $17$h$06$m$15.32$s$\pm0.01$s & $-43\degree02$'$08.79$"$\pm0.38$" \\
IGR J17379-3747 & $17$h$37$m$58.84$s$\pm0.02$s & $-37\degree46$'$18.35$"$\pm0.07$" \\
4U 1954+31 & $19$h$55$m$42.34$s$\pm0.05$s & $+32\degree05$'$48.87$"$\pm0.42$" \\
GX 1+4 & $17$h$32$m$02.13$s$\pm0.01$s & $-24\degree44$'$44.37$"$\pm0.28$" \\
Her X-1 & $16$h$57$m$49.792$s$\pm0.027$s & $+35\degree20$'$32.58$"$\pm0.23$" \\
1E 1145.1-6141 & $11$h$47$m$28.528$s$\pm0.003$s & $-61\degree57$'$13.47$"$\pm0.67$" \\
4U 1700-37 & $17$h$03$m$56.776$s$\pm0.005$s & $-37\degree50$'$38.62$"$\pm0.32$" \\
Vela X-1 & $09$h$02$m$06.836$s$\pm0.002$s & $-40\degree33$'$16.72$"$\pm0.47$" \\
IGR J16318-4848 & $16$h$31$m$48.308$s$\pm0.003$s & $-48\degree49$'$00.99$"$\pm0.30$" \\
IGR J16320-4751 & $16$h$32$m$01.758$s$\pm0.014$s & $-47\degree52$'$28.3$"$\pm1.1$" \\ 
Swift J0243.6+6124 & $02$h$43$m$40.440$s$\pm0.029$s & $+61\degree26$'$03.73$"$\pm0.10$"  \\ \hline

	\end{tabular}
\end{table*}

\section{X-ray analysis details}
\label{appendix_Xrays}

In Tables \ref{tab:weak_xrays} and \ref{tab:strong_xrays}, we list the data analysis details for the X-ray flux determination for the weakly and strongly-magnetised sources. For each source, we selected the closest X-ray observational information in time, aiming that the source was in the same state as during the radio observations. We then measured the flux either by fitting the X-ray spectrum or converting the observed count rate (upper limit) assuming a typical source spectrum or, if this was not available, the Crab spectrum. When fitting the spectrum, we attempted two single component absorbed models (i.e. \textsc{tbabs}*\textsc{powerlaw} and \textsc{tbabs}*\textsc{diskbb}) and their combination (i.e. \textsc{tbabs}*\textsc{(powerlaw+diskbb)}). We first picked the best-fitting model of the two single-component models based on their fit statistics. Then, we employed the f-test to compare the best-fitting single-component model to the combined model. We required a $\geq 5\sigma$ fit improvement in order to seleect the combined model. The keyword \textit{method} in Tables \ref{tab:weak_xrays} and \ref{tab:strong_xrays} signals whether we fitted a spectrum or converted count rates, while the listed parameters show the used model. 

There are some detailed notes for Tables \ref{tab:weak_xrays} and \ref{tab:strong_xrays}, that are relevant for a number of sources. Firstly, for the final two radio observations of IGR J17379-3747, we interpolated between two \textit{Swift} observations. For this purpose, we performed a linear interpolation in time between the logarithmic X-ray fluxes measured for the individual observations. Given the possibility of variability between the observations, we assign an enhanced systematic uncertainty of $50\%$ to the resulting interpolated X-ray flux. When multiple MJDs (\textit{MAXI}) or ObsIDs (\textit{Swift}) are shown with a dash, these observations were combined to collect enough counts for the analysis. \textit{MAXI} X-ray spectra were extracted using the on-demand online \textit{MAXI} tool for the shown MJDs (\href{http://maxi.riken.jp/mxondem/}{http://maxi.riken.jp/mxondem/}). When only errors to one side are shown in Tables \ref{tab:weak_xrays} and \ref{tab:strong_xrays}, the other error was either uncontrained or hit the physical limit of the parameter (e.g. zero for a normalisation). 

Finally, we end with analysis remarks for two individual sources. Firstly, for 4U 1954+31, no \textit{MAXI} or \textit{Swift} XRT X-ray information was available for either radio observation. However, \textit{Swift} BAT observations during the first radio epoch, show the same count rate within $1\sigma$ uncertainties as during a pointed \textit{Swift} XRT observation in 2006 \citep{masetti07a}. We therefore fitted this pointed \textit{Swift} XRT observation instead. In the 4U 1700-37 observation with \textit{Swift}, an Fe K$\alpha$ line is visible in the spectrum. Therefore, we include this line in the model as a \textsc{gaussian} component, with resulting parameters: $E=6.47\pm0.07$ keV, $\sigma = 0.3^{+0.15}_{-0.12}$ keV, and $N = (1.95^{+0.6}_{-0.4})\times10^{-2}$ photon cm$^{-2}$ s$^{-1}$.

\begin{table*}
	\centering
	\renewcommand\thetable{A4}
	\caption{Shortened source names: XMMU J174716.1 $\equiv$ XMMU J174716.1-281048 and 1RXH J173523.7 $\equiv$ 1RXH J173523.7-354013. \textit{Swift} ObsIDs are shortened by removing the starting 000. Parameters without errors were fixed in the fit or assumed for the count rate -- flux conversion, while errors are quoted at $1\sigma$.  $^a$\citet{vandeneijnden2018_igr}, $^b$\citet{kalberla05}, $^c	$\citet{sanna2018}.}
	\label{tab:weak_xrays}
	\begin{tabular}{lllllccccc} %
\multicolumn{9}{c}{Weak magnetic field neutron stars} \\ \hline
Source name & & Obs. & MJD/& Method & $N_H$ & $\Gamma$ & $N_{\rm po}$ & $kT_{\rm BB}$ & $N_{\rm BB}$ \\ 
& & & ObsID & & $[10^{22}$ & & [keV$^{-1}$cm$^{-2}$s$^{-1}$ & [keV] & [$10^{-4}$] \\ 
& & & & & cm$^{-2}$] & & at 1 keV] & & \\ 
\hline
GX 9+9 & & \textit{MAXI} & 56457 & \multicolumn{6}{l}{Count rate to flux scaling assuming the Crab spectrum and flux from \citet{madsen17}} \\
GX 9+1 & & \textit{MAXI} & 56452 & Spectrum & $4.8 \pm 0.5$ & $2.5 \pm 0.1$ & $(1.6\pm0.3)\times10^2$ & -- & -- \\
GX 3+1 & & \textit{MAXI} & 56459 & Spectrum & $6.1 \pm 1.6$ & $2.7 \pm 0.2$ & $(16.7^{+10.5}_{-6.1})\times10^2$ & -- & -- \\
GS 1826-24 & & \textit{MAXI} & 56453 & \multicolumn{6}{l}{Count rate to flux scaling assuming the Crab spectrum and flux from \citet{madsen17}} \\
4U 1702-429 & & \textit{Swift} & 88130001 & Spectrum & $3.04\pm0.05$ & $2.13\pm0.03$ & $9.0\pm0.1$ & -- & -- \\
4U 1735-44 & & \textit{MAXI} & 58063--5 & Spectrum & $2.7\pm1.0$ & $2.0\pm0.2$ & $1.2\pm0.3$ & -- & -- \\
2S 0918-549 & & \textit{Swift} & 31569013 & Spectrum & $0.6\pm0.2$ & $2.2\pm0.2$ & $(7.4\pm1.7)\times10^{-2}$ & -- & -- \\
4U 1246-588 & & \textit{Swift} & 10441001 & Spectrum & $0.5 \pm 0.1$ & $1.95 \pm 0.10$ & $(7.6\pm1.0)\times10^{-2}$ & -- & -- \\
IGR J17062-6143 & & \textit{Swift} & 34676002 & Spectrum & $0.12^a$ & $2.3 \pm 0.4$ & $(1.1\pm0.5)\times10^{-2}$ & $0.5 \pm 0.1$ & $2.3\pm1.0$ \\ 
XMMU J174716.1 & & \textit{Swift} & 30938013 & \multicolumn{6}{l}{Count rate to flux scaling assuming the Crab spectrum and flux from \citet{madsen17}} \\ \hline
1RXH J173523.7 & & \textit{Swift} & 31446002 & Spectrum & $0.7^b$ & $2.2 \pm 0.5$ & $(8 \pm 3)\times10^{-4}$ & -- & -- \\
& & & 31446003 & (averaged 3 & $0.7$ & $1.5 \pm 0.4$ & $(5 \pm 2)\times10^{-4}$ & -- & -- \\
& & & 31446005 & observations) & $0.7$ & $2.4 \pm 0.5$ & $(6 \pm 2)\times10^{-4}$ & -- & -- \\ \hline
AX J1754.2-2754 & & \textit{Swift} & 36163012 & Spectrum & $0.9^b$ & $1.7 \pm 0.4$ & $(3.7 \pm 1.5)\times10^{-4}$ & -- & -- \\ \hline
IGR J17379-3747 & 1 & \textit{Swift} & 31270002 & Spectrum & $0.9^c$ & $1.0 \pm 0.5$ & $(9 \pm 6)\times10^{-3}$ & $0.6 \pm 0.05$ & $9.6 \pm 3.1$ \\
 & 2 & \textit{Swift} & 31270005 & Spectrum & $0.9$ & $2.15 \pm 0.13$ & $(2.0\pm0.2)\times10^{-2}$ & -- & -- \\
 & 3 & \textit{Swift} & 31270007 & Rate & $0.9$ & $2.5$ & -- & -- & -- \\
 & 4 & \textit{Swift} & 31270009 & Spectrum & $0.9$ & $2.05 \pm 0.12$ & $(9 \pm 0.9)\times10^{-3}$ & -- & -- \\
 & 5 & \textit{Swift} & 31270013 & Spectrum & $0.9$ & $2.7\pm0.5$ & $(9\pm3)\times10^{-4}$ & -- & -- \\ \cline{2-10}
 & 6 & \textit{Swift} & 31270015 & Rate & $0.9$ & $2.5$ & -- & -- & -- \\
& & & 31270016 & (interpolated) & $0.9$ & $2.5$ & -- & -- & -- \\ \cline{2-10}
 & 7 & \textit{Swift} & 31270017 & Spectrum & $0.9$ & $2.3 \pm 0.2$ & $(2.5\pm0.4)\times10^{-3}$ & -- & -- \\
& & & 31270018 & (interpolated) & $0.9$ & $2.5 \pm 0.5$ & $(6 \pm 2)\times10^{-4}$ & -- & -- \\
\end{tabular}
\end{table*}

\begin{table*}
	\centering
	\renewcommand\thetable{A5}
	\caption{Shortened source name: 3XMM J181923.7 $\equiv$ 3XMM J181923.7-170616. For 2A 1822-371 and 4U 1626-67, we used the \textsc{blackbody} model as the \textsc{power law} model returned an nonphysically high absorption column and therefore unabsorbed flux. \textit{Swift} ObsIDs are shortened by removing the starting 000. Parameters without errors were fixed in the fit or assumed for the count rate -- flux conversion, while errors are quoted at $1\sigma$. $^a$\citet{hemphill19}, $^b$\citet{varun19}, $^c$\citet{corbet05}, $^d$\citet{nirmal17}, $^e$\citet{intzand01}, $^f$\citet{doroshenko17}.}
	\label{tab:strong_xrays}
	\begin{tabular}{lllllccccc} %
\multicolumn{9}{c}{Strong magnetic field neutron stars} \\ \hline
Source name & & Obs. & MJD/& Method & $N_H$ & $\Gamma$ & $N_{\rm po}$ & $kT_{\rm BB}$ & $N_{\rm BB}$ \\ 
& & & ObsID & & $[10^{22}$ & & [keV$^{-1}$cm$^{-2}$s$^{-1}$ & [keV] & [$10^{-4}$] \\ 
& & & & & cm$^{-2}$] & & at 1 keV] & & \\ \hline

GX 1+4 & 1 & \multicolumn{8}{l}{See \citet{vandeneijnden2018_gx}} \\
& 2 & \multicolumn{8}{l}{No X-ray information available close in time} \\
& 3 & \textit{MAXI} & 58488-90 & Spectrum & \multirow{2}{*}{$21.2^{+47.5}_{-20.0}$} & -- & -- & $2.7^{+1.3}_{-0.8}$ & $90^{+60}_{-20}$ \\
& 4 & \textit{MAXI} & 58490-93 & Spectrum & & -- & -- & $4.2^{+4.7}_{-1.5}$ & $135^{+310}_{50}$ \\
4U 1954+31 & 1 & \textit{Swift} & 30392001 & Spectrum & $2.5 \pm 0.3$ & -- & -- & $1.7 \pm 0.1$& $26.5 \pm 1.5$ \\
& 2 & \multicolumn{8}{l}{No X-ray information available close in time} \\
3XMM J181923.7 & & \textit{Swift} & 33498013 & Spectrum & $0.06^{+0.6}$ & $-0.4 \pm 0.4$ & $(2.5^{+2.9}_{-1.0})\times10^{-5}$ & -- & -- \\
2A 1822-371 & & \textit{MAXI} & 58103--6 & Spectrum & $2.2^{+6.7}_{-2.2}$ & -- & -- & $2.0\pm0.4$ & $116\pm25$ \\
4U 1626-67 & & \textit{MAXI} & 58097--101 & Spectrum & $0.0^{+1.5}$ & -- & -- & $1.9\pm0.2$ & $95\pm12$ \\
Her X-1 & & \multicolumn{8}{l}{See \citet{vandeneijnden2018_her}}\\
1E1145.1-6141 & & \textit{MAXI} & 58219--23 & Spectrum & $2.0\pm1.7$ & $1.5\pm0.4$ & $(6.8_{-3.5})\times10^{-2}$ & -- & -- \\
Cen X-3 & & \textit{MAXI} & 58230--80 & Spectrum & $14.8^{+13.1}_{-8.6}$ & -- & -- & $1.5\pm0.3$ & $(3\pm1)\times10^{-3}$ \\
3A 1239-599 & & \textit{Swift} & 37891007 & \multicolumn{6}{l}{Count rate to flux scaling assuming the Crab spectrum and flux from \citet{madsen17}} \\
OAO 1657-41 & & \textit{MAXI} & 58249--51 & Spectrum & $4.5\pm1.5$ & $1.6\pm0.1$ & $0.7\pm0.2$ & -- & -- \\
4U 1538-522 & & \textit{MAXI} & 58254 & Rate & $15^a$ & $1.0^b$ & -- & -- & -- \\
4U 1700-37 & & \textit{Swift} & 33631015 & Spectrum & $5.5\pm0.1$ & -- & -- & $1.92\pm0.04$ & $783 \pm 20$ \\
EXO 1722-363 & & \textit{Swift} & 10675001 & Rate & $15^c$ & $1.0^c$ & -- & -- & -- \\
Vela X-1 & & \textit{MAXI} & 58252--64 & Spectrum & $2.8_{-2.8}$ & $2.7^{+4.9}_{-1.2}$ & $0.35_{-0.35}$ & -- & -- \\
IGR J16207-5129 & & \textit{Swift} & 37888002 & Spectrum & $2.2_{-1.2}^{+1.7}$ & $0.8\pm0.7$ & $(5^{+10})\times10^{-4}$ & -- &  --\\
IGR J16318-4848 & & \textit{Swift} & 35053006 & Rate & $120^d$ & $1.0^d$ & -- & -- & -- \\
IGR J16320-4751 & & \textit{Swift} & 10676001 & Spectrum & $13.2\pm3.0$ & $0.5\pm0.4$ & $(3.6_{-1.9}^{+4.7})\times10^{-3}$ & -- & -- \\
V*V490 Cep & & \textit{Swift} & 80436005 & Spectrum & $0^{18.6}$ & $-2.4^{+2.0}$ & $(0^{1.6})\times10^{-5}$ & -- & -- \\
MXB 0656-072 & & \textit{Swift} & 80435001 & Spectrum & $0.0^{+0.3}$ & $0.8_{-0.4}^{+0.8}$ & $(1.0_{-0.3}^{+1.0})\times10^{-4}$ & -- & -- \\
SAX J2239.3+6116 & & \textit{Swift} & 10298001/2 & Rate & $3.0^e$ & $1.6^e$ & -- & -- & -- \\
V 0332+53 & & \textit{Swift} & 31293078/79 & Rate & $2.0^f$ & $2.0^f$ & -- & -- & -- \\
GRO J1744-28 & 1 & \textit{Swift} & 30898081 & Spectrum & $7.8\pm1.3$ & $1.0\pm0.2$ & $2.0_{-0.5}^{+0.8}$ & -- & -- \\
& 2 & \textit{Swift} & 30898087--91 & Rate & $7.8$ & $1.0$ & -- & -- & -- \\
Swift J0243.6+6124 & & \multicolumn{8}{l}{See \citet{vandeneijnden2018_swj0243} and \citet{vandeneijnden2019_reb}} \\

\end{tabular}
\end{table*}

\section{Details of the wind $L_R$ estimates}
\label{appendix:wind_Lr}

To estimate the stellar wind radio luminosity, we apply Equation 1 in the main paper for all sources. In this section, we briefly review the input parameters for this equation per target discussed in Section 4.2.1 of the main paper. The details for the three SyXRB are given in Section 4.4 of the main paper itself. In all cases, we assume a pure hydrogen wind with an electron temperature of $T_e = 10^4$ K, as these parameters are poorly constrained and only very weakly affect the predicted wind luminosity. In Table \ref{tab:winds}, we list the measured or otherwise assumed mass loss rates and terminal wind velocities, source distances, and resulting 5.5 GHz wind flux densities per source. Distance references can be found in Tables 1 and 2 in the main paper. 

IGR J16318-4848 is a highly absorbed SgXB, hosting a SgBe companion star \citep{filliatre04}, showing a bright IR excess in its SED from the presence of a dust shell \citep{kaplan06}. We tested whether this dust shell could contribute to the observed radio emission. Assuming a dust temperature of $1030$ K and using that the radio regime falls in the Rayleigh-Jeans tail of the dust shell's thermal SED, we find a contribution of $2\times10^{-2}$ $\mu$Jy at $5.5$ GHz. Therefore, we rule out the dust shell as the origin of the observed emission. For the estimates in Table \ref{tab:winds}, we assumed a standard $10^{-6}$ $M_{\odot}$ yr$^{-1}$ mass loss rate in the absence of literature measurements. \citet{kaplan06} argue that the IR spectrum does not show evidence for stellar wind emission, supporting the notion that the wind mass loss rate does not greatly exceed typical values.  

Finally, we turn to the BeXRBs, of which only Swift J0243.6+6124 was detected in radio. For individual BeXBRs, the stellar wind parameters are typically poorly constrained. However, the Be star winds are significantly weaker than those of the supergiant O/B donors in SgXB: typical mass loss rates are of the order $10^{-9}$ $M_{\odot}$ yr$^{-1}$ \citep{snow81}. At such mass loss rates, even a low terminal velocity (yielding the highest luminosity) of $100$ km s$^{-1}$ leads to expected 5.5 GHz flux densities below $1$ $\mu$Jy for distances larger than $1$ kpc. As all observed BeXRBs are located at larger distances (at least $3.6$ kpc), their stellar wind are not expected to be observable at current radio sensitivities.

\begin{table*}
	\centering
	\renewcommand\thetable{A6}
	\caption{Input stellar wind parameters and references for the 5.5-GHz wind radio flux density estimate. For values denoted with a \textbf{*}, no literature measurement was known to the authors -- therefore, we assumed typical values of $10^{-6}$ $M_{\odot}$ yr$^{-1}$ and $400$ km s$^{-1}$ for the wind mass loss rate and terminal velocity, respectively. The range in predicted 5.5 GHz rado flux density values for EXO 1722-363 corresponds to the range in terminal wind velocities. For IGR J16320-4751, no estimates were found in the literature and therefore no estimate is made. SyXRBs are discussed in the main text, BeXRBs in the appendix.}
	\label{tab:winds}
	\begin{tabular}{lllllll}
	
Source name & $\dot{M}$ & Reference & $v_{\infty}$ & Reference & $D$ & $S_{5.5\text{ GHz}}$ \\ 
& [$M_{\odot}$ yr$^{-1}$] & & [km s$^{-1}$] & & [kpc] & [$\mu$Jy] \\ \hline

1E 1145.1-6141 & $1.4\times10^{-6}$ & \citet{densham82} & $400$ & \citet{densham82} & 8.3 & $\sim 20$ \\
4U 1700-37 & $6\times10^{-6}$ & \citet{howarth89} & $2200$ & \citet{howarth89} & 1.5 & $\sim 200$ \\
& & \multicolumn{3}{l}{See also \citet{hammerschlag90} and \citet{heap92}} & & \\
Vela X-1 & $1\times10^{-6}$ & \citet{grinberg17} & $700$ & \citet{grinberg17} & 1.97 & $\sim 94$ \\
IGR J16318-4848 & $1\times10^{-6}$\textbf{*} & & $410$ & \citet{filliatre04} & 3.6 & $\sim 60$ \\
IGR J16320-4751 & \multicolumn{4}{l}{See caption for details} & 3.5 &  \\
Cen X-3 & $5\times10^{-6}$ & \citet{krzeminski74} & $800$ & \citet{krzeminski74} & 6.9 & $\sim 42$ \\
& & \multicolumn{3}{l}{See also \citet{krzeminski73} and \citet{clark88}} & & \\
3A 1239-599 & $1\times10^{-6}$\textbf{*} & & $400$\textbf{*} & & 4 & $\sim 50$ \\
OAO 1657-41 & $2\times10^{-6}$ & \citet{mason12} & $250$ & \citet{mason12} & 6.4 & $\sim 92$ \\
4U 1538-522 &$1\times10^{-6}$ & \citet{torrejon15} & $1300$ & \citet{hemphill14} & 5.8 & $\sim 4$ \\
EXO 1722-363 & $1.5\times10^{-6}$ & \citet{manousakis11} & $250-600$ & \citet{manousakis11} & 8 & $\sim 40-13$ \\
IGR J16207-5129 & $2\times10^{-6}$ & \citet{bodaghee10} & $400$\textbf{*} & & 6.1 & $\sim 50$ \\ \hline
	\end{tabular}
\end{table*}

\section{Spectral index upper limits}
\label{appendix:UL}

\begin{figure}
	\includegraphics[width=\columnwidth]{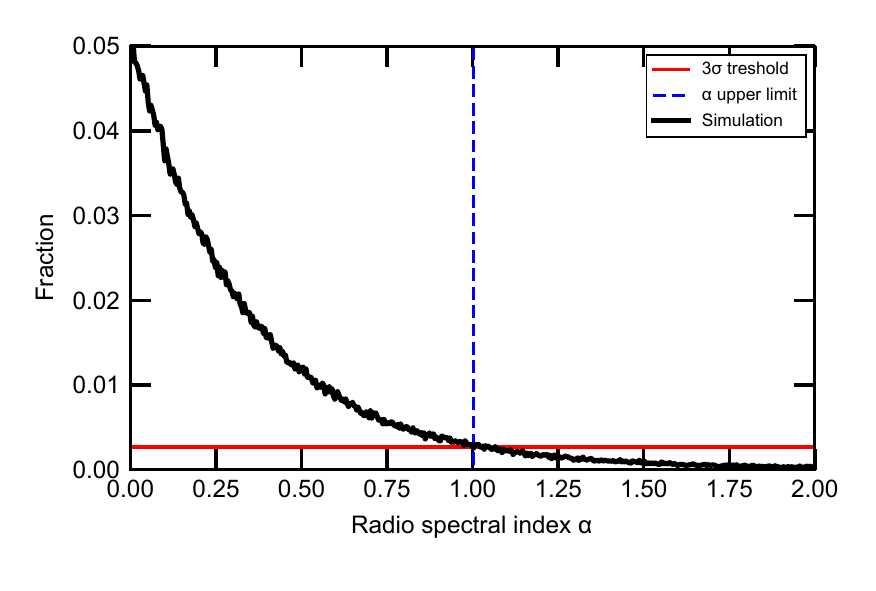}
	\renewcommand\thefigure{B1}
    \caption{The fraction of simulated data sets where IGR J16320-4751 would not be detected at 9 GHz with a $3\sigma$ significance, as a function of spectral index, assuming the observed 5.5-GHz flux density and 9-GHz sensitivity.}
    \label{fig:UL1}
\end{figure}

\begin{figure}
	\includegraphics[width=\columnwidth]{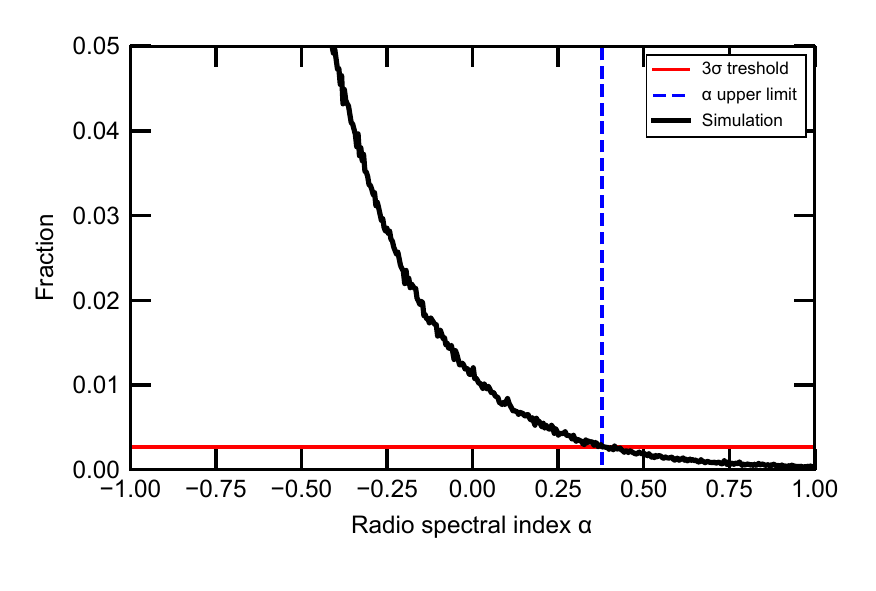}
	\renewcommand\thefigure{B2}
    \caption{The fraction of simulated data sets where 1E 1145.1-6141 would not be detected at 9 GHz with a $3\sigma$ significance, as a function of spectral index, assuming the observed 5.5-GHz flux density and 9-GHz sensitivity.}
    \label{fig:UL2}
\end{figure}

In Figures \ref{fig:UL1} and \ref{fig:UL2}, we show the estimate of the $3\sigma$ upper limit on the radio spectral index for the two sources with a detection only at 5.5 GHz but not at 9 GHz: IGR J16320-4751 and 1E 1145.1-6141. For the details of the calculation, see \citet{vandeneijnden2019_reb}.

\label{lastpage}

\end{document}

%% file: output.bbl
 \newcommand{\noop}[1]{}

%% file: ArXivVersion.bbl
\begin{thebibliography}{334}
\providecommand{\natexlab}[1]{#1}

\bibitem[{{Abada-Simon} et~al.(1993){Abada-Simon}, {Lecacheux}, {Bastian},
  {Bookbinder} \& {Dulk}}]{abada-simon93}
{Abada-Simon} M., {Lecacheux} A., {Bastian} T.~S., {Bookbinder} J.~A., {Dulk}
  G.~A., 1993, \apj, 406, 692

\bibitem[{{Ables}(1969)}]{ables69}
{Ables} J.~G., 1969, \apjl, 155, L27

\bibitem[{{Andrews} et~al.(2019){Andrews}, {Fenech}, {Prinja}, {Clark} \&
  {Hindson}}]{andrews19}
{Andrews} H., {Fenech} D., {Prinja} R.~K., {Clark} J.~S., {Hindson} L., 2019,
  \aap, 632, A38

\bibitem[{{Ankay} et~al.(2001){Ankay}, {Kaper}, {de Bruijne}, {Dewi},
  {Hoogerwerf} \& {Savonije}}]{ankay01}
{Ankay} A., {Kaper} L., {de Bruijne} J.~H.~J., {Dewi} J., {Hoogerwerf} R.,
  {Savonije} G.~J., 2001, \aap, 370, 170

\bibitem[{{Archibald} et~al.(2009)}]{archibald09}
{Archibald} A.~M. et~al., 2009, Science, 324, 1411

\bibitem[{{Arnaud}(1996)}]{arnaud96}
{Arnaud} K.~A., 1996, in G.H. {Jacoby}, J.~{Barnes}, eds, Astronomical Data
  Analysis Software and Systems V. Astronomical Society of the Pacific
  Conference Series, Vol. 101, p.~17

\bibitem[{{Atri} et~al.(2019)}]{atri2019}
{Atri} P. et~al., 2019, \mnras, 489, 3116

\bibitem[{{Baglio} et~al.(2018){Baglio}, {Russell}, {Pirbhoy}, {Bahramian},
  {Heinke}, {Roche} \& {Lewis}}]{baglio2018}
{Baglio} M.~C., {Russell} D.~M., {Pirbhoy} S., {Bahramian} A., {Heinke} C.~O.,
  {Roche} P., {Lewis} F., 2018, The Astronomer's Telegram, 12180, 1

\bibitem[{{Baglio} et~al.(2016)}]{baglio16}
{Baglio} M.~C., {D'Avanzo} P., {Campana} S., {Goldoni} P., {Masetti} N.,
  {Mu{\~n}oz-Darias} T., {Pati{\~n}o-{\'A}lvarez} V., {Chavushyan} V., 2016,
  \aap, 587, A102

\bibitem[{{Bahramian} et~al.(2017){Bahramian}, {Strader}, {Heinke}, {Sivakoff},
  {Kennea}, {Degenaar} \& {Wijnands}}]{bahramian17}
{Bahramian} A., {Strader} J., {Heinke} C.~O., {Sivakoff} G.~R., {Kennea} J.~A.,
  {Degenaar} N., {Wijnands} R., 2017, The Astronomer's Telegram, 10685, 1

\bibitem[{{Bahramian} et~al.(2014)}]{bahramian14}
{Bahramian} A. et~al., 2014, \apj, 780, 127

\bibitem[{{Bahramian} et~al.(2018)}]{bahramian18}
{Bahramian} A. et~al., 2018, \apj, 864, 28

\bibitem[{{Bailer-Jones} et~al.(2018){Bailer-Jones}, {Rybizki}, {Fouesneau},
  {Mantelet} \& {Andrae}}]{bailerjones18}
{Bailer-Jones} C.~A.~L., {Rybizki} J., {Fouesneau} M., {Mantelet} G., {Andrae}
  R., 2018, \aj, 156, 58

\bibitem[{{Bailer-Jones} et~al.(2020){Bailer-Jones}, {Rybizki}, {Fouesneau},
  {Demleitner} \& {Andrae}}]{bailerjones2020}
{Bailer-Jones} C.~A.~L., {Rybizki} J., {Fouesneau} M., {Demleitner} M.,
  {Andrae} R., 2020, arXiv e-prints, arXiv:2012.05220

\bibitem[{{Bak Nielsen} et~al.(2017){Bak Nielsen}, {Patruno} \&
  {D'Angelo}}]{baknielsen17}
{Bak Nielsen} A.~S., {Patruno} A., {D'Angelo} C., 2017, \mnras, 468, 824

\bibitem[{{Barrett} et~al.(2017){Barrett}, {Dieck}, {Beasley}, {Singh} \&
  {Mason}}]{barrett17}
{Barrett} P.~E., {Dieck} C., {Beasley} A.~J., {Singh} K.~P., {Mason} P.~A.,
  2017, \aj, 154, 252

\bibitem[{{Bassa} et~al.(2006){Bassa}, {Jonker}, {in't Zand} \&
  {Verbunt}}]{bassa06}
{Bassa} C.~G., {Jonker} P.~G., {in't Zand} J.~J.~M., {Verbunt} F., 2006, \aap,
  446, L17

\bibitem[{{Becker} et~al.(1998){Becker}, {Remillard}, {Rappaport} \&
  {McClintock}}]{becker98}
{Becker} C.~M., {Remillard} R.~A., {Rappaport} S.~A., {McClintock} J.~E., 1998,
  \apj, 506, 880

\bibitem[{{Belczy{\'n}ski} et~al.(2000){Belczy{\'n}ski}, {Miko{\l}ajewska},
  {Munari}, {Ivison} \& {Friedjung}}]{belczynski00}
{Belczy{\'n}ski} K., {Miko{\l}ajewska} J., {Munari} U., {Ivison} R.~J.,
  {Friedjung} M., 2000, \aaps, 146, 407

\bibitem[{{Benz} \& {Guedel}(1989)}]{benz89}
{Benz} A.~O., {Guedel} M., 1989, \aap, 218, 137

\bibitem[{{Benz} et~al.(1983){Benz}, {Fuerst} \& {Kiplinger}}]{benz83}
{Benz} A.~O., {Fuerst} E., {Kiplinger} A.~L., 1983, \nat, 302, 45

\bibitem[{{Bildsten} et~al.(1997)}]{bildsten97}
{Bildsten} L. et~al., 1997, \apjs, 113, 367

\bibitem[{{Blair} \& {Candy}(1985)}]{blair85}
{Blair} D.~G., {Candy} B.~N., 1985, \mnras, 212, 219

\bibitem[{{Blandford} \& {K{\"o}nigl}(1979)}]{blandford79}
{Blandford} R.~D., {K{\"o}nigl} A., 1979, \apj, 232, 34

\bibitem[{{Blandford} \& {Payne}(1982)}]{blandford82}
{Blandford} R.~D., {Payne} D.~G., 1982, \mnras, 199, 883

\bibitem[{{Blandford} \& {Znajek}(1977)}]{blandford77}
{Blandford} R.~D., {Znajek} R.~L., 1977, \mnras, 179, 433

\bibitem[{{Blay} et~al.(2005)}]{blay05}
{Blay} P., {Rib{\'o}} M., {Negueruela} I., {Torrej{\'o}n} J.~M., {Reig} P.,
  {Camero} A., {Mirabel} I.~F., {Reglero} V., 2005, \aap, 438, 963

\bibitem[{{Blomme} \& {Runacres}(1997)}]{blomme97}
{Blomme} R., {Runacres} M.~C., 1997, \aap, 323, 886

\bibitem[{{Blomme} \& {Volpi}(2014)}]{blomme2014}
{Blomme} R., {Volpi} D., 2014, \aap, 561, A18

\bibitem[{{Blomme} et~al.(2017){Blomme}, {Fenech}, {Prinja}, {Pittard} \&
  {Morford}}]{blomme2017}
{Blomme} R., {Fenech} D.~M., {Prinja} R.~K., {Pittard} J.~M., {Morford} J.~C.,
  2017, \aap, 608, A69

\bibitem[{{Bodaghee} et~al.(2010){Bodaghee}, {Tomsick}, {Rodriguez}, {Chaty},
  {Pottschmidt} \& {Walter}}]{bodaghee10}
{Bodaghee} A., {Tomsick} J.~A., {Rodriguez} J., {Chaty} S., {Pottschmidt} K.,
  {Walter} R., 2010, \apj, 719, 451

\bibitem[{{Bogdanov} et~al.(2018)}]{bogdanov18}
{Bogdanov} S. et~al., 2018, \apj, 856, 54

\bibitem[{{Bozzo} et~al.(2018)}]{bozzo18}
{Bozzo} E. et~al., 2018, \aap, 613, A22

\bibitem[{{Bright} et~al.(2020)}]{bright2020}
{Bright} J.~S. et~al., 2020, Nature Astronomy, 4, 697

\bibitem[{{Brocksopp} et~al.(2003){Brocksopp}, {Bode} \& {Eyres}}]{brocksopp03}
{Brocksopp} C., {Bode} M.~F., {Eyres} S.~P.~S., 2003, \mnras, 344, 1264

\bibitem[{{Brocksopp} et~al.(2004){Brocksopp}, {Sokoloski}, {Kaiser},
  {Richards}, {Muxlow} \& {Seymour}}]{brocksopp04}
{Brocksopp} C., {Sokoloski} J.~L., {Kaiser} C., {Richards} A.~M., {Muxlow}
  T.~W.~B., {Seymour} N., 2004, \mnras, 347, 430

\bibitem[{{Burderi} et~al.(2010){Burderi}, {Di Salvo}, {Riggio}, {Papitto},
  {Iaria}, {D'A{\`\i}} \& {Menna}}]{burderi10}
{Burderi} L., {Di Salvo} T., {Riggio} A., {Papitto} A., {Iaria} R., {D'A{\`\i}}
  A., {Menna} M.~T., 2010, \aap, 515, A44

\bibitem[{{Cackett} et~al.(2008)}]{cackett2008_iron}
{Cackett} E.~M. et~al., 2008, \apj, 674, 415-420

\bibitem[{{Campana} et~al.(2002){Campana}, {Stella}, {Israel}, {Moretti},
  {Parmar} \& {Orlandini}}]{campana02}
{Campana} S., {Stella} L., {Israel} G.~L., {Moretti} A., {Parmar} A.~N.,
  {Orlandini} M., 2002, \apj, 580, 389

\bibitem[{{Cash}(1979)}]{cash1979}
{Cash} W., 1979, \apj, 228, 939

\bibitem[{{Chakrabarty}(1998)}]{chakrabarty98}
{Chakrabarty} D., 1998, \apj, 492, 342

\bibitem[{{Chakrabarty} et~al.(2002){Chakrabarty}, {Wang}, {Juett}, {Lee} \&
  {Roche}}]{chakrabarty02}
{Chakrabarty} D., {Wang} Z., {Juett} A.~M., {Lee} J.~C., {Roche} P., 2002,
  \apj, 573, 789

\bibitem[{{Chakrabarty} et~al.(1997)}]{chakrabarty97}
{Chakrabarty} D. et~al., 1997, \apj, 474, 414

\bibitem[{{Chanmugam} \& {Dulk}(1982)}]{chanmugam82}
{Chanmugam} G., {Dulk} G.~A., 1982, \apjl, 255, L107

\bibitem[{{Chanmugam} et~al.(1987){Chanmugam}, {Rao} \&
  {Tohline}}]{chanmugam87}
{Chanmugam} G., {Rao} M., {Tohline} J.~E., 1987, \apj, 319, 188

\bibitem[{{Chapman} et~al.(1999){Chapman}, {Leitherer}, {Koribalski}, {Bouter}
  \& {Storey}}]{chapman99}
{Chapman} J.~M., {Leitherer} C., {Koribalski} B., {Bouter} R., {Storey} M.,
  1999, \apj, 518, 890

\bibitem[{{Chelovekov} \& {Grebenev}(2007)}]{chelovekov07}
{Chelovekov} I.~V., {Grebenev} S.~A., 2007, Astronomy Letters, 33, 807

\bibitem[{{Chelovekov} et~al.(2006){Chelovekov}, {Grebenev} \&
  {Sunyaev}}]{chelovekov06}
{Chelovekov} I.~V., {Grebenev} S.~A., {Sunyaev} R.~A., 2006, Astronomy Letters,
  32, 456

\bibitem[{{Churchwell} et~al.(1992){Churchwell}, {Bieging}, {van der Hucht},
  {Williams}, {Spoelstra} \& {Abbott}}]{churchwell92}
{Churchwell} E., {Bieging} J.~H., {van der Hucht} K.~A., {Williams} P.~M.,
  {Spoelstra} T.~A.~T., {Abbott} D.~C., 1992, \apj, 393, 329

\bibitem[{{Clark} et~al.(1988){Clark}, {Minato} \& {Mi}}]{clark88}
{Clark} G.~W., {Minato} J.~R., {Mi} G., 1988, \apj, 324, 974

\bibitem[{{Clark} et~al.(2002){Clark}, {Goodwin}, {Crowther}, {Kaper},
  {Fairbairn}, {Langer} \& {Brocksopp}}]{clark02}
{Clark} J.~S., {Goodwin} S.~P., {Crowther} P.~A., {Kaper} L., {Fairbairn} M.,
  {Langer} N., {Brocksopp} C., 2002, \aap, 392, 909

\bibitem[{{Clark} et~al.(2019){Clark}, {Najarro}, {Negueruela}, {Ritchie},
  {Gonz{\'a}lez-Fern{\'a}ndez} \& {Lohr}}]{clark19}
{Clark} J.~S., {Najarro} F., {Negueruela} I., {Ritchie} B.~W.,
  {Gonz{\'a}lez-Fern{\'a}ndez} C., {Lohr} M.~E., 2019, \aap, 623, A83

\bibitem[{{Contreras} et~al.(1997){Contreras}, {Rodr{\'\i}guez}, {Tapia},
  {Cardini}, {Emanuele}, {Badiali} \& {Persi}}]{contreras97}
{Contreras} M.~E., {Rodr{\'\i}guez} L.~F., {Tapia} M., {Cardini} D., {Emanuele}
  A., {Badiali} M., {Persi} P., 1997, \apjl, 488, L153

\bibitem[{{Coppejans} \& {Knigge}(2020)}]{coppejans20}
{Coppejans} D., {Knigge} C., 2020, arXiv e-prints, arXiv:2003.05953

\bibitem[{{Coppejans} et~al.(2015){Coppejans}, {K{\"o}rding}, {Miller-Jones},
  {Rupen}, {Knigge}, {Sivakoff} \& {Groot}}]{coppejans15}
{Coppejans} D.~L., {K{\"o}rding} E.~G., {Miller-Jones} J.~C.~A., {Rupen} M.~P.,
  {Knigge} C., {Sivakoff} G.~R., {Groot} P.~J., 2015, \mnras, 451, 3801

\bibitem[{{Coppejans} et~al.(2018){Coppejans}, {Miller-Jones}, {K{\"o}rding},
  {Sivakoff} \& {Rupen}}]{coppejans18}
{Coppejans} D.~L., {Miller-Jones} J.~C., {K{\"o}rding} E.~G., {Sivakoff} G.~R.,
  {Rupen} M.~P., 2018, in E.~{Murphy}, ed., Science with a Next Generation Very
  Large Array. Astronomical Society of the Pacific Conference Series, Vol. 517,
  p. 727

\bibitem[{{Coppejans} et~al.(2016)}]{coppejans16}
{Coppejans} D.~L. et~al., 2016, \mnras, 463, 2229

\bibitem[{{Corbel} \& {Fender}(2002)}]{corbel02}
{Corbel} S., {Fender} R.~P., 2002, \apjl, 573, L35

\bibitem[{{Corbel} et~al.(2000){Corbel}, {Fender}, {Tzioumis}, {Nowak},
  {McIntyre}, {Durouchoux} \& {Sood}}]{corbel00}
{Corbel} S., {Fender} R.~P., {Tzioumis} A.~K., {Nowak} M., {McIntyre} V.,
  {Durouchoux} P., {Sood} R., 2000, \aap, 359, 251

\bibitem[{{Corbel} et~al.(2003){Corbel}, {Nowak}, {Fender}, {Tzioumis} \&
  {Markoff}}]{corbel03}
{Corbel} S., {Nowak} M.~A., {Fender} R.~P., {Tzioumis} A.~K., {Markoff} S.,
  2003, \aap, 400, 1007

\bibitem[{{Corbet} et~al.(2005)}]{corbet05}
{Corbet} R. et~al., 2005, The Astronomer's Telegram, 649, 1

\bibitem[{{Corbet} et~al.(2019)}]{corbet19}
{Corbet} R.~H.~D. et~al., 2019, \apj, 884, 93

\bibitem[{{Coriat} et~al.(2019){Coriat}, {Fender}, {Tasse}, {Smirnov},
  {Tzioumis} \& {Broderick}}]{coriat2019}
{Coriat} M., {Fender} R.~P., {Tasse} C., {Smirnov} O., {Tzioumis} A.~K.,
  {Broderick} J.~W., 2019, \mnras, 484, 1672

\bibitem[{{Coriat} et~al.(2011)}]{coriat11}
{Coriat} M. et~al., 2011, \mnras, 414, 677

\bibitem[{{Corradi} et~al.(2008)}]{corradi08}
{Corradi} R.~L.~M. et~al., 2008, \aap, 480, 409

\bibitem[{{Corradi} et~al.(2010)}]{corradi10}
{Corradi} R.~L.~M. et~al., 2010, \aap, 509, A41

\bibitem[{{Court} et~al.(2018){Court}, {Altamirano} \& {Sanna}}]{court18}
{Court} J.~M.~C., {Altamirano} D., {Sanna} A., 2018, \mnras, 477, L106

\bibitem[{{Cowley} et~al.(1998){Cowley}, {Schmidtke}, {Crampton} \&
  {Hutchings}}]{cowley98}
{Cowley} A.~P., {Schmidtke} P.~C., {Crampton} D., {Hutchings} J.~B., 1998,
  \apj, 504, 854

\bibitem[{{Crampton} et~al.(1996){Crampton}, {Hutchings}, {Cowley},
  {Schmidtke}, {McGrath}, {O'Donoghue} \& {Harrop-Allin}}]{crampton96}
{Crampton} D., {Hutchings} J.~B., {Cowley} A.~P., {Schmidtke} P.~C., {McGrath}
  T.~K., {O'Donoghue} D., {Harrop-Allin} M.~K., 1996, \apj, 456, 320

\bibitem[{{Cui}(1997)}]{cui97}
{Cui} W., 1997, \apjl, 482, L163

\bibitem[{{Cui} \& {Smith}(2004)}]{cui04}
{Cui} W., {Smith} B., 2004, \apj, 602, 320

\bibitem[{{Cutler} et~al.(1986){Cutler}, {Dennis} \& {Dolan}}]{cutler86}
{Cutler} E.~P., {Dennis} B.~R., {Dolan} J.~F., 1986, \apj, 300, 551

\bibitem[{{D'Angelo} \& {Spruit}(2012)}]{dangelo12}
{D'Angelo} C.~R., {Spruit} H.~C., 2012, \mnras, 420, 416

\bibitem[{{Degenaar} et~al.(2011){Degenaar}, {Wijnands} \& {Kaur}}]{degenaar11}
{Degenaar} N., {Wijnands} R., {Kaur} R., 2011, \mnras, 414, L104

\bibitem[{{Degenaar} et~al.(2014){Degenaar}, {Miller}, {Harrison}, {Kennea},
  {Kouveliotou} \& {Younes}}]{degenaar14}
{Degenaar} N., {Miller} J.~M., {Harrison} F.~A., {Kennea} J.~A., {Kouveliotou}
  C., {Younes} G., 2014, \apjl, 796, L9

\bibitem[{{Degenaar} et~al.(2010)}]{degenaar10}
{Degenaar} N. et~al., 2010, \mnras, 404, 1591

\bibitem[{{Degenaar} et~al.(2017)}]{degenaar17}
{Degenaar} N., {Pinto} C., {Miller} J.~M., {Wijnands} R., {Altamirano} D.,
  {Paerels} F., {Fabian} A.~C., {Chakrabarty} D., 2017, \mnras, 464, 398

\bibitem[{{Deller} et~al.(2015)}]{deller2015}
{Deller} A.~T. et~al., 2015, \apj, 809, 13

\bibitem[{{Densham} \& {Charles}(1982)}]{densham82}
{Densham} R.~H., {Charles} P.~A., 1982, \mnras, 201, 171

\bibitem[{{Dhawan} et~al.(2000){Dhawan}, {Mirabel} \&
  {Rodr{\'\i}guez}}]{dhawan2000}
{Dhawan} V., {Mirabel} I.~F., {Rodr{\'\i}guez} L.~F., 2000, \apj, 543, 373

\bibitem[{{D{\'\i}az Trigo} et~al.(2017){D{\'\i}az Trigo}, {Migliari},
  {Miller-Jones}, {Rahoui}, {Russell} \& {Tudor}}]{diaztrigo17}
{D{\'\i}az Trigo} M., {Migliari} S., {Miller-Jones} J.~C.~A., {Rahoui} F.,
  {Russell} D.~M., {Tudor} V., 2017, \aap, 600, A8

\bibitem[{{D{\'{\i}}az Trigo} et~al.(2018)}]{diaztrigo18}
{D{\'{\i}}az Trigo} M. et~al., 2018, \aap, 616, A23

\bibitem[{{Dieball} et~al.(2005)}]{dieball05}
{Dieball} A., {Knigge} C., {Zurek} D.~R., {Shara} M.~M., {Long} K.~S.,
  {Charles} P.~A., {Hannikainen} D.~C., {van Zyl} L., 2005, \apjl, 634, L105

\bibitem[{{Din{\c{c}}er} et~al.(2014){Din{\c{c}}er}, {Kalemci}, {Tomsick},
  {Buxton} \& {Bailyn}}]{dincer14}
{Din{\c{c}}er} T., {Kalemci} E., {Tomsick} J.~A., {Buxton} M.~M., {Bailyn}
  C.~D., 2014, \apj, 795, 74

\bibitem[{{Doroshenko} et~al.(2017){Doroshenko}, {Tsygankov}, {Mushtukov},
  {Lutovinov}, {Santangelo}, {Suleimanov} \& {Poutanen}}]{doroshenko17}
{Doroshenko} V., {Tsygankov} S.~S., {Mushtukov} A.~e.~A., {Lutovinov} A.~A.,
  {Santangelo} A., {Suleimanov} V.~F., {Poutanen} J., 2017, \mnras, 466, 2143

\bibitem[{{Dougherty} \& {Williams}(2000)}]{dougherty00}
{Dougherty} S.~M., {Williams} P.~M., 2000, \mnras, 319, 1005

\bibitem[{{Dougherty} et~al.(1995){Dougherty}, {Bode}, {Lloyd}, {Davis} \&
  {Eyres}}]{dougherty95}
{Dougherty} S.~M., {Bode} M.~F., {Lloyd} H.~M., {Davis} R.~J., {Eyres} S.~P.,
  1995, \mnras, 272, 843

\bibitem[{{Dougherty} et~al.(1996){Dougherty}, {Williams}, {van der Hucht},
  {Bode} \& {Davis}}]{dougherty96}
{Dougherty} S.~M., {Williams} P.~M., {van der Hucht} K.~A., {Bode} M.~F.,
  {Davis} R.~J., 1996, \mnras, 280, 963

\bibitem[{{Drappeau} et~al.(2015){Drappeau}, {Malzac}, {Belmont}, {Gandhi} \&
  {Corbel}}]{drappeau15}
{Drappeau} S., {Malzac} J., {Belmont} R., {Gandhi} P., {Corbel} S., 2015,
  \mnras, 447, 3832

\bibitem[{{Dubus}(2013)}]{dubus13}
{Dubus} G., 2013, \aapr, 21, 64

\bibitem[{{Duldig} et~al.(1979){Duldig}, {Greenhill}, {Thomas}, {Haynes},
  {Simons} \& {Murdin}}]{duldig79}
{Duldig} M.~L., {Greenhill} J.~G., {Thomas} R.~M., {Haynes} R.~F., {Simons}
  L.~W.~J., {Murdin} P.~G., 1979, \mnras, 187, 567

\bibitem[{{Dulk} et~al.(1983){Dulk}, {Bastian} \& {Chanmugam}}]{dulk83}
{Dulk} G.~A., {Bastian} T.~S., {Chanmugam} G., 1983, \apj, 273, 249

\bibitem[{{Enoto} et~al.(2014)}]{enoto14}
{Enoto} T. et~al., 2014, \apj, 786, 127

\bibitem[{{Espey} \& {Crowley}(2008)}]{espey08}
{Espey} B.~R., {Crowley} C., 2008, in A.~{Evans}, M.F. {Bode}, T.J. {O'Brien},
  M.J. {Darnley}, eds, RS Ophiuchi (2006) and the Recurrent Nova Phenomenon.
  Astronomical Society of the Pacific Conference Series, Vol. 401, p. 166

\bibitem[{{Espinasse} \& {Fender}(2018)}]{espinasse18}
{Espinasse} M., {Fender} R., 2018, \mnras, 473, 4122

\bibitem[{{Fabian}(2012)}]{fabian12}
{Fabian} A.~C., 2012, \araa, 50, 455

\bibitem[{{Falanga} et~al.(2015){Falanga}, {Bozzo}, {Lutovinov},
  {Bonnet-Bidaud}, {Fetisova} \& {Puls}}]{falanga15}
{Falanga} M., {Bozzo} E., {Lutovinov} A., {Bonnet-Bidaud} J.~M., {Fetisova} Y.,
  {Puls} J., 2015, \aap, 577, A130

\bibitem[{{Falcke} et~al.(2004){Falcke}, {K{\"o}rding} \& {Markoff}}]{falcke04}
{Falcke} H., {K{\"o}rding} E., {Markoff} S., 2004, \aap, 414, 895

\bibitem[{{Fender}(2016)}]{fender16}
{Fender} R., 2016, Astronomische Nachrichten, 337, 381

\bibitem[{{Fender} \& {Mu{\~n}oz-Darias}(2016)}]{fender16b}
{Fender} R., {Mu{\~n}oz-Darias} T., 2016, {The Balance of Power: Accretion and
  Feedback in Stellar Mass Black Holes}, Vol. 905. p.~65

\bibitem[{{Fender} et~al.(1998){Fender}, {Spencer}, {Tzioumis}, {Wu}, {van der
  Klis}, {van Paradijs} \& {Johnston}}]{fender98}
{Fender} R., {Spencer} R., {Tzioumis} T., {Wu} K., {van der Klis} M., {van
  Paradijs} J., {Johnston} H., 1998, \apjl, 506, L121

\bibitem[{{Fender} et~al.(2019){Fender}, {Bright}, {Mooley} \&
  {Miller-Jones}}]{fender19}
{Fender} R., {Bright} J., {Mooley} K., {Miller-Jones} J., 2019, \mnras, 490,
  L76

\bibitem[{{Fender} et~al.(1999)}]{fender99}
{Fender} R. et~al., 1999, \apjl, 519, L165

\bibitem[{{Fender} \& {Hendry}(2000)}]{fender00}
{Fender} R.~P., {Hendry} M.~A., 2000, \mnras, 317, 1

\bibitem[{{Fender} \& {Kuulkers}(2001)}]{fender01}
{Fender} R.~P., {Kuulkers} E., 2001, \mnras, 324, 923

\bibitem[{{Fender} et~al.(2004){Fender}, {Belloni} \& {Gallo}}]{fender04}
{Fender} R.~P., {Belloni} T.~M., {Gallo} E., 2004, \mnras, 355, 1105

\bibitem[{{Fender} et~al.(2005){Fender}, {Maccarone} \& {van
  Kesteren}}]{fender05}
{Fender} R.~P., {Maccarone} T.~J., {van Kesteren} Z., 2005, \mnras, 360, 1085

\bibitem[{Fender et~al.(2009)Fender, Homan \& Belloni}]{fender09}
Fender R.~P., Homan J., Belloni T.~M., 2009, MNRAS, 396, 1370

\bibitem[{{Fenech} et~al.(2018)}]{fenech18}
{Fenech} D.~M. et~al., 2018, \aap, 617, A137

\bibitem[{{Ferrigno} et~al.(2007){Ferrigno}, {Segreto}, {Santangelo}, {Wilms},
  {Kreykenbohm}, {Denis} \& {Staubert}}]{ferrigno07}
{Ferrigno} C., {Segreto} A., {Santangelo} A., {Wilms} J., {Kreykenbohm} I.,
  {Denis} M., {Staubert} R., 2007, \aap, 462, 995

\bibitem[{{Filliatre} \& {Chaty}(2004)}]{filliatre04}
{Filliatre} P., {Chaty} S., 2004, \apj, 616, 469

\bibitem[{{Finger} et~al.(1996){Finger}, {Koh}, {Nelson}, {Prince}, {Vaughan}
  \& {Wilson}}]{finger96}
{Finger} M.~H., {Koh} D.~T., {Nelson} R.~W., {Prince} T.~A., {Vaughan} B.~A.,
  {Wilson} R.~B., 1996, \nat, 381, 291

\bibitem[{{Fragos} et~al.(2013{\natexlab{a}}){Fragos}, {Lehmer}, {Naoz},
  {Zezas} \& {Basu-Zych}}]{fragos13}
{Fragos} T., {Lehmer} B.~D., {Naoz} S., {Zezas} A., {Basu-Zych} A.,
  2013{\natexlab{a}}, \apjl, 776, L31

\bibitem[{{Fragos} et~al.(2013{\natexlab{b}})}]{fragos12}
{Fragos} T. et~al., 2013{\natexlab{b}}, \apj, 764, 41

\bibitem[{{Gaensler} et~al.(2000){Gaensler}, {Stappers}, {Frail}, {Moffett},
  {Johnston} \& {Chatterjee}}]{gaensler00}
{Gaensler} B.~M., {Stappers} B.~W., {Frail} D.~A., {Moffett} D.~A., {Johnston}
  S., {Chatterjee} S., 2000, \mnras, 318, 58

\bibitem[{{Gaia Collaboration} et~al.(2020){Gaia Collaboration}, {Brown},
  {Vallenari}, {Prusti}, {de Bruijne}, {Babusiaux} \& {Biermann}}]{gaia2020}
{Gaia Collaboration}, {Brown} A.~G.~A., {Vallenari} A., {Prusti} T., {de
  Bruijne} J.~H.~J., {Babusiaux} C., {Biermann} M., 2020, arXiv e-prints,
  arXiv:2012.01533

\bibitem[{{Gallo} et~al.(2003){Gallo}, {Fender} \& {Pooley}}]{gallo03}
{Gallo} E., {Fender} R.~P., {Pooley} G.~G., 2003, \mnras, 344, 60

\bibitem[{{Gallo} et~al.(2005){Gallo}, {Fender}, {Kaiser}, {Russell},
  {Morganti}, {Oosterloo} \& {Heinz}}]{gallo05}
{Gallo} E., {Fender} R., {Kaiser} C., {Russell} D., {Morganti} R., {Oosterloo}
  T., {Heinz} S., 2005, \nat, 436, 819

\bibitem[{{Gallo} et~al.(2018){Gallo}, {Degenaar} \& {van den
  Eijnden}}]{gallo18}
{Gallo} E., {Degenaar} N., {van den Eijnden} J., 2018, \mnras, 478, L132

\bibitem[{{Gallo} et~al.(2014)}]{gallo2014}
{Gallo} E. et~al., 2014, \mnras, 445, 290

\bibitem[{{Galloway} et~al.(2008){Galloway}, {{\"O}zel} \&
  {Psaltis}}]{galloway2008}
{Galloway} D., {{\"O}zel} F., {Psaltis} D., 2008, \mnras, 387, 268

\bibitem[{{Galloway} et~al.(2002){Galloway}, {Chakrabarty}, {Morgan} \&
  {Remillard}}]{galloway02}
{Galloway} D.~K., {Chakrabarty} D., {Morgan} E.~H., {Remillard} R.~A., 2002,
  \apjl, 576, L137

\bibitem[{{Gandhi} et~al.(2017)}]{gandhi17}
{Gandhi} P. et~al., 2017, Nature Astronomy, 1, 859

\bibitem[{{Gehrels} et~al.(2004)}]{gehrels04}
{Gehrels} N. et~al., 2004, \apj, 611, 1005

\bibitem[{{Ghosh} \& {Lamb}(1978)}]{ghosh1978}
{Ghosh} P., {Lamb} F.~K., 1978, \apjl, 223, L83

\bibitem[{{Gim{\'e}nez-Garc{\'\i}a} et~al.(2015){Gim{\'e}nez-Garc{\'\i}a},
  {Torrej{\'o}n}, {Eikmann}, {Mart{\'\i}nez-N{\'u}{\~n}ez}, {Oskinova},
  {Rodes-Roca} \& {Bernab{\'e}u}}]{gimenez15}
{Gim{\'e}nez-Garc{\'\i}a} A., {Torrej{\'o}n} J.~M., {Eikmann} W.,
  {Mart{\'\i}nez-N{\'u}{\~n}ez} S., {Oskinova} L.~M., {Rodes-Roca} J.~J.,
  {Bernab{\'e}u} G., 2015, \aap, 576, A108

\bibitem[{{Grinberg} et~al.(2015)}]{grinberg_cygx1}
{Grinberg} V. et~al., 2015, \aap, 576, A117

\bibitem[{{Grinberg} et~al.(2017)}]{grinberg17}
{Grinberg} V. et~al., 2017, \aap, 608, A143

\bibitem[{{G{\"u}del}(2002)}]{gudel02}
{G{\"u}del} M., 2002, \araa, 40, 217

\bibitem[{{Guedel} \& {Benz}(1993)}]{gudel93}
{Guedel} M., {Benz} A.~O., 1993, \apjl, 405, L63

\bibitem[{{Gusinskaia} et~al.(2017)}]{gusinskaia17}
{Gusinskaia} N.~V. et~al., 2017, \mnras, 470, 1871

\bibitem[{{Gusinskaia} et~al.(2020{\natexlab{a}})}]{gusinskaia20}
{Gusinskaia} N.~V. et~al., 2020{\natexlab{a}}, \mnras, 492, 2858

\bibitem[{{Gusinskaia} et~al.(2020{\natexlab{b}})}]{gusinskaia20_igrj17591}
{Gusinskaia} N.~V. et~al., 2020{\natexlab{b}}, \mnras, 492, 1091

\bibitem[{{Hammerschlag-Hensberge} et~al.(1990){Hammerschlag-Hensberge},
  {Howarth} \& {Kallman}}]{hammerschlag90}
{Hammerschlag-Hensberge} G., {Howarth} I.~D., {Kallman} T.~R., 1990, \apj, 352,
  698

\bibitem[{{Hannikainen} et~al.(1998){Hannikainen}, {Hunstead},
  {Campbell-Wilson} \& {Sood}}]{hannikainen98}
{Hannikainen} D.~C., {Hunstead} R.~W., {Campbell-Wilson} D., {Sood} R.~K.,
  1998, \aap, 337, 460

\bibitem[{{Harmon} et~al.(1995)}]{harmon95}
{Harmon} B.~A. et~al., 1995, \nat, 374, 703

\bibitem[{{Hasinger} \& {van der Klis}(1989)}]{hasinger89}
{Hasinger} G., {van der Klis} M., 1989, \aap, 225, 79

\bibitem[{{Heap} \& {Corcoran}(1992)}]{heap92}
{Heap} S.~R., {Corcoran} M.~F., 1992, \apj, 387, 340

\bibitem[{{Heinke} et~al.(2015){Heinke}, {Bahramian}, {Degenaar} \&
  {Wijnands}}]{heinke15}
{Heinke} C.~O., {Bahramian} A., {Degenaar} N., {Wijnands} R., 2015, \mnras,
  447, 3034

\bibitem[{{Heinz} et~al.(2007){Heinz}, {Schulz}, {Brandt} \&
  {Galloway}}]{heinz07}
{Heinz} S., {Schulz} N.~S., {Brandt} W.~N., {Galloway} D.~K., 2007, \apjl, 663,
  L93

\bibitem[{{Heinz} et~al.(2013)}]{heinz13}
{Heinz} S. et~al., 2013, \apj, 779, 171

\bibitem[{{Hemphill} et~al.(2014){Hemphill}, {Rothschild}, {Markowitz},
  {F{\"u}rst}, {Pottschmidt} \& {Wilms}}]{hemphill14}
{Hemphill} P.~B., {Rothschild} R.~E., {Markowitz} A., {F{\"u}rst} F.,
  {Pottschmidt} K., {Wilms} J., 2014, \apj, 792, 14

\bibitem[{{Hemphill} et~al.(2019)}]{hemphill19}
{Hemphill} P.~B. et~al., 2019, \apj, 873, 62

\bibitem[{{Hern{\'a}ndez Santisteban} et~al.(2019)}]{hernandezsantisteban19}
{Hern{\'a}ndez Santisteban} J.~V. et~al., 2019, \mnras, 488, 4596

\bibitem[{{Hinkle} et~al.(2006){Hinkle}, {Fekel}, {Joyce}, {Wood}, {Smith} \&
  {Lebzelter}}]{hinkle06}
{Hinkle} K.~H., {Fekel} F.~C., {Joyce} R.~R., {Wood} P.~R., {Smith} V.~V.,
  {Lebzelter} T., 2006, \apj, 641, 479

\bibitem[{{Hjellming} et~al.(1990{\natexlab{a}}){Hjellming}, {Han}, {Cordova}
  \& {Hasinger}}]{hjellming1990_cygx2}
{Hjellming} R.~M., {Han} X.~H., {Cordova} F.~A., {Hasinger} G.,
  1990{\natexlab{a}}, \aap, 235, 147

\bibitem[{{Hjellming} et~al.(1990{\natexlab{b}})}]{hjellming1990_scox1}
{Hjellming} R.~M. et~al., 1990{\natexlab{b}}, \apj, 365, 681

\bibitem[{{Homan} et~al.(2010)}]{homan10}
{Homan} J. et~al., 2010, \apj, 719, 201

\bibitem[{{Homer} et~al.(1996){Homer}, {Charles}, {Naylor}, {van Paradijs},
  {Auriere} \& {Koch-Miramond}}]{homer96}
{Homer} L., {Charles} P.~A., {Naylor} T., {van Paradijs} J., {Auriere} M.,
  {Koch-Miramond} L., 1996, \mnras, 282, L37

\bibitem[{{Howarth} \& {Prinja}(1989)}]{howarth89}
{Howarth} I.~D., {Prinja} R.~K., 1989, \apjs, 69, 527

\bibitem[{{Iaria} et~al.(2016)}]{iaria16}
{Iaria} R. et~al., 2016, ArXiv e-prints

\bibitem[{{Illarionov} \& {Sunyaev}(1975)}]{illarionov75}
{Illarionov} A.~F., {Sunyaev} R.~A., 1975, \aap, 39, 185

\bibitem[{{Ilovaisky} et~al.(1982){Ilovaisky}, {Chevalier} \&
  {Motch}}]{ilovaisky82}
{Ilovaisky} S.~A., {Chevalier} C., {Motch} C., 1982, \aap, 114, L7

\bibitem[{{in't Zand} et~al.(2000){in't Zand}, {Halpern}, {Eracleous},
  {McCollough}, {Augusteijn}, {Remillard} \& {Heise}}]{intzand00}
{in't Zand} J.~J.~M., {Halpern} J., {Eracleous} M., {McCollough} M.,
  {Augusteijn} T., {Remillard} R.~A., {Heise} J., 2000, \aap, 361, 85

\bibitem[{{in't Zand} et~al.(2001){in't Zand}, {Swank}, {Corbet} \&
  {Markwardt}}]{intzand01}
{in't Zand} J.~J.~M., {Swank} J., {Corbet} R.~H.~D., {Markwardt} C.~B., 2001,
  \aap, 380, L26

\bibitem[{{in't Zand} et~al.(2005){in't Zand}, {Cumming}, {van der Sluys},
  {Verbunt} \& {Pols}}]{intzand05}
{in't Zand} J.~J.~M., {Cumming} A., {van der Sluys} M.~V., {Verbunt} F., {Pols}
  O.~R., 2005, \aap, 441, 675

\bibitem[{{in't Zand} et~al.(2008){in't Zand}, {Bassa}, {Jonker}, {Keek},
  {Verbunt}, {M{\'e}ndez} \& {Markwardt}}]{intzand08}
{in't Zand} J.~J.~M., {Bassa} C.~G., {Jonker} P.~G., {Keek} L., {Verbunt} F.,
  {M{\'e}ndez} M., {Markwardt} C.~B., 2008, \aap, 485, 183

\bibitem[{{Iyer} \& {Paul}(2017)}]{nirmal17}
{Iyer} N., {Paul} B., 2017, \mnras, 471, 355

\bibitem[{{Jaodand} et~al.(2018){Jaodand}, {Hessels} \&
  {Archibald}}]{jaodand18}
{Jaodand} A., {Hessels} J.~W.~T., {Archibald} A., 2018, in P.~{Weltevrede},
  B.B.P. {Perera}, L.L. {Preston}, S.~{Sanidas}, eds, Pulsar Astrophysics the
  Next Fifty Years. IAU Symposium, Vol. 337, pp. 47--51

\bibitem[{{Jenke} \& {Wilson-Hodge}(2017)}]{jenke17}
{Jenke} P., {Wilson-Hodge} C.~A., 2017, The Astronomer's Telegram, 10812, 1

\bibitem[{{Jones} et~al.(1973){Jones}, {Forman}, {Tananbaum}, {Schreier},
  {Gursky}, {Kellogg} \& {Giacconi}}]{jones73}
{Jones} C., {Forman} W., {Tananbaum} H., {Schreier} E., {Gursky} H., {Kellogg}
  E., {Giacconi} R., 1973, \apjl, 181, L43

\bibitem[{{Juett} et~al.(2001){Juett}, {Psaltis} \& {Chakrabarty}}]{juett01}
{Juett} A.~M., {Psaltis} D., {Chakrabarty} D., 2001, \apjl, 560, L59

\bibitem[{{Justham} \& {Schawinski}(2012)}]{justham12}
{Justham} S., {Schawinski} K., 2012, \mnras, 423, 1641

\bibitem[{{Kalberla} et~al.(2005){Kalberla}, {Burton}, {Hartmann}, {Arnal},
  {Bajaja}, {Morras} \& {P{\"o}ppel}}]{kalberla05}
{Kalberla} P.~M.~W., {Burton} W.~B., {Hartmann} D., {Arnal} E.~M., {Bajaja} E.,
  {Morras} R., {P{\"o}ppel} W.~G.~L., 2005, \aap, 440, 775

\bibitem[{{Kaplan} et~al.(2006){Kaplan}, {Moon} \& {Reach}}]{kaplan06}
{Kaplan} D.~L., {Moon} D.~S., {Reach} W.~T., 2006, \apjl, 649, L107

\bibitem[{{Karovska} et~al.(2010){Karovska}, {Gaetz}, {Carilli}, {Hack},
  {Raymond} \& {Lee}}]{karovska10}
{Karovska} M., {Gaetz} T.~J., {Carilli} C.~L., {Hack} W., {Raymond} J.~C.,
  {Lee} N.~P., 2010, \apjl, 710, L132

\bibitem[{{Keek} et~al.(2017){Keek}, {Iwakiri}, {Serino}, {Ballantyne}, {in 't
  Zand} \& {Strohmayer}}]{keek17}
{Keek} L., {Iwakiri} W., {Serino} M., {Ballantyne} D.~R., {in 't Zand}
  J.~J.~M., {Strohmayer} T.~E., 2017, \apj, 836, 111

\bibitem[{{Kennea} et~al.(2017){Kennea}, {Lien}, {Krimm}, {Cenko} \&
  {Siegel}}]{kennea17}
{Kennea} J.~A., {Lien} A.~Y., {Krimm} H.~A., {Cenko} S.~B., {Siegel} M.~H.,
  2017, The Astronomer's Telegram, 10809

\bibitem[{{Kennel} \& {Coroniti}(1984)}]{kennel84}
{Kennel} C.~F., {Coroniti} F.~V., 1984, \apj, 283, 694

\bibitem[{{King} et~al.(2013){King}, {Miller}, {G{\"u}ltekin}, {Walton},
  {Fabian}, {Reynolds} \& {Nandra}}]{king13}
{King} A.~L., {Miller} J.~M., {G{\"u}ltekin} K., {Walton} D.~J., {Fabian}
  A.~C., {Reynolds} C.~S., {Nandra} K., 2013, \apj, 771, 84

\bibitem[{{Koljonen} \& {Russell}(2019)}]{koljonen19}
{Koljonen} K.~I.~I., {Russell} D.~M., 2019, \apj, 871, 26

\bibitem[{{Kong} et~al.(2006){Kong}, {Charles}, {Homer}, {Kuulkers} \&
  {O'Donoghue}}]{kong06}
{Kong} A.~K.~H., {Charles} P.~A., {Homer} L., {Kuulkers} E., {O'Donoghue} D.,
  2006, \mnras, 368, 781

\bibitem[{{K{\"o}rding} et~al.(2008){K{\"o}rding}, {Rupen}, {Knigge}, {Fender},
  {Dhawan}, {Templeton} \& {Muxlow}}]{kording08}
{K{\"o}rding} E., {Rupen} M., {Knigge} C., {Fender} R., {Dhawan} V.,
  {Templeton} M., {Muxlow} T., 2008, Science, 320, 1318

\bibitem[{{Kouroubatzakis} et~al.(2017){Kouroubatzakis}, {Reig}, {Andrews} \&
  {)}}]{kouroubatzakis17}
{Kouroubatzakis} K., {Reig} P., {Andrews} J., {)} A.~Z., 2017, The Astronomer's
  Telegram, 10822, 1

\bibitem[{{Krzeminski}(1973)}]{krzeminski73}
{Krzeminski} W., 1973, \iaucirc, 2612, 1

\bibitem[{{Krzeminski}(1974)}]{krzeminski74}
{Krzeminski} W., 1974, \apjl, 192, L135

\bibitem[{{Kuiper} et~al.(2020){Kuiper}, {Tsygankov}, {Falanga}, {Mereminskij},
  {Galloway}, {Poutanen} \& {Li}}]{kuiper20}
{Kuiper} L., {Tsygankov} S.~S., {Falanga} M., {Mereminskij} I.~A., {Galloway}
  D.~K., {Poutanen} J., {Li} Z., 2020, arXiv e-prints, arXiv:2002.12154

\bibitem[{{Kuranov} \& {Postnov}(2015)}]{kuranov15}
{Kuranov} A.~G., {Postnov} K.~A., 2015, Astronomy Letters, 41, 114

\bibitem[{{Kurapati} et~al.(2017)}]{kurapati17}
{Kurapati} S. et~al., 2017, \mnras, 465, 2160

\bibitem[{{Lamers}(1998{\natexlab{a}})}]{lamers98}
{Lamers} H.~J.~G.~L.~M., 1998{\natexlab{a}}, \apss, 260, 63

\bibitem[{{Lamers}(1998{\natexlab{b}})}]{lamers98b}
{Lamers} H.~J.~G.~L.~M., 1998{\natexlab{b}}, \apss, 260, 81

\bibitem[{{Lampton} et~al.(1971){Lampton}, {Bowyer}, {Welch} \&
  {Grasdalen}}]{lampton71}
{Lampton} M., {Bowyer} S., {Welch} J., {Grasdalen} G., 1971, \apjl, 164, L61

\bibitem[{{Leahy} \& {Abdallah}(2014)}]{leahy14}
{Leahy} D.~A., {Abdallah} M.~H., 2014, \apj, 793, 79

\bibitem[{{Lewin} et~al.(1971){Lewin}, {Ricker} \& {McClintock}}]{lewin71}
{Lewin} W.~H.~G., {Ricker} G.~R., {McClintock} J.~E., 1971, \apjl, 169, L17

\bibitem[{{Li} et~al.(2020){Li}, {Strader}, {Miller-Jones}, {Heinke} \&
  {Chomiuk}}]{li20}
{Li} K.~L., {Strader} J., {Miller-Jones} J.~C.~A., {Heinke} C.~O., {Chomiuk}
  L., 2020, \apj, 895, 89

\bibitem[{{Lin} et~al.(2009){Lin}, {Remillard} \& {Homan}}]{lin09}
{Lin} D., {Remillard} R.~A., {Homan} J., 2009, \apj, 696, 1257

\bibitem[{{Livio}(1997)}]{livio97}
{Livio} M., 1997, {The Formation Of Astrophysical Jets}, Astronomical Society
  of the Pacific Conference Series, Vol. 121. p. 845

\bibitem[{{Livio}(1999)}]{livio99}
{Livio} M., 1999, \physrep, 311, 225

\bibitem[{{Longair}(1992)}]{longair92}
{Longair} M.~S., 1992, {High energy astrophysics. Vol.1: Particles, photons and
  their detection}

\bibitem[{{Lucy} et~al.(2019)}]{lucy19}
{Lucy} A.~B. et~al., 2019, The Astronomer's Telegram, 13152, 1

\bibitem[{{Ludlam} et~al.(2019)}]{ludlam2019}
{Ludlam} R.~M. et~al., 2019, \apj, 873, 99

\bibitem[{{Maccarone} et~al.(2020){Maccarone}, {van den Eijnden}, {Russell} \&
  {Degenaar}}]{maccarone20}
{Maccarone} T.~J., {van den Eijnden} J., {Russell} T.~D., {Degenaar} N., 2020,
  \mnras, 499, 957

\bibitem[{{Madej} et~al.(2013){Madej}, {Jonker}, {Groot}, {van Haaften},
  {Nelemans} \& {Maccarone}}]{madej13}
{Madej} O.~K., {Jonker} P.~G., {Groot} P.~J., {van Haaften} L.~M., {Nelemans}
  G., {Maccarone} T.~J., 2013, \mnras, 429, 2986

\bibitem[{{Madsen} et~al.(2017){Madsen}, {Forster}, {Grefenstette}, {Harrison}
  \& {Stern}}]{madsen17}
{Madsen} K.~K., {Forster} K., {Grefenstette} B.~W., {Harrison} F.~A., {Stern}
  D., 2017, \apj, 841, 56

\bibitem[{{Malzac}(2013)}]{malzac13}
{Malzac} J., 2013, \mnras, 429, L20

\bibitem[{{Malzac}(2014)}]{malzac14}
{Malzac} J., 2014, \mnras, 443, 299

\bibitem[{{Manousakis} \& {Walter}(2011)}]{manousakis11}
{Manousakis} A., {Walter} R., 2011, \aap, 526, A62

\bibitem[{{Markoff} et~al.(2001){Markoff}, {Falcke} \& {Fender}}]{markoff01}
{Markoff} S., {Falcke} H., {Fender} R., 2001, \aap, 372, L25

\bibitem[{{Markoff} et~al.(2005){Markoff}, {Nowak} \& {Wilms}}]{markoff05}
{Markoff} S., {Nowak} M.~A., {Wilms} J., 2005, \apj, 635, 1203

\bibitem[{{Markwardt} et~al.(1999){Markwardt}, {Strohmayer} \&
  {Swank}}]{markwardt99}
{Markwardt} C.~B., {Strohmayer} T.~E., {Swank} J.~H., 1999, \apjl, 512, L125

\bibitem[{{Mart{\'\i}nez-N{\'u}{\~n}ez} et~al.(2017)}]{martineznunez17}
{Mart{\'\i}nez-N{\'u}{\~n}ez} S. et~al., 2017, \ssr, 212, 59

\bibitem[{{Masetti} et~al.(2007)}]{masetti07a}
{Masetti} N., {Rigon} E., {Maiorano} E., {Cusumano} G., {Palazzi} E.,
  {Orlandini} M., {Amati} L., {Frontera} F., 2007, \aap, 464, 277

\bibitem[{{Mason} et~al.(2012)}]{mason12}
{Mason} A.~B., {Clark} J.~S., {Norton} A.~J., {Crowther} P.~A., {Tauris} T.~M.,
  {Langer} N., {Negueruela} I., {Roche} P., 2012, \mnras, 422, 199

\bibitem[{{Mason} \& {Gray}(2007)}]{mason07}
{Mason} P.~A., {Gray} C.~L., 2007, \apj, 660, 662

\bibitem[{{Massi} \& {Kaufman Bernad{\'o}}(2008)}]{massi08}
{Massi} M., {Kaufman Bernad{\'o}} M., 2008, \aap, 477, 1

\bibitem[{{Matsuoka} et~al.(2009)}]{matsuoka09}
{Matsuoka} M. et~al., 2009, \pasj, 61, 999

\bibitem[{{McClintock} et~al.(2014){McClintock}, {Narayan} \&
  {Steiner}}]{mcclintock14}
{McClintock} J.~E., {Narayan} R., {Steiner} J.~F., 2014, \ssr, 183, 295

\bibitem[{{McMullin} et~al.(2007){McMullin}, {Waters}, {Schiebel}, {Young} \&
  {Golap}}]{mcmullin07}
{McMullin} J.~P., {Waters} B., {Schiebel} D., {Young} W., {Golap} K., 2007, in
  R.A. {Shaw}, F.~{Hill}, D.J. {Bell}, eds, Astronomical Data Analysis Software
  and Systems XVI. Astronomical Society of the Pacific Conference Series, Vol.
  376, p. 127

\bibitem[{{Merloni} et~al.(2003){Merloni}, {Heinz} \& {di Matteo}}]{merloni03}
{Merloni} A., {Heinz} S., {di Matteo} T., 2003, \mnras, 345, 1057

\bibitem[{{Meshcheryakov} et~al.(2010){Meshcheryakov}, {Revnivtsev},
  {Pavlinsky}, {Khamitov} \& {Bikmaev}}]{meshcheryakov10}
{Meshcheryakov} A.~V., {Revnivtsev} M.~G., {Pavlinsky} M.~N., {Khamitov} I.,
  {Bikmaev} I.~F., 2010, Astronomy Letters, 36, 738

\bibitem[{{Meyer-Hofmeister} \& {Meyer}(2014)}]{meyer14}
{Meyer-Hofmeister} E., {Meyer} F., 2014, \aap, 562, A142

\bibitem[{{Michel}(1982)}]{michel82}
{Michel} F.~C., 1982, Reviews of Modern Physics, 54, 1

\bibitem[{{Middleditch} et~al.(1981){Middleditch}, {Mason}, {Nelson} \&
  {White}}]{middleditch81}
{Middleditch} J., {Mason} K.~O., {Nelson} J.~E., {White} N.~E., 1981, \apj,
  244, 1001

\bibitem[{{Migliari}(2011)}]{migliari11c}
{Migliari} S., 2011, in G.E. {Romero}, R.A. {Sunyaev}, T.~{Belloni}, eds, Jets
  at All Scales. IAU Symposium, Vol. 275, pp. 233--241

\bibitem[{{Migliari} \& {Fender}(2006)}]{migliari06}
{Migliari} S., {Fender} R.~P., 2006, \mnras, 366, 79

\bibitem[{{Migliari} et~al.(2003){Migliari}, {Fender}, {Rupen}, {Jonker},
  {Klein-Wolt}, {Hjellming} \& {van der Klis}}]{migliari03}
{Migliari} S., {Fender} R.~P., {Rupen} M., {Jonker} P.~G., {Klein-Wolt} M.,
  {Hjellming} R.~M., {van der Klis} M., 2003, \mnras, 342, L67

\bibitem[{{Migliari} et~al.(2004){Migliari}, {Fender}, {Rupen}, {Wachter},
  {Jonker}, {Homan} \& {van der Klis}}]{migliari04}
{Migliari} S., {Fender} R.~P., {Rupen} M., {Wachter} S., {Jonker} P.~G.,
  {Homan} J., {van der Klis} M., 2004, \mnras, 351, 186

\bibitem[{{Migliari} et~al.(2011{\natexlab{a}}){Migliari}, {Miller-Jones} \&
  {Russell}}]{migliari11b}
{Migliari} S., {Miller-Jones} J.~C.~A., {Russell} D.~M., 2011{\natexlab{a}},
  \mnras, 415, 2407

\bibitem[{{Migliari} et~al.(2011{\natexlab{b}}){Migliari}, {Tudose},
  {Miller-Jones}, {Kuulkers}, {Nakajima} \& {Yamaoka}}]{migliari11}
{Migliari} S., {Tudose} V., {Miller-Jones} J.~C.~A., {Kuulkers} E., {Nakajima}
  M., {Yamaoka} K., 2011{\natexlab{b}}, The Astronomer's Telegram, 3198

\bibitem[{{Migliari} et~al.(2010)}]{migliari10}
{Migliari} S. et~al., 2010, \apj, 710, 117

\bibitem[{{Miller} et~al.(2012)}]{miller12}
{Miller} J.~M., {Pooley} G.~G., {Fabian} A.~C., {Nowak} M.~A., {Reis} R.~C.,
  {Cackett} E.~M., {Pottschmidt} K., {Wilms} J., 2012, \apj, 757, 11

\bibitem[{{Miller-Jones} et~al.(2011){Miller-Jones}, {Sivakoff}, {Heinke},
  {Altamirano}, {Kuulkers} \& {Morii}}]{millerjones11b}
{Miller-Jones} J.~C.~A., {Sivakoff} G.~R., {Heinke} C.~O., {Altamirano} D.,
  {Kuulkers} E., {Morii} M., 2011, The Astronomer's Telegram, 3378, 1

\bibitem[{{Miller-Jones} et~al.(2010)}]{millerjones10}
{Miller-Jones} J.~C.~A. et~al., 2010, \apjl, 716, L109

\bibitem[{{Miller-Jones} et~al.(2012)}]{millerjones12}
{Miller-Jones} J.~C.~A. et~al., 2012, \mnras, 421, 468

\bibitem[{{Mirabel} et~al.(2011){Mirabel}, {Dijkstra}, {Laurent}, {Loeb} \&
  {Pritchard}}]{mirabel11}
{Mirabel} I.~F., {Dijkstra} M., {Laurent} P., {Loeb} A., {Pritchard} J.~R.,
  2011, \aap, 528, A149

\bibitem[{{Mondal} et~al.(2019){Mondal}, {Dewangan} \&
  {Raychaudhuri}}]{mondal19}
{Mondal} A.~S., {Dewangan} G.~C., {Raychaudhuri} B., 2019, \mnras, 487, 5441

\bibitem[{{Moran} et~al.(1989)}]{moran89}
{Moran} J.~P., {Davis} R.~J., {Bode} M.~F., {Taylor} A.~R., {Spencer} R.~E.,
  {Argue} A.~N., {Irwin} M.~J., {Shanklin} J.~D., 1989, \nat, 340, 449

\bibitem[{{Motch}(1998)}]{motch98}
{Motch} C., 1998, \aap, 338, L13

\bibitem[{{Motta} \& {Fender}(2019)}]{motta19}
{Motta} S.~E., {Fender} R.~P., 2019, \mnras, 483, 3686

\bibitem[{{Motta} et~al.(2018){Motta}, {Casella} \& {Fender}}]{motta18}
{Motta} S.~E., {Casella} P., {Fender} R.~P., 2018, \mnras, 478, 5159

\bibitem[{{Mukherjee} et~al.(2015){Mukherjee}, {Bult}, {van der Klis} \&
  {Bhattacharya}}]{mukherjee15}
{Mukherjee} D., {Bult} P., {van der Klis} M., {Bhattacharya} D., 2015, \mnras,
  452, 3994

\bibitem[{{Negueruela} et~al.(2010){Negueruela}, {Clark} \&
  {Ritchie}}]{negueruela10}
{Negueruela} I., {Clark} J.~S., {Ritchie} B.~W., 2010, \aap, 516, A78

\bibitem[{{Nelemans} \& {Jonker}(2010)}]{nelemans10}
{Nelemans} G., {Jonker} P.~G., 2010, \nar, 54, 87

\bibitem[{{Nelemans} et~al.(2004){Nelemans}, {Jonker}, {Marsh} \& {van der
  Klis}}]{nelemans04}
{Nelemans} G., {Jonker} P.~G., {Marsh} T.~R., {van der Klis} M., 2004, \mnras,
  348, L7

\bibitem[{{Nelson} \& {Spencer}(1988)}]{nelson88}
{Nelson} R.~F., {Spencer} R.~E., 1988, \mnras, 234, 1105

\bibitem[{{Ogley} et~al.(2002)}]{ogley02}
{Ogley} R.~N., {Chaty} S., {Crocker} M., {Eyres} S.~P.~S., {Kenworthy} M.~A.,
  {Richards} A.~M.~S., {Rodr{\'{\i}}guez} L.~F., {Stirling} A.~M., 2002,
  \mnras, 330, 772

\bibitem[{{Olnon}(1975)}]{olnon1975}
{Olnon} F.~M., 1975, \aap, 39, 217

\bibitem[{{Ortiz-Le{\'o}n} et~al.(2011){Ortiz-Le{\'o}n}, {Loinard},
  {Rodr{\'\i}guez}, {Mioduszewski} \& {Dzib}}]{ortiz2011}
{Ortiz-Le{\'o}n} G.~N., {Loinard} L., {Rodr{\'\i}guez} L.~F., {Mioduszewski}
  A.~J., {Dzib} S.~A., 2011, \apj, 737, 30

\bibitem[{{Padin} et~al.(1985){Padin}, {Davis} \& {Bode}}]{padin85}
{Padin} S., {Davis} R.~J., {Bode} M.~F., 1985, \nat, 315, 306

\bibitem[{{Panagia} \& {Felli}(1975)}]{panagia1975}
{Panagia} N., {Felli} M., 1975, \aap, 39, 1

\bibitem[{{Papitto} et~al.(2013)}]{papitto13}
{Papitto} A. et~al., 2013, \mnras, 429, 3411

\bibitem[{{Paredes} \& {Bordas}(2019)}]{paredes19}
{Paredes} J.~M., {Bordas} P., 2019, arXiv e-prints, arXiv:1901.03624

\bibitem[{{Parfrey} et~al.(2016){Parfrey}, {Spitkovsky} \&
  {Beloborodov}}]{parfrey16}
{Parfrey} K., {Spitkovsky} A., {Beloborodov} A.~M., 2016, \apj, 822, 33

\bibitem[{{Parfrey} et~al.(2017){Parfrey}, {Spitkovsky} \&
  {Beloborodov}}]{parfrey17}
{Parfrey} K., {Spitkovsky} A., {Beloborodov} A.~M., 2017, \mnras, 469, 3656

\bibitem[{{Patruno}(2012)}]{patruno12c}
{Patruno} A., 2012, \apjl, 753, L12

\bibitem[{{Patruno} \& {Watts}(2012)}]{patruno12}
{Patruno} A., {Watts} A.~L., 2012, ArXiv e-prints

\bibitem[{{Patruno} et~al.(2017){Patruno}, {Haskell} \&
  {Andersson}}]{patruno17}
{Patruno} A., {Haskell} B., {Andersson} N., 2017, ArXiv e-prints

\bibitem[{{Pavelin} et~al.(1994){Pavelin}, {Spencer} \& {Davis}}]{pavelin94}
{Pavelin} P.~E., {Spencer} R.~E., {Davis} R.~J., 1994, \mnras, 269, 779

\bibitem[{{Penninx}(1989)}]{penninx89}
{Penninx} W., 1989, in J.~{Hunt}, B.~{Battrick}, eds, Two Topics in X-Ray
  Astronomy, Volume 1: X Ray Binaries. Volume 2: AGN and the X Ray Background.
  ESA Special Publication, Vol.~1, p. 185

\bibitem[{{Penninx} et~al.(1988){Penninx}, {Lewin}, {Zijlstra}, {Mitsuda} \&
  {van Paradijs}}]{penninx88}
{Penninx} W., {Lewin} W.~H.~G., {Zijlstra} A.~A., {Mitsuda} K., {van Paradijs}
  J., 1988, \nat, 336, 146

\bibitem[{{Pestalozzi} et~al.(2009){Pestalozzi}, {Torkelsson}, {Hobbs} \&
  {L{\'o}pez-S{\'a}nchez}}]{pestalozzi09}
{Pestalozzi} M., {Torkelsson} U., {Hobbs} G., {L{\'o}pez-S{\'a}nchez}
  {\'A}.~R., 2009, \aap, 506, L21

\bibitem[{{Plotkin} et~al.(2013){Plotkin}, {Gallo} \& {Jonker}}]{plotkin13}
{Plotkin} R.~M., {Gallo} E., {Jonker} P.~G., 2013, \apj, 773, 59

\bibitem[{{Puls} et~al.(2006){Puls}, {Markova}, {Scuderi}, {Stanghellini},
  {Taranova}, {Burnley} \& {Howarth}}]{puls2006}
{Puls} J., {Markova} N., {Scuderi} S., {Stanghellini} C., {Taranova} O.~G.,
  {Burnley} A.~W., {Howarth} I.~D., 2006, \aap, 454, 625

\bibitem[{{Qiu} et~al.(2017){Qiu}, {Zhou}, {Yu}, {Li} \& {Xu}}]{qiu17}
{Qiu} H., {Zhou} P., {Yu} W., {Li} X., {Xu} X., 2017, \apj, 847, 44

\bibitem[{{Rappaport} \& {Joss}(1997)}]{rappaport97}
{Rappaport} S., {Joss} P.~C., 1997, \apj, 486, 435

\bibitem[{{Rappaport} et~al.(1987){Rappaport}, {Nelson}, {Ma} \&
  {Joss}}]{rappaport87}
{Rappaport} S., {Nelson} L.~A., {Ma} C.~P., {Joss} P.~C., 1987, \apj, 322, 842

\bibitem[{{Rea} et~al.(2005)}]{rea05}
{Rea} N., {Stella} L., {Israel} G.~L., {Matt} G., {Zane} S., {Segreto} A.,
  {Oosterbroek} T., {Orlandini} M., 2005, \mnras, 364, 1229

\bibitem[{{Rees} \& {Gunn}(1974)}]{rees74}
{Rees} M.~J., {Gunn} J.~E., 1974, \mnras, 167, 1

\bibitem[{{Reig}(2011)}]{reig11}
{Reig} P., 2011, \apss, 332, 1

\bibitem[{{Reig} et~al.(2017){Reig}, {Blay} \& {Blinov}}]{reig17}
{Reig} P., {Blay} P., {Blinov} D., 2017, \aap, 598, A16

\bibitem[{{Reynolds} et~al.(1997){Reynolds}, {Quaintrell}, {Still}, {Roche},
  {Chakrabarty} \& {Levine}}]{reynolds97}
{Reynolds} A.~P., {Quaintrell} H., {Still} M.~D., {Roche} P., {Chakrabarty} D.,
  {Levine} S.~E., 1997, \mnras, 288, 43

\bibitem[{{Rib{\'o}} et~al.(2017)}]{ribo17}
{Rib{\'o}} M. et~al., 2017, \apjl, 835, L33

\bibitem[{{Rodriguez} et~al.(2006)}]{rodriguez06}
{Rodriguez} J. et~al., 2006, \mnras, 366, 274

\bibitem[{{Romanova} et~al.(2009){Romanova}, {Ustyugova}, {Koldoba} \&
  {Lovelace}}]{romanova09}
{Romanova} M.~M., {Ustyugova} G.~V., {Koldoba} A.~V., {Lovelace} R.~V.~E.,
  2009, \mnras, 399, 1802

\bibitem[{{Romero} et~al.(2017){Romero}, {Boettcher}, {Markoff} \&
  {Tavecchio}}]{romero17}
{Romero} G.~E., {Boettcher} M., {Markoff} S., {Tavecchio} F., 2017, \ssr, 207,
  5

\bibitem[{{Rubin} et~al.(1996)}]{rubin96}
{Rubin} B.~C. et~al., 1996, \apj, 459, 259

\bibitem[{{Russell} et~al.(2006){Russell}, {Fender}, {Hynes}, {Brocksopp},
  {Homan}, {Jonker} \& {Buxton}}]{russellD06}
{Russell} D.~M., {Fender} R.~P., {Hynes} R.~I., {Brocksopp} C., {Homan} J.,
  {Jonker} P.~G., {Buxton} M.~M., 2006, \mnras, 371, 1334

\bibitem[{{Russell} et~al.(2007){Russell}, {Fender} \& {Jonker}}]{russellD07}
{Russell} D.~M., {Fender} R.~P., {Jonker} P.~G., 2007, \mnras, 379, 1108

\bibitem[{{Russell} et~al.(2013{\natexlab{a}})}]{russell13b}
{Russell} D.~M. et~al., 2013{\natexlab{a}}, \apjl, 768, L35

\bibitem[{{Russell} et~al.(2013{\natexlab{b}})}]{russell13}
{Russell} D.~M. et~al., 2013{\natexlab{b}}, \mnras, 429, 815

\bibitem[{{Russell} et~al.(2017){Russell}, {Degenaar}, {Miller-Jones} \&
  {Tudor}}]{russell17}
{Russell} T., {Degenaar} N., {Miller-Jones} J., {Tudor} V., 2017, The
  Astronomer's Telegram, No.~10106, 10106

\bibitem[{{Russell} et~al.(2014){Russell}, {Soria}, {Miller-Jones}, {Curran},
  {Markoff}, {Russell} \& {Sivakoff}}]{russell14}
{Russell} T.~D., {Soria} R., {Miller-Jones} J.~C.~A., {Curran} P.~A., {Markoff}
  S., {Russell} D.~M., {Sivakoff} G.~R., 2014, \mnras, 439, 1390

\bibitem[{{Russell} et~al.(2018){Russell}, {Degenaar}, {Wijnands}, {van den
  Eijnden}, {Gusinskaia}, {Hessels} \& {Miller-Jones}}]{russell18}
{Russell} T.~D., {Degenaar} N., {Wijnands} R., {van den Eijnden} J.,
  {Gusinskaia} N.~V., {Hessels} J.~W.~T., {Miller-Jones} J.~C.~A., 2018, \apjl,
  869, L16

\bibitem[{{Russell} et~al.(2015)}]{russellT15}
{Russell} T.~D. et~al., 2015, \mnras, 450, 1745

\bibitem[{{Russell} et~al.(2016)}]{russell16}
{Russell} T.~D. et~al., 2016, \mnras, 460, 3720

\bibitem[{{Russell} et~al.(2019)}]{russell2019_1535}
{Russell} T.~D. et~al., 2019, \apj, 883, 198

\bibitem[{{Russell} et~al.(2020)}]{russell2020}
{Russell} T.~D. et~al., 2020, \mnras, 498, 5772

\bibitem[{{Rutledge} et~al.(1998){Rutledge}, {Moore}, {Fox}, {Lewin} \& {van
  Paradijs}}]{rutledge98}
{Rutledge} R., {Moore} C., {Fox} D., {Lewin} W., {van Paradijs} J., 1998, The
  Astronomer's Telegram, 8, 1

\bibitem[{{Rybicki} \& {Lightman}(1979)}]{rybicki79}
{Rybicki} G.~B., {Lightman} A.~P., 1979, {Radiative processes in astrophysics}

\bibitem[{{Sanna} et~al.(2017)}]{sanna17}
{Sanna} A. et~al., 2017, \mnras, 471, 463

\bibitem[{{Sanna} et~al.(2018)}]{sanna2018}
{Sanna} A. et~al., 2018, \aap, 616, L17

\bibitem[{{Savolainen} et~al.(2009){Savolainen}, {Hannikainen}, {Vilhu},
  {Paizis}, {Nevalainen} \& {Hakala}}]{savolainen09}
{Savolainen} P., {Hannikainen} D.~C., {Vilhu} O., {Paizis} A., {Nevalainen} J.,
  {Hakala} P., 2009, \mnras, 393, 569

\bibitem[{{Schulz} et~al.(2020){Schulz}, {Kallman}, {Heinz}, {Sell}, {Jonker}
  \& {Brandt}}]{schulz20}
{Schulz} N.~S., {Kallman} T.~E., {Heinz} S., {Sell} P., {Jonker} P., {Brandt}
  W.~N., 2020, \apj, 891, 150

\bibitem[{{Seaquist} \& {Taylor}(1990)}]{seaquist90}
{Seaquist} E.~R., {Taylor} A.~R., 1990, \apj, 349, 313

\bibitem[{{Seaquist} et~al.(1984){Seaquist}, {Taylor} \& {Button}}]{seaquist84}
{Seaquist} E.~R., {Taylor} A.~R., {Button} S., 1984, \apj, 284, 202

\bibitem[{{Seaquist} et~al.(1993){Seaquist}, {Krogulec} \&
  {Taylor}}]{seaquist93}
{Seaquist} E.~R., {Krogulec} M., {Taylor} A.~R., 1993, \apj, 410, 260

\bibitem[{{Selina} et~al.(2018)}]{selina18}
{Selina} R.~J. et~al., 2018, {The ngVLA Reference Design}, Astronomical Society
  of the Pacific Conference Series, Vol. 517. p.~15

\bibitem[{{Shaw} et~al.(2020)}]{shaw20}
{Shaw} A.~W. et~al., 2020, \mnras, 492, 4344

\bibitem[{{Snow}(1981)}]{snow81}
{Snow} T.~P. J., 1981, \apj, 251, 139

\bibitem[{{Sokoloski} et~al.(2017){Sokoloski}, {Lawrence}, {Crotts} \&
  {Mukai}}]{sokoloski17}
{Sokoloski} J.~L., {Lawrence} S., {Crotts} A.~P.~S., {Mukai} K., 2017, arXiv
  e-prints, arXiv:1702.05898

\bibitem[{{Soleri} \& {Fender}(2011)}]{soleri11}
{Soleri} P., {Fender} R., 2011, \mnras, 413, 2269

\bibitem[{{Soleri} et~al.(2009){Soleri}, {Tudose}, {Fender}, {van der Klis} \&
  {Jonker}}]{soleri09}
{Soleri} P., {Tudose} V., {Fender} R., {van der Klis} M., {Jonker} P.~G., 2009,
  \mnras, 399, 453

\bibitem[{{Staubert} et~al.(2019)}]{staubert19}
{Staubert} R. et~al., 2019, \aap, 622, A61

\bibitem[{{Stewart} et~al.(1993){Stewart}, {Caswell}, {Haynes} \&
  {Nelson}}]{stewart93}
{Stewart} R.~T., {Caswell} J.~L., {Haynes} R.~F., {Nelson} G.~J., 1993, \mnras,
  261, 593

\bibitem[{{Stirling} et~al.(2001){Stirling}, {Spencer}, {de la Force},
  {Garrett}, {Fender} \& {Ogley}}]{stirling2001}
{Stirling} A.~M., {Spencer} R.~E., {de la Force} C.~J., {Garrett} M.~A.,
  {Fender} R.~P., {Ogley} R.~N., 2001, \mnras, 327, 1273

\bibitem[{{Strohmayer} et~al.(2018{\natexlab{a}})}]{strohmayer2018_1737}
{Strohmayer} T.~E. et~al., 2018{\natexlab{a}}, The Astronomer's Telegram,
  11507, 1

\bibitem[{{Strohmayer} et~al.(2018{\natexlab{b}})}]{strohmayer2018}
{Strohmayer} T.~E. et~al., 2018{\natexlab{b}}, \apjl, 858, L13

\bibitem[{{Tananbaum} et~al.(1972){Tananbaum}, {Gursky}, {Kellogg}, {Giacconi}
  \& {Jones}}]{tananbaum72b}
{Tananbaum} H., {Gursky} H., {Kellogg} E., {Giacconi} R., {Jones} C., 1972,
  \apjl, 177, L5

\bibitem[{{Tetarenko} et~al.(2015)}]{tetarenko15}
{Tetarenko} A.~J. et~al., 2015, \apj, 805, 30

\bibitem[{{Tetarenko} et~al.(2018{\natexlab{a}})}]{tetarenko18_igrj16597}
{Tetarenko} A.~J. et~al., 2018{\natexlab{a}}, \apj, 854, 125

\bibitem[{{Tetarenko} et~al.(2016){Tetarenko}, {Sivakoff}, {Heinke} \&
  {Gladstone}}]{tetarenko2016_watchdog}
{Tetarenko} B.~E., {Sivakoff} G.~R., {Heinke} C.~O., {Gladstone} J.~C., 2016,
  \apjs, 222, 15

\bibitem[{{Tetarenko} et~al.(2018{\natexlab{b}}){Tetarenko}, {Lasota},
  {Heinke}, {Dubus} \& {Sivakoff}}]{tetarenko18_winds}
{Tetarenko} B.~E., {Lasota} J.~P., {Heinke} C.~O., {Dubus} G., {Sivakoff}
  G.~R., 2018{\natexlab{b}}, \nat, 554, 69

\bibitem[{{Thompson} et~al.(2005){Thompson}, {Rothschild}, {Tomsick} \&
  {Marshall}}]{thompson05}
{Thompson} T.~W.~J., {Rothschild} R.~E., {Tomsick} J.~A., {Marshall} H.~L.,
  2005, \apj, 634, 1261

\bibitem[{{Thompson} et~al.(2007){Thompson}, {Tomsick}, {in 't Zand },
  {Rothschild} \& {Walter}}]{thompson07}
{Thompson} T.~W.~J., {Tomsick} J.~A., {in 't Zand } J.~J.~M., {Rothschild}
  R.~E., {Walter} R., 2007, \apj, 661, 447

\bibitem[{{Torrej{\'o}n} et~al.(2015){Torrej{\'o}n}, {Schulz}, {Nowak},
  {Oskinova}, {Rodes-Roca}, {Shenar} \& {Wilms}}]{torrejon15}
{Torrej{\'o}n} J.~M., {Schulz} N.~S., {Nowak} M.~A., {Oskinova} L.,
  {Rodes-Roca} J.~J., {Shenar} T., {Wilms} J., 2015, \apj, 810, 102

\bibitem[{{Tsygankov} et~al.(2017){Tsygankov}, {Mushtukov}, {Suleimanov},
  {Doroshenko}, {Abolmasov}, {Lutovinov} \& {Poutanen}}]{tsygankov17}
{Tsygankov} S.~S., {Mushtukov} A.~A., {Suleimanov} V.~F., {Doroshenko} V.,
  {Abolmasov} P.~K., {Lutovinov} A.~A., {Poutanen} J., 2017, \aap, 608, A17

\bibitem[{{Tsygankov} et~al.(2018){Tsygankov}, {Doroshenko}, {Mushtukov},
  {Lutovinov} \& {Poutanen}}]{tsygankov18}
{Tsygankov} S.~S., {Doroshenko} V., {Mushtukov} A.~A., {Lutovinov} A.~A.,
  {Poutanen} J., 2018, \mnras, 479, L134

\bibitem[{{Tudor} et~al.(2017)}]{tudor17}
{Tudor} V. et~al., 2017, \mnras, 470, 324

\bibitem[{{Tudose} et~al.(2006){Tudose}, {Fender}, {Kaiser}, {Tzioumis}, {van
  der Klis} \& {Spencer}}]{tudose06}
{Tudose} V., {Fender} R.~P., {Kaiser} C.~R., {Tzioumis} A.~K., {van der Klis}
  M., {Spencer} R.~E., 2006, \mnras, 372, 417

\bibitem[{{van den Berg} \& {Homan}(2017)}]{vandenberg17}
{van den Berg} M., {Homan} J., 2017, \apj, 834, 71

\bibitem[{{van den Berg} et~al.(2014){van den Berg}, {Homan}, {Fridriksson} \&
  {Linares}}]{vandenberg14}
{van den Berg} M., {Homan} J., {Fridriksson} J.~K., {Linares} M., 2014, \apj,
  793, 128

\bibitem[{{van den Eijnden} et~al.(2018{\natexlab{a}}){van den Eijnden},
  {Degenaar}, {Russell}, {Wijnand s}, {Miller-Jones}, {Sivakoff} \&
  {Hern{\'a}ndez Santisteban}}]{vandeneijnden2018_swj0243}
{van den Eijnden} J., {Degenaar} N., {Russell} T.~D., {Wijnand s} R.,
  {Miller-Jones} J.~C.~A., {Sivakoff} G.~R., {Hern{\'a}ndez Santisteban} J.~V.,
  2018{\natexlab{a}}, \nat, 562, 233

\bibitem[{{van den Eijnden} et~al.(2018{\natexlab{b}})}]{vandeneijnden2018_gx}
{van den Eijnden} J., {Degenaar} N., {Russell} T.~D., {Miller-Jones} J.~C.~A.,
  {Wijnands} R., {Miller} J.~M., {King} A.~L., {Rupen} M.~P.,
  2018{\natexlab{b}}, \mnras, 474, L91

\bibitem[{{van den Eijnden} et~al.(2018{\natexlab{c}})}]{vandeneijnden2018_her}
{van den Eijnden} J., {Degenaar} N., {Russell} T.~D., {Miller-Jones} J.~C.~A.,
  {Wijnands} R., {Miller} J.~M., {King} A.~L., {Rupen} M.~P.,
  2018{\natexlab{c}}, \mnras, 473, L141

\bibitem[{{van den Eijnden} et~al.(2018{\natexlab{d}})}]{vandeneijnden2018_igr}
{van den Eijnden} J. et~al., 2018{\natexlab{d}}, \mnras, 475, 2027

\bibitem[{{van den Eijnden} et~al.(2019)}]{vandeneijnden2019_reb}
{van den Eijnden} J., {Degenaar} N., {Russell} T.~D., {Hern{\'a}ndez
  Santisteban} J.~V., {Wijnands} R., {Miller-Jones} J.~C.~A., {Rouco Escorial}
  A., {Sivakoff} G.~R., 2019, \mnras, 483, 4628

\bibitem[{{van der Meer} et~al.(2004){van der Meer}, {di Salvo}, {Kaper},
  {M{\'e}ndez} \& {van der Klis}}]{vandermeer04}
{van der Meer} A., {di Salvo} T., {Kaper} L., {M{\'e}ndez} M., {van der Klis}
  M., 2004, Nuclear Physics B Proceedings Supplements, 132, 624

\bibitem[{{van Loo}(2007)}]{vanloo2007}
{van Loo} S., 2007, {Non-Thermal Radio Emission from O Stars: Binary Versus
  Single}, Astronomical Society of the Pacific Conference Series, Vol. 367. p.
  187

\bibitem[{{Varun} et~al.(2019){Varun}, {Maitra}, {Pragati}, {Harsha} \&
  {Biswajit}}]{varun19}
{Varun}, {Maitra} C., {Pragati} P., {Harsha} R., {Biswajit} P., 2019, \mnras,
  484, L1

\bibitem[{{Verner} et~al.(1996){Verner}, {Ferland}, {Korista} \&
  {Yakovlev}}]{verner96}
{Verner} D.~A., {Ferland} G.~J., {Korista} K.~T., {Yakovlev} D.~G., 1996, \apj,
  465, 487

\bibitem[{{Weiler} \& {Panagia}(1978)}]{weiler78}
{Weiler} K.~W., {Panagia} N., 1978, \aap, 70, 419

\bibitem[{{Weston} et~al.(2016{\natexlab{a}})}]{weston16a}
{Weston} J.~H.~S. et~al., 2016{\natexlab{a}}, \mnras, 457, 887

\bibitem[{{Weston} et~al.(2016{\natexlab{b}})}]{weston16b}
{Weston} J.~H.~S. et~al., 2016{\natexlab{b}}, \mnras, 460, 2687

\bibitem[{{White} et~al.(1976){White}, {Mason}, {Sanford} \&
  {Murdin}}]{white76}
{White} N.~E., {Mason} K.~O., {Sanford} P.~W., {Murdin} P., 1976, \mnras, 176,
  201

\bibitem[{{White} et~al.(1980){White}, {Pravdo}, {Becker}, {Boldt}, {Holt} \&
  {Serlemitsos}}]{white80}
{White} N.~E., {Pravdo} S.~H., {Becker} R.~H., {Boldt} E.~A., {Holt} S.~S.,
  {Serlemitsos} P.~J., 1980, \apj, 239, 655

\bibitem[{{Wijnands}(2008)}]{wijnands08}
{Wijnands} R., 2008, in R.M. {Bandyopadhyay}, S.~{Wachter}, D.~{Gelino}, C.R.
  {Gelino}, eds, A Population Explosion: The Nature \& Evolution of X-ray
  Binaries in Diverse Environments. American Institute of Physics Conference
  Series, Vol. 1010, pp. 382--386

\bibitem[{{Wijnands} \& {Degenaar}(2016)}]{wijnands16}
{Wijnands} R., {Degenaar} N., 2016, \mnras, 463, L46

\bibitem[{{Williams} et~al.(1997){Williams}, {Dougherty}, {Davis}, {van der
  Hucht}, {Bode} \& {Setia Gunawan}}]{williams97}
{Williams} P.~M., {Dougherty} S.~M., {Davis} R.~J., {van der Hucht} K.~A.,
  {Bode} M.~F., {Setia Gunawan} D.~Y.~A., 1997, \mnras, 289, 10

\bibitem[{{Wilms} et~al.(2000){Wilms}, {Allen} \& {McCray}}]{wilms00}
{Wilms} J., {Allen} A., {McCray} R., 2000, \apj, 542, 914

\bibitem[{{Wilson-Hodge} et~al.(2018)}]{wilson18}
{Wilson-Hodge} C.~A. et~al., 2018, \apj, 863, 9

\bibitem[{{Wright} \& {Barlow}(1975)}]{wright75}
{Wright} A.~E., {Barlow} M.~J., 1975, \mnras, 170, 41

\bibitem[{{Wright} et~al.(1988){Wright}, {Cropper}, {Stewart}, {Nelson} \&
  {Slee}}]{wright88}
{Wright} A.~E., {Cropper} M., {Stewart} R.~T., {Nelson} G.~J., {Slee} O.~B.,
  1988, \mnras, 231, 319

\bibitem[{{Younes} et~al.(2015)}]{younes15}
{Younes} G. et~al., 2015, \apj, 804, 43

\bibitem[{{Zhong} \& {Wang}(2011)}]{zhong11}
{Zhong} J., {Wang} Z., 2011, \apj, 729, 8

\end{thebibliography}
